\documentclass[fleqn,usenatbib]{mnras}

\usepackage{newtxtext,newtxmath}

\usepackage[T1]{fontenc}
\usepackage{ae,aecompl}
\usepackage[toc,page]{appendix}
\usepackage{enumitem}

\usepackage{graphicx}	
\usepackage{amsmath}	
\usepackage{tabularx}
\usepackage[dvipsnames]{xcolor}
\usepackage{ulem}
\newcommand{\kms}{\mbox{$\>{\rm km\, s^{-1}}$}}
\newcommand{\pc}{\>{\rm pc}}
\newcommand{\kpc}{\mbox{$\>{\rm kpc}$}}
 
\newcommand{\kmsk}{\mbox{$\>{\rm kpc\, km\, s^{-1}}$}}
\newcommand{\Gyr}{\mbox{$\>{\rm Gyr}$}}
\newcommand{\Myr}{\mbox{$\>{\rm Myr}$}}

\newcommand{\Msun}{\>{\rm M_{\odot}}}

\newcommand\degrees{^\circ}
\newcommand\gaia{{\it Gaia}}
\newcommand{\avg}[1]{\mbox{$\overline{#1}$}}
\newcommand{\avgbin}[1]{\mbox{$\left<{#1}\right>$}}

\newcommand{\ie}{{\it i.e.}}

\newcommand{\tf}{\mbox{$t_\mathrm{form}$}}

\newcommand{\thetaf}{\mbox{$\theta_\mathrm{form}$}}
\newcommand{\zf}{\mbox{$z_\mathrm{form}$}}
\newcommand{\Rmin}{\mbox{$R_\mathrm{min}$}}

\newcommand{\Rmax}{\mbox{$R_\mathrm{max}$}}
\newcommand{\zmin}{\mbox{$z_\mathrm{min}$}}
\newcommand{\zmax}{\mbox{$z_\mathrm{max}$}}
\newcommand{\Rpeak}{\mbox{$R_\mathrm{peak}$}}
\newcommand{\zpeak}{\mbox{$z_\mathrm{peak}$}}

\newcommand{\thetae}{\mbox{$\theta_\mathrm{end}$}}
\newcommand{\thetajg}{\mbox{$\theta_{L,\,\mathrm{gas}}$}}
\newcommand{\re}{\mbox{$r_\mathrm{end}$}}
\newcommand{\thetaj}{\mbox{$\theta_{L}$}}
\newcommand{\Dthetaj}{\mbox{$\dot{\avg{\theta_{L}}}$}}
\newcommand{\thetajN}{\mbox{$\theta_{\overline{L}}$}}
\newcommand{\DthetajN}{\mbox{$\dot{\theta}_{\overline{L}}$}}
\newcommand{\tform}{\mbox{$t_\mathrm{form}$}}
\newcommand{\ttilt}{\mbox{$\tau_\mathrm{tilt}$}}
\newcommand{\dt}{\mbox{$\delta t$}}
\newcommand{\tpm}{\mbox{$\tau_\mathrm{pm}$}}

\newcommand{\rform}{\mbox{$r_\mathrm{form}$}}
\newcommand{\rxyf}{\mbox{$R_\mathrm{form}$}}
\newcommand{\dentropy}{\mbox{$\dot{\hat{S}}(t)$}}
\newcommand{\entropy}{\mbox{$\hat{S}(t)$}}

\newcommand{\Spm}{\mbox{$S_\mathrm{pm}$}}
\newcommand{\taupm}{\mbox{$\tau_\mathrm{pm}$}}

\title[Warp star settling]{How stars formed in warps settle into (and contaminate) thick discs}

\author[Khachaturyants et al.]{Tigran Khachaturyants$^{1}$\thanks{E-mail: astrotkh@gmail.com}, Leandro {Beraldo e Silva}$^{1}$,
	Victor P. Debattista$^{1}$ \\
	$^{1}$ Jeremiah Horrocks Institute, University of Central Lancashire, Preston, PR1 2HE, UK \\
}

\date{Accepted XXX. Received YYY; in original form ZZZ}

\pubyear{2021}

\begin{document}
	\label{firstpage}
	\maketitle
	
	\begin{abstract}
In recent years star formation has been discovered in the Milky Way's warp. These stars formed in the warp (warp stars) must eventually settle into the plane of the disc. We use an $N$-body$+$smooth particle hydrodynamics model of a warped galaxy to study how warp stars settle into the disc. By following warp stars in angular momentum space, we show that they first tilt to partially align with the main disc in a time scale of $\sim1 \Gyr$. Then, once differential precession halts this process, they phase mix into an axisymmetric distribution on a time scale of $\sim 6 \Gyr$. The warp stars end up contaminating the geometric thick disc. Because the warp in our fiducial simulation is growing, the {\it warp stars} settle to a distribution with a negative vertical age gradient as younger stars settle further from the mid-plane. While vertically extended, warp star orbits are still nearly circular and they are therefore subject to radial migration, with a net movement inwards. As a result warp stars can be found throughout the disc. The density distribution of a given population of warp stars evolves from a torus to an increasingly centrally filled-in density distribution. Therefore we argue that, in the Milky Way, warp stars should be found in the Solar Neighbourhood. Moreover, settled warp stars may constitute part of the young flaring population seen in the Milky Way's outskirts.
	\end{abstract}
	
	

\section{Introduction}
\label{sec:Introduction}

Warps are common features in disc galaxies, both in their HI gas \citep{sancisi, bosma, garciaruiz02} and, to a lesser extent, in their stars \citep{reshetnikov}. In the Milky Way (MW), a warp in the H{\sc i} has long been known \citep{Kerr, weaver_williams, levine06, Kalberla}, while subsequently a stellar warp was also observed \citep{Djorgovski+89,Porcel+95,Freudenreich+98,Drimmel+01}. The extent of the stellar warp, traced by red clump (RC) stars, in the MW was measured by \cite{rcwarp} and found to reach a maximum height of $|z|\sim1.5 \kpc$ at $R\leq14 \kpc$ on both sides of the disc.

The cause of warping in galactic discs is still not definitively established, with several mechanisms proposed \citep[see the reviews of][]{binney92,kuijken+01, Sellwood+13}. These include tidal interactions, direct gas accretion, and disc-halo interactions. In the MW, tidal interaction with the Large Magellanic Cloud (LMC) and the Sagittarius dwarf galaxy (Sgr) have been considered possible sources of the warping \citep{Weinberg98, 1999MNRAS.303L...7J, bailin03, 2011Natur.477..301P, 2013MNRAS.429..159G, 2018MNRAS.473.1218L}. Alternatively, misaligned cold gas accretion has been proposed \citep{1989MNRAS.237..785O} and found in cosmological simulations, particularly in MW-like models \citep{roskar+10,Stewart+11,vandeVoort+15,Gomez+17,Starkenburg+19,Duckworth+20}, to be the cause of galactic warps in a significant number of cases. In the TNG100 run of the IllustrisTNG cosmological simulation suite, \cite{Semczuk+20} showed that $16\%$ of galaxies had S-shape warps and only a third of them had their warps tidally induced by other galaxies. The hot gas corona is a component that is thought to encompass galaxies such as the MW and may be the main source of accreting gas. Moreover, cosmological simulations have long shown that the angular momentum of hot gas coronae is usually misaligned with the halo and stellar discs embedded within them \citep{bosch+02,Chen+03,Bailin+05,Sharma+05,roskar+10,Gomez+17,earp+19}. Additionally, this misaligned infall of cooling gas has been shown in isolated \citep{Debattista+15} and cosmological simulations \citep{earp+19} to tilt the stellar disc and maintain its misalignment with the halo. Such misalignments between halos and their embedded stellar discs have been inferred in large extragalactic surveys \citep[e.g.][]{Wang+08,Wang+10,Nierenberg+11,Li+13} and proposed to be occurring in the MW \citep{Debattista+13}. Misaligned gas accretion can enable the persistence of warps as gas is continuously accreted onto the outskirts of discs. 

Evidence of cold gas accretion has been inferred in external galaxies via large complexes of H{\sc i} at the outskirts of spiral galaxies \citep{Fraternali, 2008A&ARv..15..189S, westmeier, Zschaechner}. Cosmological simulations have shown that highly misaligned cold accretion along filaments can cause polar-ring galaxies \citep{2006ApJ...636L..25M}, and measurements of the metallicity of polar rings supports this scenario \citep{Spavone+10}. In the case of the MW, the gas accretion scenario not only provides an explanation for the origin of the warp but also for the near-constant star formation rate \citep{1980ApJ...242..242T, 2000MNRAS.318..658B}. In principle, both the gas accretion and tidal interaction mechanisms of warp formation can be active in any one galaxy. However, direct observational evidence of ongoing gas accretion is hard to obtain in the MW.

Using data from \gaia-DR2 \citep{2018A&A...616A...1G} and from the {\it Wide-field Infrared Survey Explorer} ({\it WISE}) catalogue of periodic variables \citep{2018ApJS..237...28C}, \cite{chen} compiled a sample of classical Cepheids, finding that the Galactic warp is also traced by these stars. Of all the warp tracers they considered (including dust, pulsars, and red clump stars), they found that the H{\sc i} gas and Cepheids have the most similar distributions. They showed that the disparity between the H{\sc i} and other warp tracers is significant in both phase and amplitude, while the Cepheids appear to mirror the H{\sc i} warp, implying that they formed in-situ in the warp. Evidence of star formation in the outskirts of galactic discs has also been inferred in external galaxies via UV-bright stellar complexes \citep{2005ApJ...619L..79T, zaritsky+07, zaritsky, mondal}. These stellar complexes are observed far outside the optical discs, which are usually warped, and in one case \citep{2005ApJ...619L..79T} they were directly associated with the warped H{\sc i} disc. In NGC~4565, \cite{Radburn+14} used {\it Hubble Space Telescope} resolved stellar populations to show that the H{\sc i} warp is indeed traced by young (age $< 600 \Myr$) populations, while older ($> 1\Gyr$) populations are symmetrically distributed around the mid-plane. The results of \cite{Radburn+14} and of \cite{chen} in the MW support the view that some star formation occurs in warps, and that older stars do not trace the gas warp. These results therefore suggest that the presence of young stars in the warp is not due to bending waves, which would produce a similar warp signature in all stellar populations. Understanding where stars that formed in the warp end up can shed light on the formation and evolution of the warp, and consequently on the evolution of the MW as a whole.

\cite{roskar+10} presented a fully cosmological simulation of a Milky Way-like galaxy in which its hot gaseous corona has angular momentum misaligned with that of the disc. The gas cools and sinks toward the stellar disc, forming a warp. Stars formed in this warp settle into the disc and populate the geometric thick disc \citep[see Fig. 13 in][]{roskar+10}. In this paper, we use a warped $N$-body+SPH (Smooth Particle Hydrodynamics) simulation to investigate, in further detail, the settling of stars formed in the gas accreting along a warp. 
The paper is organised as follows: in Section~\ref{section:Simulation} we describe the warped simulation, the pre-processing of the simulation snapshots, and how stars formed in the warp (hereafter, `warp stars') are defined in the simulation. In Section~\ref{section:dyn_evo} we analyse different warp populations separated by their time of formation and track the changes of their angular momenta throughout the simulation's evolution. In Section~\ref{section:disc_structure} we turn our attention to the resulting density distribution of warp stars in the disc. In Section~\ref{section:discussion} we present our conclusions, before ending with a summary of our results.
	
\section{Simulation}
\label{section:Simulation}
\subsection{Fiducial simulation}
The warped simulation is produced via the method of \citet{vpd2015}, which constructs triaxial dark matter models with gas angular momentum misaligned with the principal axes of the halo. The resulting misalignment mirrors that found in cosmological simulations \citep{bosch+02,roskar+10,Gomez+17,earp+19}. 
As shown by \citet{AumerWhite2013}, inserting a rotating gas corona into a non-spherical dark matter halo leads to a substantial loss of gas angular momentum. To produce a non-spherical system, we use adiabatic gas in merging haloes.  We merge two identical spherical Navarro–Frenk–White (NFW) \citep{NFW1996} dark matter haloes, each with a co-spatial gas corona comprising 10 per cent of the total mass.
	
Each dark matter halo has a mass $M_{200} = 8.7 \times 10^{11}\Msun$ and virial radius $r_{200} = 196 \kpc$. The gas is in pressure
equilibrium within the global potential. Gas velocities are initialised to give a spin parameter of $\lambda = 0.16$ \citep{Bullock+2001}, with specific angular momentum $j \propto R$, where $R$ is the cylindrical radius. Both the dark matter halo and the gas corona are comprised of $10^6$ particles. Gas particles initially have masses $1.4 \times 10^5 \Msun$ and softening $\epsilon = 20 \pc$, while dark matter particles come in two mass flavours ($10^6 \Msun$ and $3.6 \times 10^6 \Msun$ inside and outside $200 \kpc$, respectively) and $\epsilon = 100 \pc$.  The two halos are placed $500 \kpc$ apart and approach each other head-on at $100 \kms$.  If the direction of the separation vector (and the relative velocity) is the $x$-axis, we tilt one of the halos about the $y$-axis so that the final system will be prolate with long axis along the $x$-axis and a gas angular momentum tilted with respect to the axes of the halo.
	
This simulation is evolved with the smooth particle hydrodynamics code {\sc gasoline} \citep{Wadsley+2004}, with a base time-step $\dt = 10 \Myr$ which, for individual particles, is refined such that each particle satisfies the condition $\dt = \dt/2^n < \eta
\sqrt{\epsilon/a_g}$, where $a_g$ is the acceleration at the particle's current position, with $\eta = 0.175$, and the opening angle of the tree code calculation set to $\theta = 0.7$.
	
At the end of this setup, the dark matter halo has $r_{200} = 238 \kpc$ and $M_{200} = 1.6\times10^{12}\Msun$, while the gas has $\lambda = 0.11$.  At this point we turn on gas cooling, star formation and stellar feedback using the blastwave prescriptions of
\citet{Stinson+2006}. Gas particles form stars with efficiency 0.1 if a gas particle has number density $n > 1$ cm$^{-3}$, temperature $T < 15,000$ K and is part of a converging flow. We refer to this density criterion as the \textit{star formation threshold}, and, in our fiducial simulation, this threshold is relatively low, which increases the total star formation in the warp, providing us with a statistically significant number of warp stars to follow as they settle into the disc. Conversely the amount of star formation in the warp is higher than would be expected in real galaxies, including the Milky Way.
	
Star particles form with an initial mass of 1/3 that of the initial gas particle masses, which at our resolution corresponds to $4.6 \times 10^4\Msun$. The star particles all have $\epsilon = 20 \pc$. Once the mass of a gas particle drops below $1/5$ of its initial mass, the remaining mass is distributed amongst the nearest
neighbouring gas particles, leading to a decreasing number of gas particles. Each star particle represents an entire stellar population with a Miller–Scalo \citep{MillerScalo1979} initial mass function. The evolution of star particles includes asymptotic giant branch stellar
winds and feedback from Type II and Type Ia supernovae, with their energy injected into the interstellar medium (ISM). Each supernova
releases $10^{50}$ erg into the ISM. The time-step of gas particles also satisfies the condition $\dt_{gas} = h \eta_{courant}/[(1 +
\alpha)c + \beta \mu_{max}]$, where $h$ is
the SPH smoothing length, $\eta_{courant} = 0.4$, $\alpha = 1$ is the shear coefficient,
$\beta = 2$ is the viscosity coefficient and $\mu_{max}$ is described in \citet{Wadsley+2004}. The SPH kernel uses the 32 nearest neighbours. Gas cooling takes into account the gas metallicity using the prescriptions of \citet{Shen+2010}; in order to prevent the cooling from dropping below our resolution, we set a pressure floor on gas particles of $p_{floor} = 3G\epsilon^2\rho^2$, where $G$ is Newton’s gravitational constant, and $\rho$ is the gas particle’s density \citep{Agertz+2009}.

\begin{figure}
\includegraphics[width=1\linewidth]{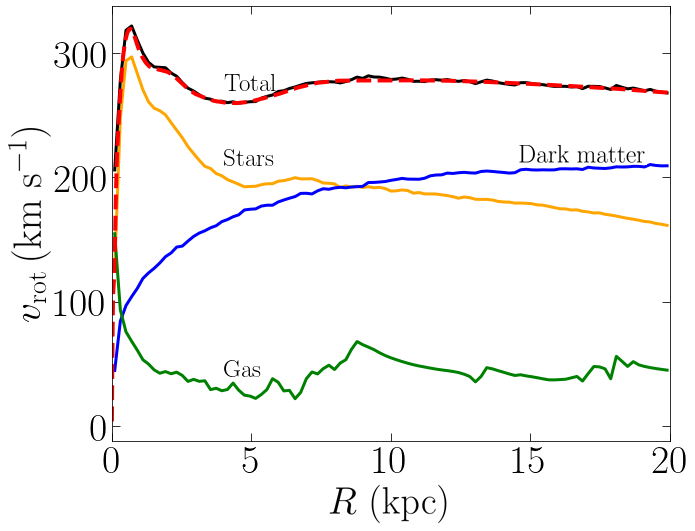}  

	\caption{
Rotation curve of the simulation at $12 \Gyr$. Solid lines represent the contribution of each galactic component (annotated above each curve) along with a total rotation curve, while the rotation curve of the interpolated total potential is represented by the dashed red line.
}
\label{fig:rot_curve}
\end{figure}

In the following, we refer to $t = 0$ as the time gas cooling and star formation are switched on. Fig.~\ref{fig:rot_curve} shows the rotation curve of the simulation at $t=12 \Gyr$ (last timestep). We interpolate the potential of the simulation with the {\sc agama} software library \citep{agama}, using a single multipole approximation for the stellar, gas, and dark matter particles combined. The rotation curve of the interpolated potential is presented in Fig.~\ref{fig:rot_curve} as a dashed red line. As in the MW, we observe a relatively flat rotation curve. 

\subsection{Supplemental simulation}
\label{subsection:supp_sim}
To demonstrate that the low star formation threshold and, therefore, increased star formation in the warp, do not affect the main conclusions of this work, we perform the same analysis on a second, supplemental simulation. The supplemental simulation has the same initial conditions as the fiducial simulation, however, it uses sub-grid physics prescriptions that create less favourable conditions for star formation in low-density regions, such as the warp. Firstly, the gas cooling in the supplemental simulation does not take into account the gas metallicity, meaning that gas cools less efficiently in the warp. Secondly, more energy from the stellar feedback is coupled to the gas than in the fiducial simulation, with supernovae releasing $4\times10^{50}\rm{erg}$ into the interstellar medium. Lastly, the star formation threshold in the supplemental simulation is higher by two orders of magnitude than in the fiducial simulation with gas particles only forming stars when their number density exceeds $100~\rm{cm}^{-3}$. We present the results of the supplemental simulation analysis in Appendix~\ref{sec:appendix}.

\subsection{Pre-processing the simulation}
\label{subsection:preproc}
The snapshots of the fiducial and supplemental simulations are processed through our custom {\sc Python} library suite that centres the galactic disc and then rotates it into the $(x,y)$ plane based on the angular momentum of the inner stellar disc. The inner stellar disc is defined by a radial upper limit of $r\leq5 \kpc$. We compute the angular momentum of the misaligned cold gas ($T_{gas}<50,000$K) at the outer edge of the galactic disc ($15 \leq R/\kpc \leq 20$) to determine the orientation of the gas warp. Each snapshot is rotated by the cylindrical angle of the warp's angular momentum, $\varphi_{_{L}}$, so that the warp's major axis is on the $x$-axis and, consequently, the line of nodes is on the $y$-axis. The disc is finally rotated by $180\degrees$ about the $y$-axis, resulting in a negative angular momentum, which matches the sense of rotation and warp orientation of the Milky Way \citep{chen}. As a result of these rotations, the south side of the gas warp (below the mid-plane) is along the positive $x$-axis. This orientation is implied in any plots throughout this paper.

\begin{figure}
\includegraphics[width=1\linewidth]{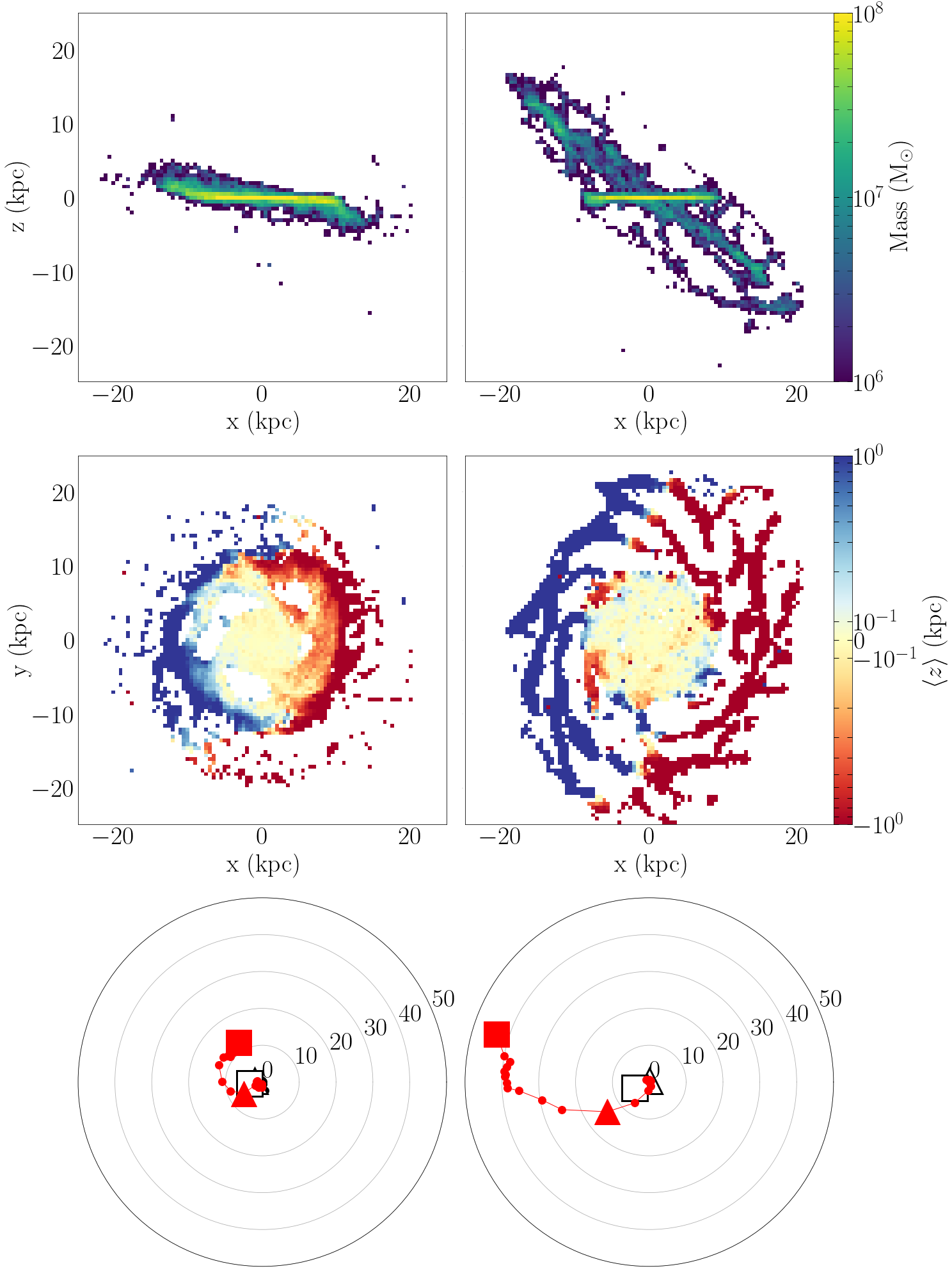}  
\caption{The structure of the gas warp at $2 \Gyr$ (left column) and $12 \Gyr$ (right column). Top row: The edge-on column density distribution of cold gas (T$\leq 50,000$ K) in the simulation. In the span of $10 \Gyr$, the gas warp can be traced to larger $R$ (from $R \sim 13 \kpc$ to $R\sim 20 \kpc$) and $|z|$ (from $\left| z \right| \sim 5 \kpc$ up to $\left| z \right| \sim 20 \kpc$). Middle row: the face-on mean height, $\left<z\right>$, distribution of cold gas (T$\leq 50,000$ K) in the simulation.
Bottom row: The Briggs figures for the cold gas (red) and stellar (black) discs. There are two distinct markers that show values at $R=10 \kpc$ (triangle marker) and at $R=20 \kpc$ (square marker). The Briggs figures show that the gas disc becomes significantly more warped between the two times, while the stellar disc is initially slightly warped ($\thetaj\sim 2^{\circ}$ at $R=10\kpc$) and becomes even less so by the end ($\thetaj\sim 0.3^{\circ}$ at $R=10\kpc$).}
\label{fig:warp_briggs}
\end{figure}
	
\begin{figure}
\includegraphics[width=1\linewidth]{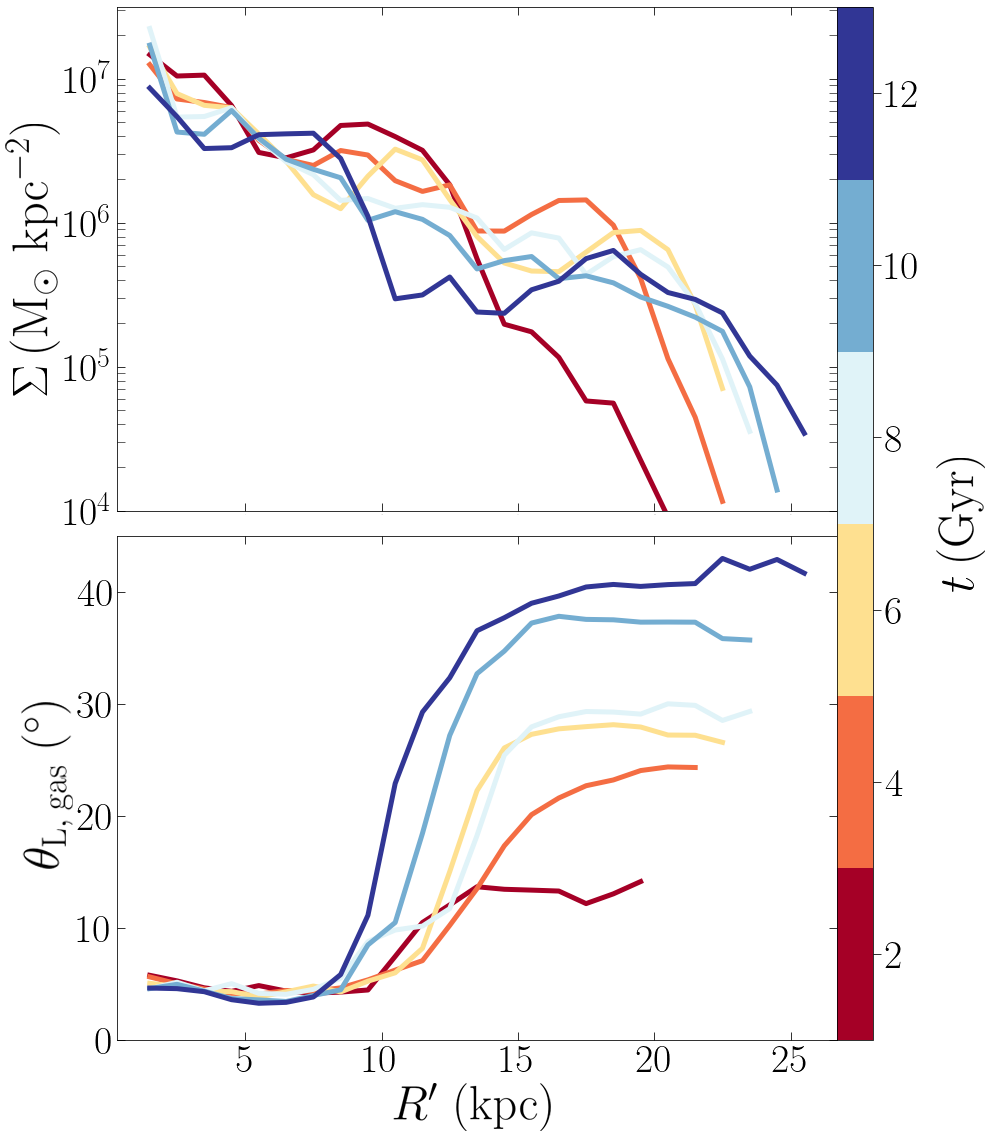}  
	\caption{Profiles of the surface density, $\Sigma$, (top) and $\thetajg$ (bottom) in the cold gas at different times (colour), where $R^{\prime}$ is defined as the cylindrical radius in the cold gas plane at each annulus. The gas warp grows horizontally and becomes more inclined with time.}
\label{fig:inc_evo}
\end{figure}	

The extent of the simulation's gas warp is shown in the top row of Fig.~\ref{fig:warp_briggs} where we present the edge-on column density of cold gas at $2 \Gyr$ (left) and $12 \Gyr$ (right). In the span of $10 \Gyr$ the warp grows significantly in radial extent, and becomes more inclined relative to the disc. To quantify the inclination and orientation of the warp, for each component (stars and cold gas), we measure the spherical angles $\thetaj$ (polar) and $\phi_L$ (azimuthal) between their angular momenta, measured within spherical annuli, and the inner stellar disc. As the angular momentum vector of the stellar disc has been realigned along the $z$-axis for all times, $\thetaj$ and $\phi_L$ are simply:
\begin{equation}
\label{eq:phi_L}
\phi_{L}=\arctan(L_y/L_x),
\end{equation}
and
\begin{equation}
\label{eq:thet_j}
\theta_{L}=\arccos(L_z/|L|),
\end{equation}
where $L_x$, $L_y$, $L_z$, and $|L|$ are the three Cartesian components and magnitude of the angular momentum, respectively.
In the bottom row of Fig.~\ref{fig:warp_briggs} we present Briggs figures \citep{Briggs90} for the stellar (black) and cold gas (red) discs at $2 \Gyr$ (left) and $12 \Gyr$ (right), where the triangle (square) marker represents $R=10 \kpc$ ($R=20 \kpc$). Briggs figures are cylindrical polar plots where $\thetaj$ and $\phi_L$ are represented by the polar $r$ and $\phi$ coordinates, respectively. The $\thetaj$ and $\phi_L$ angles are calculated for the mean angular momentum vector in each bin of a cylindrical grid with $0 \leq R / \kpc \leq 20 \kpc$ and $\Delta R=1 \kpc$. The cold gas warp grows significantly over the $10 \Gyr$  interval, while the stellar warp decreases in extent, and then flattens over the same time interval.
In Fig.~\ref{fig:inc_evo} we show the profiles of the surface density, $\Sigma$, (top) and of $\thetajg$ (bottom) for the cold gas disc at different times (colour), where $R^{\prime}$ is defined as the cylindrical radius in the cold gas plane at each annulus. Over the model's evolution, the inclination of the cold gas warp beyond $10 \kpc$ increases by a factor $\sim 4$, reaching $\thetaj\sim40\degrees$.
The warp also grows in mass and size as the $\Sigma$ profile increases beyond $15 \kpc$ and reaches $R^{\prime} \sim 25 \kpc$ by the end of the simulation.

\subsection{Defining warp stars}
\label{subsection:warp_stars}
	
\begin{figure}
\includegraphics[width=1\hsize]{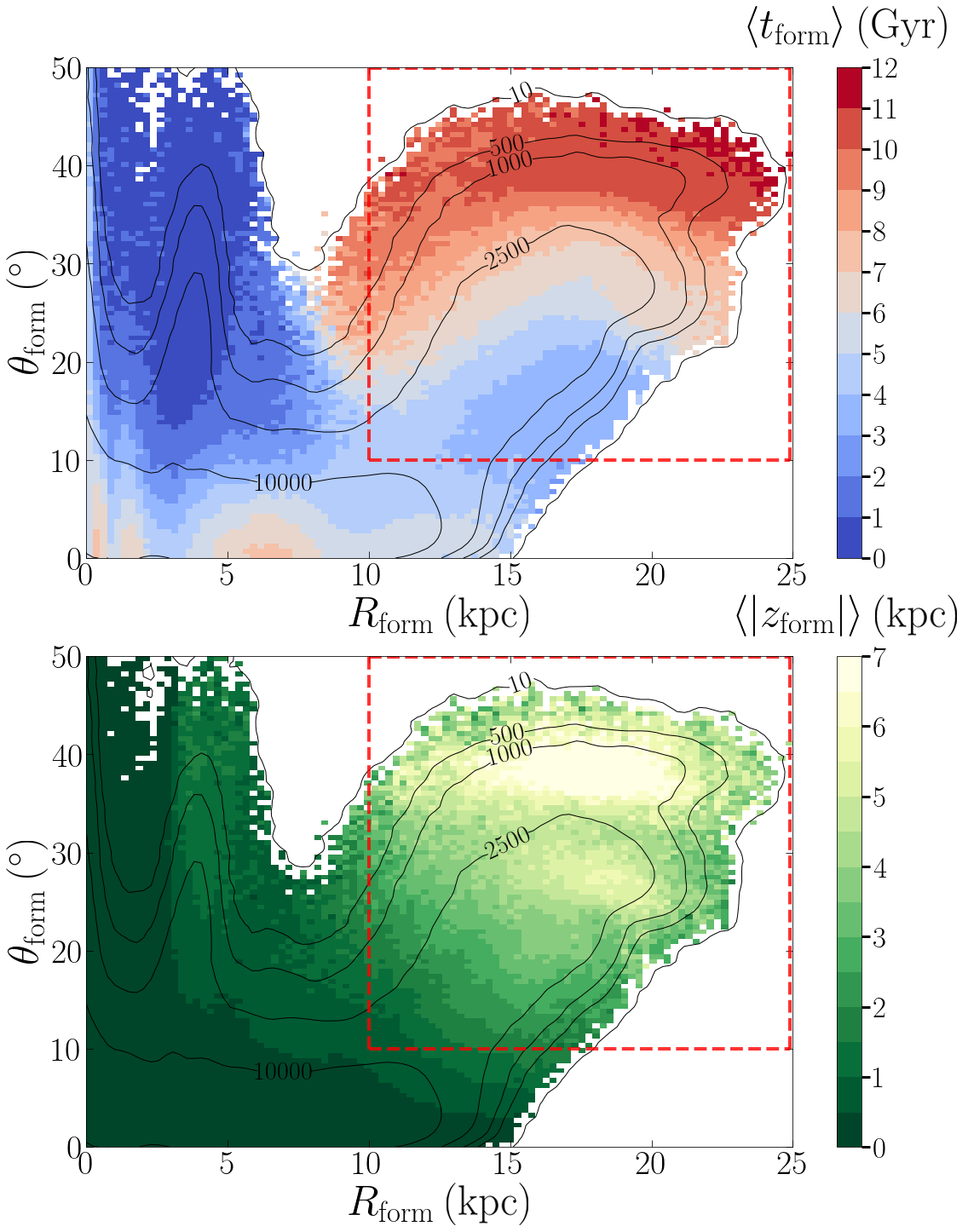}  
	\caption{The distribution of stars in the $\rxyf$-$\thetaf$ space (formation space), coloured by the mean time of formation (top) and by the mean absolute formation height, $\avgbin{|\zf|}$, (bottom). Bins that contain less than 10 stellar particles are not shown. The black lines show the number counts in the formation space for both panels. We define stars formed in the warp as those with $\thetaf\geq 10\degrees$ and $\rxyf \geq 10 \kpc$, the "tail-like" region outlined by the red square. A population of stars that was formed in an early, transient warp at low radii ($\rxyf \leq 7 \kpc$) and high inclinations relative to the disc ($\thetaf\geq 10\degrees$) is not included in our warp star population.}
\label{fig:formation_space}
\end{figure}

We record the phase-space coordinates and time at formation, $\tform$, for every star in the simulation. The phase-space coordinates need to be centred and reoriented relative to the disc at their respective $\tform$. Using our \textit{\sc Python} library suite, we create an interpolating function that takes into account the centre of mass and angular momentum vector of the galactic disc at each $100 \Myr$ saved snapshot. We calculate the location of the centre of mass and orientation of the galactic disc for each star by interpolating to their individual $\tform$. This procedure gives the formation location in galaxy-centred coordinates and the inclination of the star's angular momentum at formation relative to that of the galactic disc ($\thetaf$). For all stars, we extract the cylindrical galactocentric formation radius, $\rxyf$, and the angular momentum inclination, $\thetaf$, which we define as
\begin{equation}
\label{eq:thet_f}
\thetaf=\arccos{\frac{L_{z,\rm{form}}}{|L_{\rm{form}}|}},
\end{equation}
where $L_{z,\rm{form}}$ and $|L_{\rm{form}}|$ are the vertical component and magnitude of a star's angular momentum at formation, respectively. Throughout this work we define multiple different angles and use them in the analysis of warp populations; these angles and their respective equations are presented in Table~\ref{tab:thetas}.

\begin{table}
\centering
\resizebox{\columnwidth}{!}{%
{\renewcommand{\arraystretch}{1.4}
\begin{tabular}{cc}
\hline
Angle                 & Definition                                                                 \\ \hline
$\phi_L$              & The azimuth of the stellar angular momentum relative to the disc (Eq.~\ref{eq:phi_L})     \\
$\theta_L$            & The inclination of the stellar angular momentum relative to the disc (Eq.~\ref{eq:thet_j}) \\

$\theta_{\rm{form}}$  & The $\theta_L$ at formation (Eq.~\ref{eq:thet_f})         \\
\multicolumn{1}{l}{$\theta_{\rm{end}}$} & The $\theta_L$ at the last timestep (Eq.~\ref{eq:thetae})   \\
$\overline{\theta_L}$ & The average $\theta_L$ of stars in a mono-age population (Eq.~\ref{eq:avg_theta_L})\\
$\theta_{\overline{L}}$                 & The $\theta_L$ of a mono-age population's \avg{L} (Eq.~\ref{eq:theta_avg_L})\\ \hline
\end{tabular}%
}}
\caption{The angles defined and used throughout this work.}
\label{tab:thetas}
\end{table}

To identify warp stars, we plot the distribution of all stars in $\thetaf$-$\rxyf$ space (hereafter \textit{formation space}).
In Fig.~\ref{fig:formation_space} we present the distribution of the mean time of formation, $\avgbin{\tform}$, (top) and the mean absolute height of formation, $\avgbin{|\zf|}$, (bottom) in the formation space.  The "tail-like" region at $\rxyf > 10\kpc$ (outlined by a red square) is comprised of stars that formed at relatively high $|\zf|$, which increases with $\tform$. These stars are formed throughout the model's evolution starting from $2 \Gyr$ and lasting till the end of the simulation, at $12 \Gyr$. This population is highly inclined ($\thetaf>10\degrees$) and is formed on the outskirts of the disc; thus we define the primary warp population as stars with $\rxyf\geq10 \kpc$ and $\thetaf\geq 10 \degrees$. There are $\sim 6\times 10^5$ warp stars in the simulation and they comprise $18\%$ of all stars. The other significant populations that we observe are the in-situ main disc population ($\rxyf \leq 10 \kpc$ and $\thetaf \leq 10 \degrees$), and a "hump-like" region containing an old warp population ($2 \leq \rxyf / \kpc \leq 5$ and $\thetaf \geq 15 \degrees$).  This early warp population derives from a short-lived warp epoch when the model is still settling, and we therefore do not include it in our analysis of the warp. Neglecting this population does not change any of the following results.


\section{Dynamical evolution of warp populations}
\label{section:dyn_evo}

We study how warp stars settle into the disc by considering mono-age populations. Our goal is to unravel the mechanisms by which they settle and reach equilibrium within the main disc, the timescale for settling, and the (evolving) density distribution they settle to.


\subsection{Overall evolution}
\label{subsection:considerations}

\begin{figure*}
\includegraphics[width=.95\linewidth]{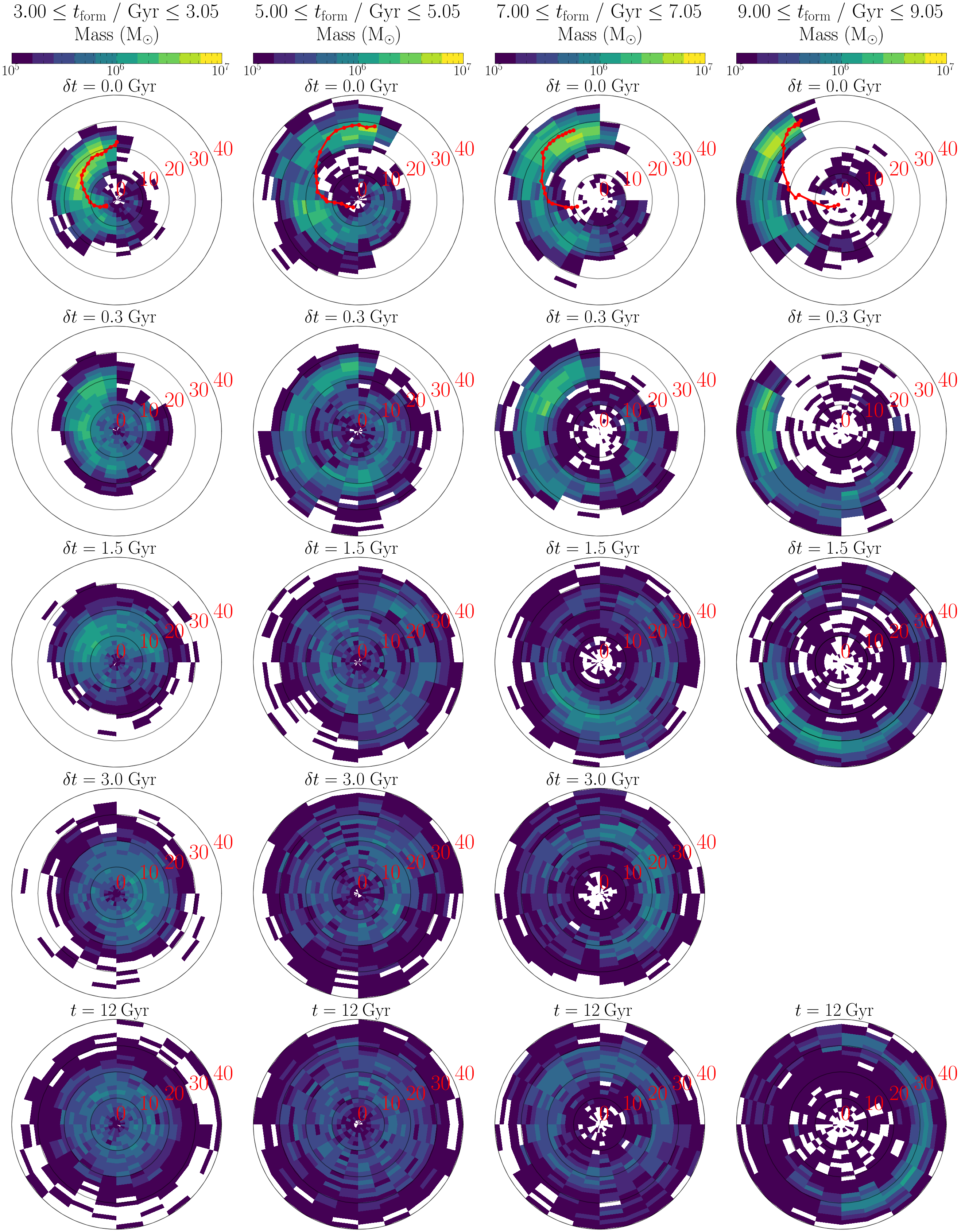}  

	\caption{Mass distribution in Briggs figures for 4 mono-age warp populations. (Briggs figures plot the spherical polar orientation angles of the angular momentum vector, $\thetaj$ and $\phi_L$, as the cylindrical polar $r$ and $\phi$ coordinates, respectively.) The populations are formed over $50 \Myr$ intervals (indicated at the top of each column) and we follow them at different later times, $\delta t$ (rows). The bottom row shows the populations at the end of the simulation, at $t=12 \Gyr$. At $\dt = 0 \Myr$ (first row), the red line represents the Briggs figure of the gas warp at the formation of each population. Some stars appear to start out with $\thetaj<10\degrees$, but this is due to them drifting already during the first $50\Myr$. In the two left columns, the initial $m=1$ LON spiral distributions phase mix into uniform distributions by the end of the simulation, whereas in the right two columns phase mixing is incomplete at $t=12\Gyr$.
} 
\label{fig:briggs_evo}
\end{figure*}
	
Fig.~\ref{fig:briggs_evo} presents the mass distribution in Briggs figures for 4 representative mono-age warp populations (columns) at various times after their formation, denoted by $\dt$. 
The Briggs figures provide a clear picture of how warp stars start out heavily inclined relative to the disc (outer regions in the diagrams) and end up phase-mixing into a homogeneous distribution.
All populations form along the gas warp, indicated by the solid red lines in the first row. The gas warp traces a leading spiral shape (the sense of disc rotation in these figures is clockwise), which is one of the characteristics of gas warps \citep{Briggs90}\footnote{We remind the reader that this is a spiral in the orientation of the angular momentum vector of different shells.  In coordinate space this represents a winding of the intersection of each annulus with the main plane of the galaxy, \ie\ the spiral can be thought of as the radial locus of the line-of-nodes (LON). For this reason, we will refer to this spiral as the LON spiral.}. Warp stars formed at different times have different ranges of $\thetaj$, with the earlier-forming population ($\tform = 3 \Gyr$) centred on $\thetaj=20\degrees$ and the later ($\tform = 9 \Gyr$) centred on $\thetaj=35\degrees$. The phase mixing of warp stars in $\phi_L$ is already visible $300 \Myr$ after formation for all 4 mono-age populations, as the spiral structure winds up. This winding represents the differential precession of different annuli of the chosen warp population. The higher the initial radius of formation, \rxyf, of the stars, the slower is the precession of the population, and the longer is the time required for the LON spiral to wind up. For instance, after $\dt= 1.5\Gyr$, the warp population formed at $t\simeq 3 \Gyr$ is well on its way to being uniform in $\phi_L$, whereas the warp populations forming at $t\simeq 9 \Gyr$ are considerably less wound up. By the end of the simulation, the later-forming populations have still not fully phase-mixed in $\phi_L$, as evident by the horseshoe distribution for the population formed at $\sim 9\Gyr$. The phase mixing indicates that the warp populations settle into nearly-axisymmetric discs or tori -- see Sec.~\ref{subsection:phasemixing}. They remain relatively thick, as can be seen by the large $\thetaj$ values of most of the stars, corresponding to stars which avoid having an angular momentum directed along the $z$-axis. 
	
A weaker evolution that can be discerned from the Briggs figures is a rapid early decline in the values of $\thetaj$.	This is easiest to see directly for the population formed at $3\Gyr$, but is present to different extents in all 4 populations. This process represents a tilting of each warp population. Lastly, the Briggs figures show that there is a tendency for some stars to move to larger \thetaj; we quantify this in Sec.~\ref{subsection:tilting} by analysing the difference between the \thetaj\ at formation and at the last timestep. In Sec.~\ref{subsection:migration} we demonstrate that the increase in \thetaj is caused by stars migrating to smaller radii, while preserving their vertical motions so that the net orbital plane of each star becomes more tilted. In the following subsections we study in greater detail the tilting of warp populations, their phase mixing and finally their radial migration.


\subsection{Orbital tilting}
\label{subsection:tilting}

\begin{figure*}
\includegraphics[width=1\linewidth]{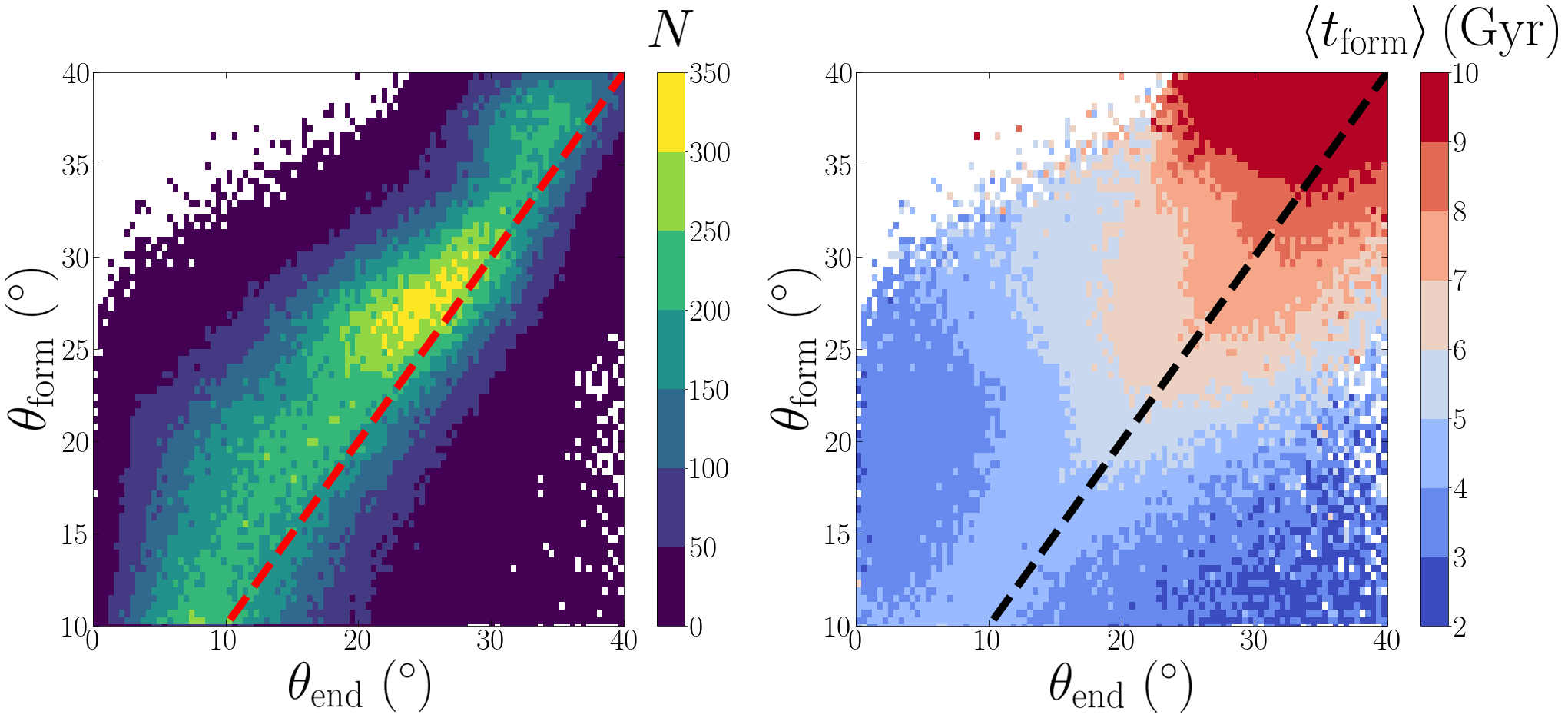}  

	\caption{Distribution of $\thetaf$ versus $\thetaj$ at the end of the simulation, $\thetae$, for individual warp stars formed before $10\Gyr$ coloured by the number (left) and the mean time of formation (right). The diagonal dashed line indicates $\thetaf=\thetae$ and it shows that $\sim27\%$ of warp stars experience an increase in their tilt, while most warp stars tilt to align with the galactic disc.
}
\label{fig:thetaf_v_thetaj}
\end{figure*}	

\begin{figure}
\includegraphics[width=1\linewidth]{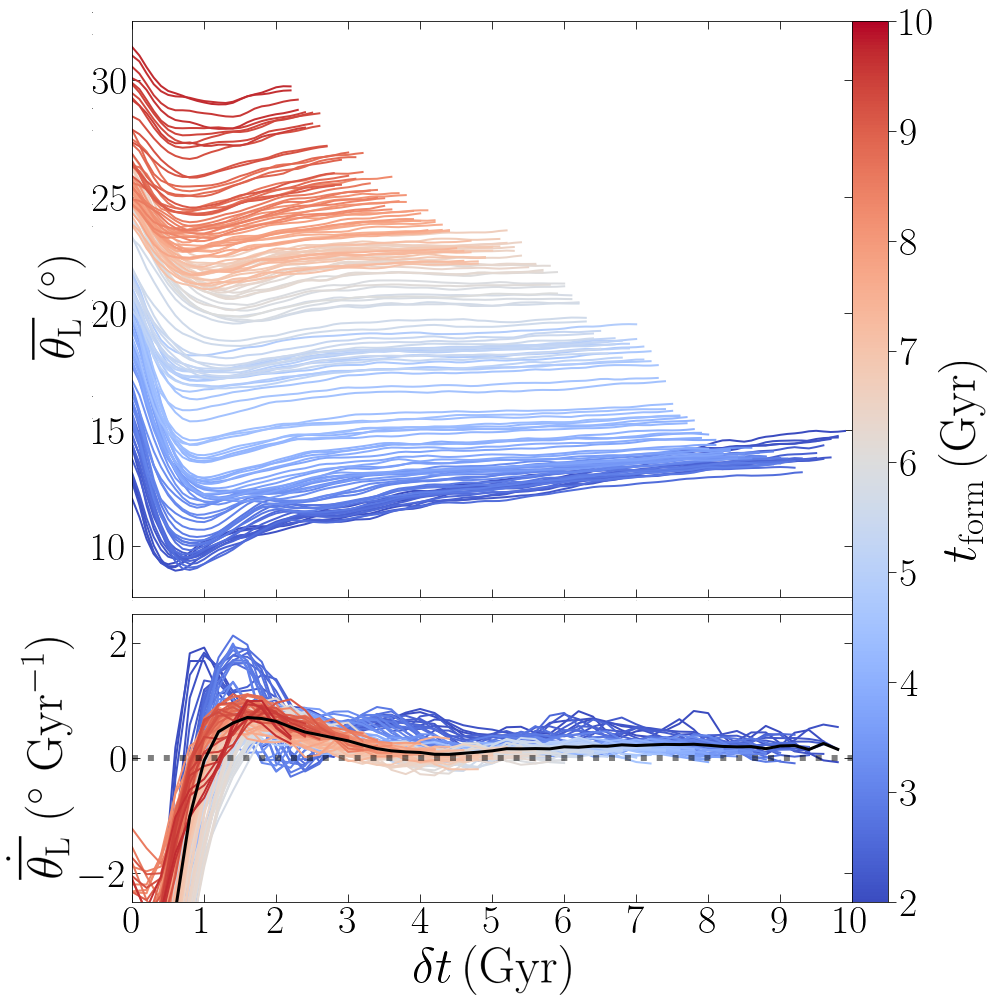}  
    	\caption{Top: evolution of the population-averaged $\thetaj$, $\avg{\thetaj}$, of all mono-age warp populations formed in the simulation before $\tform\leq 10 \Gyr$, where $\dt$ is the time since a population's formation. Each curve is coloured by $\tform$. A 1D Gaussian filter with a mask size of $w=0.5\Gyr$ and standard deviation of $\sigma=0.1\Gyr$ is applied to the evolution at each $\dt$. Bottom: evolution of the rate of change of $\avg{\thetaj}$, $\Dthetaj$, for the same mono-age warp populations. The rates of change are calculated from the smoothed evolution curves. The solid black line represents the median rate of change between all mono-age populations which has a tilting time of $\ttilt\sim0.9\Gyr$. The dotted horizontal line indicates $\Dthetaj=0\degrees\Gyr^{-1}$.
}
\label{fig:Mthetaj_evo}
\end{figure}

Fig.~\ref{fig:briggs_evo} suggested that warp populations reach lower $\thetaj$ as they realign with the galactic disc, which henceforth we will refer to as tilting. We now study the tilting of mono-age warp populations in more detail.
We start by showing that tilting is indeed taking place by comparing the \thetaf\ of all warp stars versus their \thetaj\ at the end of the simulation, \thetae, which we define as
\begin{equation}
\label{eq:thetae}
\thetae=\arccos{\frac{L_{z,\rm{end}}}{|L_{\rm{end}}|}},
\end{equation}
where $L_{z,\rm{end}}$ and $|L_{\rm{end}}|$ are the vertical component and magnitude of a star's angular momentum at the end of the simulation ($t=12\Gyr$), respectively. Fig.~\ref{fig:thetaf_v_thetaj} presents the distribution of warp stars in the $(\thetae$, $\thetaf)$ space. The diagonal lines in both panels indicate $\thetaf=\thetae$. Overall, warp stars experience some degree of tilting. A majority of warp stars ($\sim73\%$) are located above the $\thetaf=\thetae$ line, indicative of an increasing alignment with the disc, and experience, on average, a shift of $\left<\thetae-\thetaf\right>\simeq -5.2\degrees$. The remaining warp stars become more misaligned with the disc and experience, on average, a shift of $\left<\thetae-\thetaf\right>\simeq +2.7\degrees$. The right panel of Fig.~\ref{fig:thetaf_v_thetaj} shows the distribution of average time of formation, $\avgbin{\tform}$ in the $(\thetae,\thetaf)$ space. All warp stars, regardless of $\tform$, undergo some tilting, with the median tilt being $med(\thetae-\thetaf) = - 3.5 \degrees$.

The top panel of Fig.~\ref{fig:Mthetaj_evo} presents the evolution of the population-averaged $\thetaj$, $\avg{\thetaj}$, for all mono-age warp populations formed during $2\leq \tform / \Gyr\leq 10$, in bins of $\Delta \tform=50 \Myr$. The average is over all $N$ star particles in a given population:
\begin{equation}
\label{eq:avg_theta_L}
\avg{\theta_{L}}=\frac{\sum_{i}^{N}\theta_{L_{i}}}{N},
\end{equation}
where $\theta_{L_{i}}$ is the angular momentum inclination of a star in the population. All warp populations experience a rapid drop in $\avg{\thetaj}$ by $\dt \sim 1 \Gyr$, followed by a smaller and gentler rise. The decrease in $\avg{\thetaj}$ varies from $\sim 5\degrees$ for the oldest population to about half that for younger populations. The bottom panel shows the rate of change of $\avg{\thetaj}$, $\Dthetaj$, for the same populations. The horizontal dotted line represents $\Dthetaj=0\degrees \Gyr^{-1}$. We observe that $\Dthetaj$ starts out negative for all populations and quickly plateaus at a nearly constant value of $\Dthetaj\sim0.5\degrees \Gyr^{-1}$. The initial negative tilt rate is due to the bulk tilting warp populations experience as they settle into the disc. This is produced by the torquing from the main disc and persists so long as the warp populations remain more or less coherent before differential precession destroys a relatively coherent plane for each population. The Briggs figures of Fig.~\ref{fig:briggs_evo} show that, for a wide range of \tform, by $\dt = 300\Myr$ the warp populations have precessed differentially enough that the innermost populations are then tilted in the opposite sense as the outermost ones ($\delta \phi \sim 180\degrees$). At this point the global tilting of a population becomes less efficient and their evolution is dominated by precession, which we study in Section \ref{subsection:phasemixing}.

\begin{figure}
\includegraphics[width=1\linewidth]{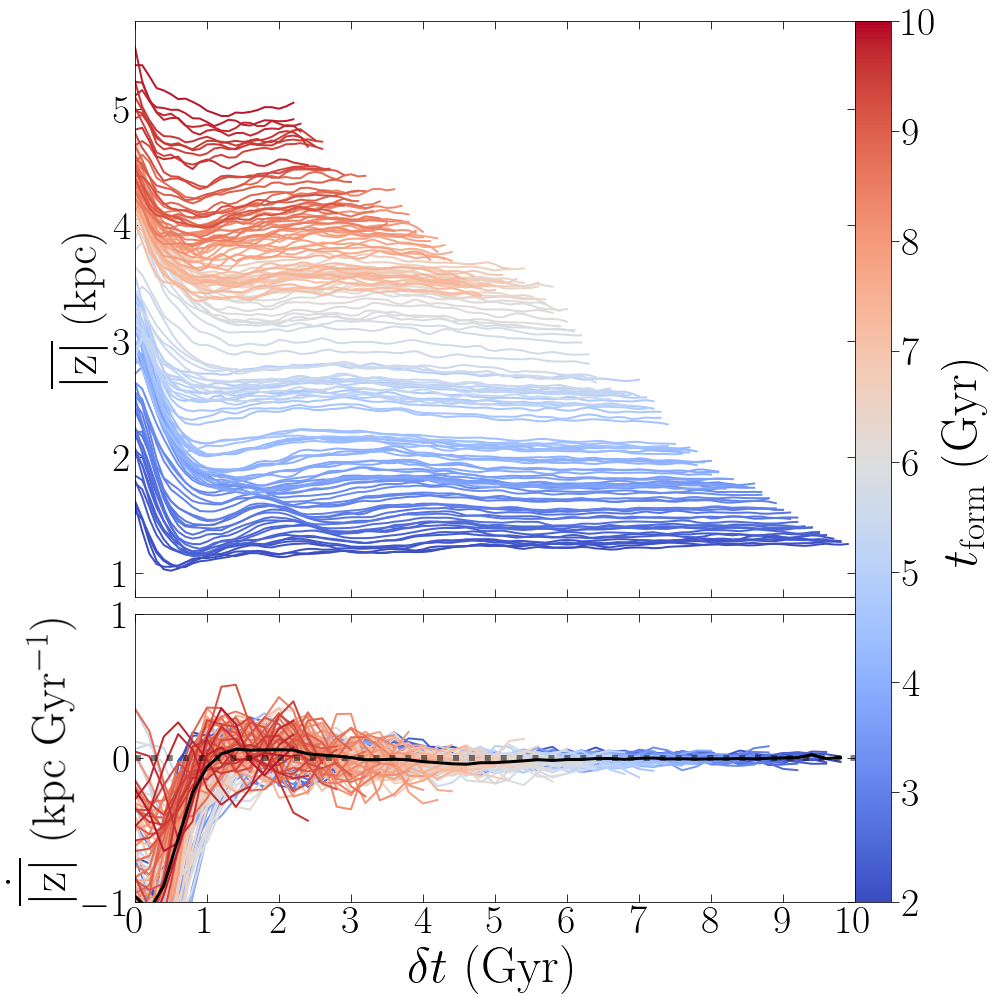}
    	\caption{Top: evolution of the population-averaged absolute $z$, $\avg{|z|}$ of different mono-age warp populations before $\tform\leq 10 \Gyr$, where $\dt$ is time since a population's formation. Each curve is coloured by $\tform$. A 1D Gaussian filter with a mask size of $w=0.5\Gyr$ and standard deviation of $\sigma=0.1\Gyr$ is applied to the evolution at each $\dt$. For each population the value of $\avg{|z|}$ starts to flatten after $\dt \sim 1 \Gyr$, reaching a stable configuration. The value of $\avg{|z|}$ for each population increases with $\tform$ as older populations form at higher $|z|$, similar to how older populations form at higher $\thetaf$ (Fig.~\ref{fig:formation_space}). Bottom: evolution of the $\avg{|z|}$ rate of change, $\dot{\avg{|z|}}$, for the same mono-age warp populations. The rates of change are calculated from the smoothed evolution curves. A rapid decrease in $\dot{\avg{|z|}}$ happens during the first $1 \Gyr$ and then settles about $\dot{\avg{|z|}}=0 \kpc\Gyr^{-1}$ (dashed horizontal line). The solid black line represents the median rate of change between all mono-age populations; this has a tilting time of $\ttilt\sim1\Gyr$.
}
\label{fig:z_settle}
\end{figure}

Fig.~\ref{fig:z_settle} shows the evolution and rate of change of the population-averaged $|z|$, $\avg{|z|}$, for the same mono-age populations.
The evolution of $\avg{|z|}$ is shown in the top panel; all of the warp populations plateau after just $1\Gyr$. The rate of change of $\avg{|z|}$, $\dot{\avg{|z|}}$, shown in the bottom panel, starts out mostly negative and quickly drops to $0 \kpc \Gyr^{-1}$ in less than $1 \Gyr$, a timescale similar to that in the first part of the $\Dthetaj$ evolution. As with the evolution of $\avg{\thetaj}$, we note a correlation between the age of the population and $\avg{|z|}$, with younger populations being formed further away from the mid-plane, and therefore settling to a thicker distribution. While $\avg{|z|}$ declines by $\la 1~\kpc$ during the tilting interval, Fig.~\ref{fig:z_settle} also shows that the thickness does not change much after tilting ends.

Given the similarity in the evolution of $\Dthetaj$ and $\dot{\avg{|z|}}$, we measure a timescale for the bulk tilting of warp populations. In order to measure the tilting times, \ttilt, for both $\avg{\thetaj}$ and $\avg{|z|}$ we set as a criterion the first time the rate of change reaches values of $\Dthetaj\geq0 \degrees \Gyr^{-1}$ and $\dot{\avg{|z|}}\geq0 \kpc \Gyr^{-1}$, respectively. We find that $\sim50\%$ of mono-age warp populations experience bulk tilting by $\dt=1 \Gyr$ using either the $\Dthetaj$ or the $\dot{\avg{|z|}}$ criterion. In both cases the longest tilting time is $\ttilt\sim1.8\Gyr$.


\subsection{Phase mixing}
\label{subsection:phasemixing}

The Briggs figures of mono-age warp populations in Fig.~\ref{fig:briggs_evo} show that their LON spirals wind up. This winding represents a phase-mixing so that eventually no trace of a warp remains and a warp population becomes axisymmetric. In this Section we study the phase mixing using three separate observables: the $m=1$ Fourier amplitude, the total angular momentum, and the entropy of each mono-age population.

\subsubsection{Winding of the LON spiral}
\label{subsubsection:winding}
\begin{figure}
\includegraphics[width=1\linewidth]{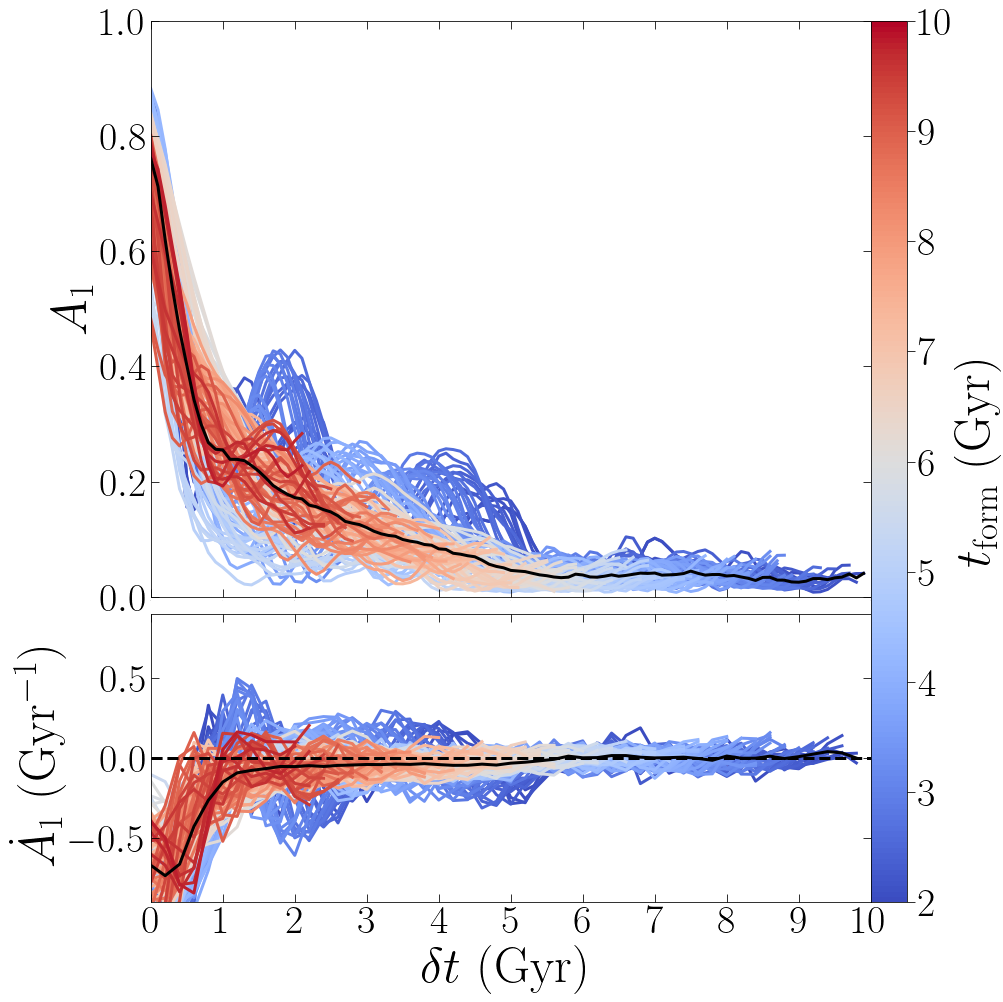}  
	\caption{Top: Evolution of the $A_1$ amplitude for all mono-age warp populations formed in the simulation before $\tform\leq 10 \Gyr$, where $\dt$ is the times since a population's formation. Each curve is coloured by $\tform$. A 1D Gaussian filter with a mask size of $w=0.5\Gyr$ and standard deviation of $\sigma=0.1\Gyr$ is applied to the evolution at each $\dt$. The black solid line is the median value of $A_1$ between all mono-age populations. Bottom: evolution of the rate of change of $A_1$, $\dot{A}_1$, for the same mono-age warp populations. The rates of change are calculated from the smoothed evolution curves. The solid black line represents the median rate of change between all mono-age populations. The dotted horizontal line indicates $\dot{A}_1=0 \Gyr^{-1}$.
}
\label{fig:tau}
\end{figure}

The distributions of angular momenta at formation for each mono-age warp population traces an $m=1$ spiral in Fig.~\ref{fig:briggs_evo}. By measuring the evolution of the amplitude of the Fourier $m=1$, $A_1$, in the angular momentum space in the Briggs figures we can follow the phase-mixing of each population. In the top panel of Fig.~\ref{fig:tau} we plot the evolution of $A_1$ for all mono-age warp populations. The peak $A_1$ for each warp population is at formation, and rapidly declines during the first $1 \Gyr$. The decline in most warp populations is not monotonic, with the oldest warp populations having multiple peaks of decreasing amplitude lasting up to $\dt=5 \Gyr$ after which the decrease is smoother. For younger populations $A_1$ declines more smoothly, though still not monotonically. However populations with $\tform\geq9.5 \Gyr$ exhibit a similar second peak as in the older populations. The bottom panel shows the rate of change of $A_1$, $\dot{A}_1$, for the same populations, with $\dot{A}_1=0 \Gyr^{-1}$ indicated by a dashed horizontal line. For all mono-age populations a significant oscillation in $\dot{A}_1$ is observed. The median curves of $A_1$ and $\dot{A}_1$ across all mono-age populations are shown as solid black lines in the top and bottom panels of Fig.~\ref{fig:tau}, respectively.

Fitting an exponential decay to $A_1$ as a function of time leads to exponential times $0.9<\tau/\Gyr<2.3$. The phase-mixing timescales, $\taupm$, can be estimated by taking the median of the time derivatives between all mono-age populations and measuring when it reaches $0 \Gyr^{-1}$. In the bottom panel of Fig.~\ref{fig:tau}, the median of the time derivative reaches the zero-line around $\taupm\sim 6 \Gyr$, a timescale that is longer than the tilting times computed in Section~\ref{subsection:tilting}.

\subsubsection{Phase mixing from the average angular momentum vector}
\label{subsubsection:angmom_phasemixing}

\begin{figure}
\includegraphics[width=1\linewidth]{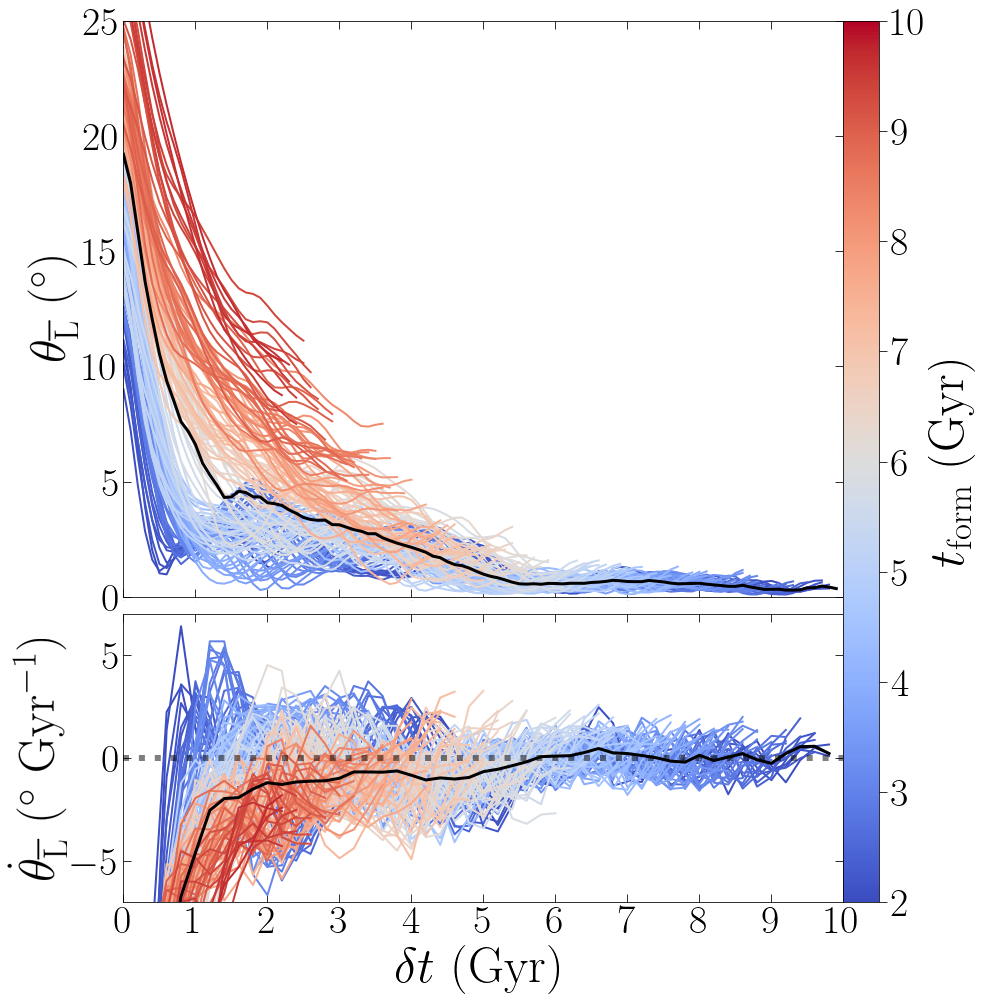}  
	\caption{Top: evolution of the population-averaged angular momentum inclination, $\thetajN$, for all mono-age warp populations formed in the simulation before $\tform\leq 10 \Gyr$, where $\dt$ is the time since a population's formation. Each curve is coloured by $\tform$. A 1D Gaussian filter with a mask size of $w=0.5\Gyr$ and standard deviation of $\sigma=0.1\Gyr$ is applied to the evolution at each $\dt$. The black solid line is the median value of $\thetajN$ between all mono-age populations. Bottom: evolution of the rate of change of $\thetajN$, $\DthetajN$, for the same mono-age warp populations. The rates of change are calculated from the smoothed evolution curves. The solid black line represents the median rate of change between all mono-age populations. The dotted horizontal line indicates $\DthetajN=0\degrees\Gyr^{-1}$.}
\label{fig:thetajM_evo}
\end{figure}

The uniform distribution of $\phi_L$ in the Briggs figures (Fig.~\ref{fig:briggs_evo}) of settled populations implies that if the angular momentum of each warp population were vector-averaged over all stars then the resulting mean angular momentum would be along the $z$ axis, with inclination $\theta=0\degrees$. To estimate the phase-mixing timescale differently, we analyse the inclination of the average angular momentum vector, $\theta_{\avg{L}}$, which we define for a given mono-age population as
\begin{equation}
\label{eq:theta_avg_L}
\theta_{\avg{L}}=\arccos\left({\frac{\sum_{i}^{N}L_{z,i}}{\lVert\sum_{i}^{N}L_i\rVert}}\right),
\end{equation}
where $L_{z,i}$ and $L_{i}$ are the vertical angular momentum and the angular momentum vector of a star in a given population, respectively.
In the top panel of Fig.~\ref{fig:thetajM_evo} we
present the evolution of the inclination of the average angular momentum, $\thetajN$, for mono-age populations. The evolution of $\thetajN$ shows that warp populations with $\tform\geq 6\Gyr$ do not reach $\sim 0 \degrees$, indicating that they are still phase-mixing, in agreement with Fig.~\ref{fig:briggs_evo}. Older populations with $\tform<6\Gyr$ settle to $\thetajN=0\degrees$ on different timescales, with the oldest population presenting multiple peaks, as in Fig.~\ref{fig:tau}. In the bottom panel, the evolution of the rate of change, $\DthetajN$, shows that the phase-mixing process is much more rapid for the older populations but then $\thetajN$ rises again at $\sim2\Gyr$ and then oscillates about $\DthetajN=0\degrees\Gyr^{-1}$. Younger populations show a slower and smoother increase towards $\DthetajN=0\degrees\Gyr^{-1}$ in their rate of change. The median curves for $\thetajN$ and $\DthetajN$ between all mono-age populations are shown as solid black lines.

Based on the $\thetajN$ evolution we estimate the phase-mixing time, $\tpm$, using the time when the median of the time derivatives reaches $0\degrees\Gyr^{-1}$. In the bottom panel of Fig.~\ref{fig:thetajM_evo} this occurs at around $\taupm\sim 6 \Gyr$. This timescale is again longer than the tilting times computed in Section~\ref{subsection:tilting}.

\begin{figure}
\includegraphics[width=1\linewidth]{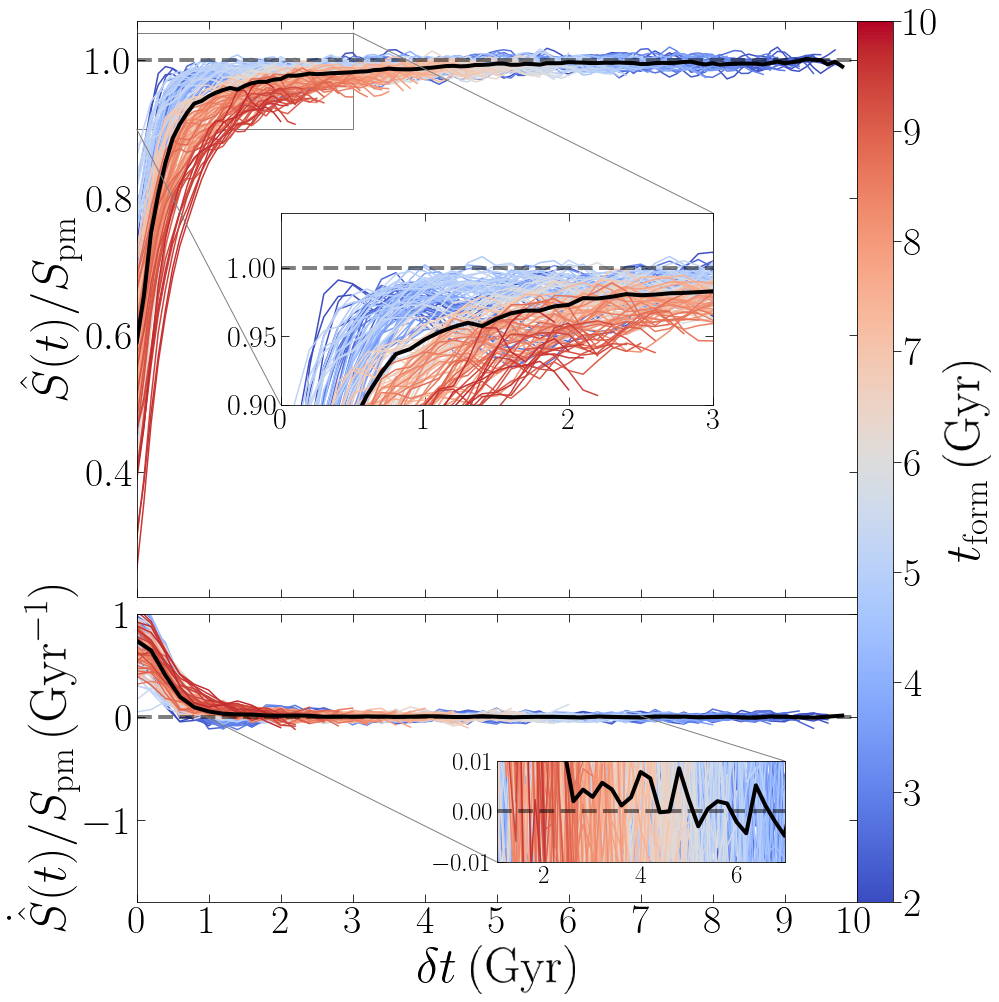} 
	\caption{Top: evolution of the entropy estimate, $\entropy$, normalised by $\Spm$ for all mono-age warp populations formed in the simulation before $\tform\leq 10 \Gyr$, where $\dt$ is the time since a population's formation. Each curve is coloured by $\tform$. A 1D Gaussian filter with a mask size of $w=0.5\Gyr$ and standard deviation of $\sigma=0.1\Gyr$ is applied to each population. The black solid line is the median value of $\entropy$ between all mono-age populations. The dotted horizontal line indicates $\entropy/\Spm=1$.  The inset shows an expanded version of the region indicated by the rectangle at top left. Bottom: evolution of the rate of change of $\entropy$, $\dentropy$, normalised by $\Spm$ for the same mono-age warp populations. The rates of change are calculated from the smoothed evolution curves. The black solid line is the median rate of change between all mono-age populations. The dotted horizontal line indicates $\dentropy=0$. The inset shows an expanded version of the region around the zero-line.}
\label{fig:entropy}
\end{figure}

\subsubsection{Entropy-based phase-mixing timescale}

Lastly, we also characterise the phase mixing process using a non-parametric entropy estimator. The entropy definition we adopt is:
\begin{equation}
\label{eq:S_def}
    S = -\int f(\phi_L)\ln f(\phi_L) \,\mathrm{d}\phi_L,
\end{equation}
where $f(\phi_L)$ is the probability density function. This functional form is chosen primarily because of its desirable mathematical properties, including that it can be estimated as
\begin{equation}
\label{eq:S_hat}
    \hat{S} = -\frac{1}{N}\sum_{i=1}^N \ln \hat{f}_i,
\end{equation}
where the sum runs over the warp stars of a given population and $\hat{f}_i$ is the estimate of $f(\phi_L)$ for each star particle. Eq.~\ref{eq:S_hat} converges to Eq.~\ref{eq:S_def} if $\hat{f}_i$ is calculated with specific recipes \cite[see][and references therein]{beraldoesilva+19a, beraldoesilva+19b}. Here we adopt the Nearest Neighbour method, where, in one dimension, the distribution is estimated as:
\begin{equation}
\label{eq:f_hat}
    \hat{f}_i = \frac{1}{2(N-1)e^\gamma D_{in}},
\end{equation}
where $\gamma \simeq 0.577$ is the Euler-Mascheroni constant and $D_{in}$ is the distance of particle $i$ to its nearest neighbour \cite[see][for more general expressions]{10.5555/2875145, beraldoesilva+19a, beraldoesilva+19b}. Since ${-\upi \leq \phi_L \leq \upi}$, for a fully-mixed population the phase-mixed distribution is ${f_\mathrm{pm}=1/(2\upi)}$, and from Eq.~\ref{eq:S_def}, the phase-mixed entropy is ${\Spm = \ln (2\upi)}$.

We use Eqs.~\ref{eq:S_hat}-\ref{eq:f_hat} to estimate the entropy of the same mono-age warp populations defined before at different times. The entropy evolution, normalised by the phase-mixed value $\Spm$, is shown in the top panel of Fig.~\ref{fig:entropy}, colour-coded by the formation times. The fluctuations around $\hat{S}(t)/S_\mathrm{pm}=1$ for long times provide a sense of the uncertainty level on the entropy estimate. All mono-age warp populations show a rapid increase in entropy, on a time-scale of $\delta t\sim 1$ Gyr, after which the system asymptotically evolves to the phase-mixed state, on a longer time-scale. Populations born after $\tf \gtrsim 6 \Gyr$ do not have time to completely phase-mix, in good agreement with the Briggs figures in Fig.~\ref{fig:briggs_evo}. Young populations are born with smaller entropies, which is a result of the larger radius at which they are forming, resulting in a narrower range of $\phi_L$ values. The median curves for $\hat{S}/S_{pm}$ and $\dot{\hat{S}}/S_{pm}$ across all mono-age populations are shown as solid black lines in the top and bottom panels of Fig.~\ref{fig:entropy}, respectively.

We estimate the phase-mixing timescale using the time when the median of the time derivatives reaches $0 \Gyr^{-1}$. Though the median of the time derivatives fluctuates as it approaches the zero-line (see bottom inset), we estimate that $\taupm\sim 5-6 \Gyr$.
This phase-mixing timescale is in agreement with the results from Sections \ref{subsubsection:winding} \& \ref{subsubsection:angmom_phasemixing}, reaffirming that phase-mixing continues long after the tilting has concluded.


\subsection{Inward migration of warp populations}
\label{subsection:migration}

In our definition, a warp population must have formed at $r > 10 \kpc$. Fig.~\ref{fig:Mthetaj_evo} showed that, at $\dt \gtrsim 2\Gyr$, 
$\Dthetaj \sim 0.5\degrees$  for many warp populations. A naive interpretation of this result is that the warp populations continue to heat vertically after they settle. Aside from the fact that thick populations do not heat vertically efficiently since they spend most of their time away from the thin disc, where most of the perturbers that can heat them reside, Fig.~\ref{fig:z_settle} contradicts this interpretation, because it shows that $\avg{|z|}$ is not increasing at the same time. A different interpretation is therefore needed. Here we show that warp populations migrate inwards; with $\avg{|z|}$ constant, the inward migration must result in an increasing $\thetaj$ and a positive $\Dthetaj$.

\begin{figure}
\includegraphics[width=1\linewidth]{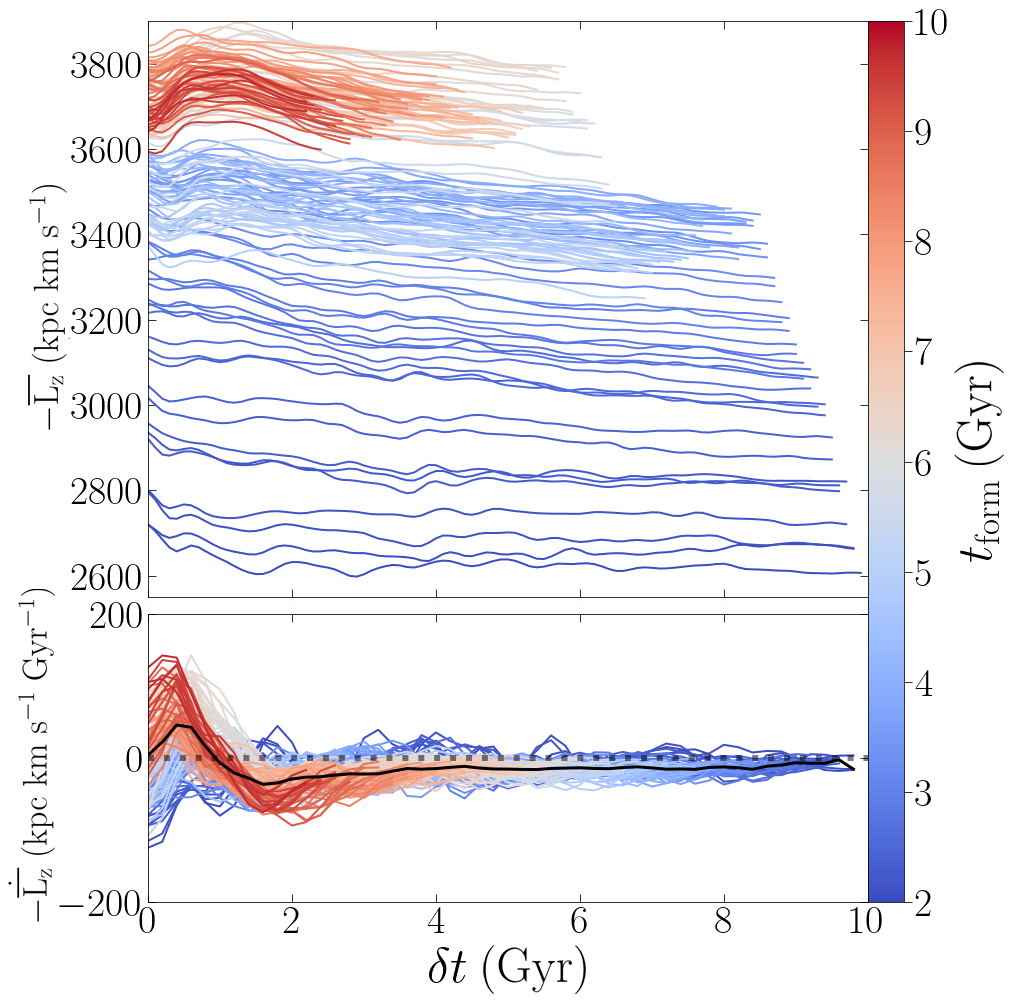}  
	\caption{Top: evolution of the population-averaged vertical angular momentum, $\avg{L_z}$, for different mono-age warp populations in the simulation, where $\dt$ is the time since a population's formation. Each curve is coloured by $\tform$. A 1D Gaussian filter with a mask size of $w=0.5\Gyr$ and standard deviation of $\sigma=0.1\Gyr$ is applied to the evolution at each $\dt$. Most of the change in the vertical angular momentum happens during the first $1\Gyr$ after which they decline slowly or remain flat. Bottom: evolution of the $\avg{L_z}$ rate of change, $\dot{\avg{L_z}}$, for the same mono-age warp populations. The rates of change are calculated from the smoothed evolution curves. The solid black line represents the median rate of change between all mono-age populations. We observe that $\avg{L_z}$ is continuously decreasing as the rate of change remains mostly below $\dot{\avg{L_z}}=0 \kmsk \Gyr^{-1}$ (dashed horizontal line).
}
\label{fig:L_settle}
\end{figure}
    
Fig.~\ref{fig:L_settle} considers the evolution and rate of change of the population-averaged vertical angular momentum, $\avg{L_z}$. The evolution of $\avg{L_z}$ (top panel) shows an increase in the first $1 \Gyr$ for all populations with $\tform>4.5\Gyr$, while older populations show a decrease. These changes subsequently slow down significantly as all populations plateau with only a weak negative gradient. The rate of change, $\dot{\avg{L_z}}$, (bottom panel) shows that after $\dt\sim1\Gyr$ all populations have a negative $\dot{\avg{L_z}}$, though there is an initial spike for populations with $\tform\geq4.5\Gyr$. The solid black line indicates the median over all rates of change. The vertical angular momentum correlates with $\tform$ of the warp populations, because of the growing radius of the warp. The initial spike in younger populations is related to the growing warp as younger populations have larger $\thetaf$ and due to the projection of $L_z$, even small tilts translate to larger changes in $L_z$. Older populations form in a younger, shallower warp and do not experience the same initial spike. In spite of this difference, all warp populations have comparable values of $\dot{\avg{L_z}}$ for $\dt > 1 \Gyr$. The net decrease in vertical angular momentum of warp populations well after they formed represents either a radial heating of each population, or an inward migration.

\begin{figure}
\includegraphics[width=1\linewidth]{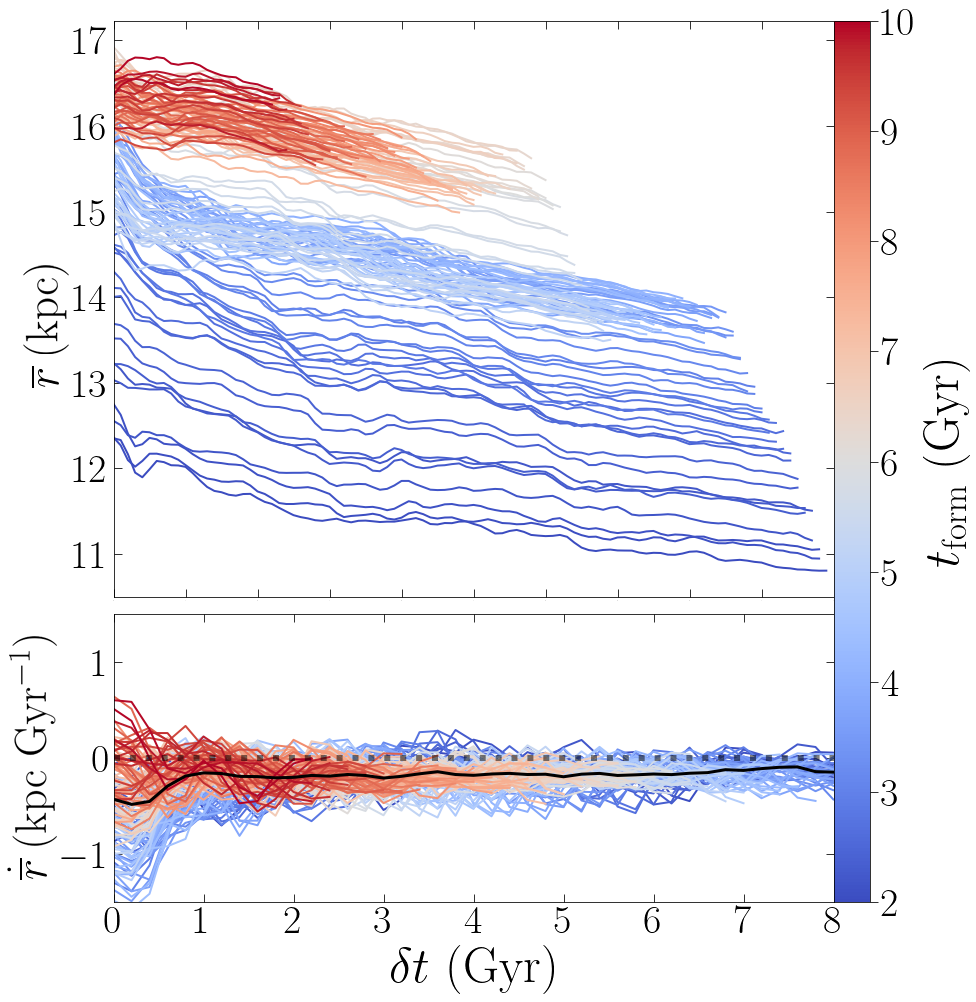}  
	\caption{Top: evolution of the population-averaged spherical radius, $\avg{r}$, of different mono-age warp populations, where $\dt$ is the time since a population's formation. Each curve is coloured by $\tform$. A 1D Gaussian filter with a mask size of $w=0.5\Gyr$ and standard deviation of $\sigma=0.1\Gyr$ is applied to the evolution at each $\dt$. The decrease of $\avg{r}$ is different for all populations and is strongest for $\tform \sim 3 \Gyr$ at $15\%$ with the weakest decrease for $\tform > 6 \Gyr$ at $5\%$. Bottom: evolution of the $\avg{r}$ rate of change, $\dot{\avg{r}}$, for the same mono-age warp populations. The rates of change are calculated from the smoothed evolution curves. The solid black line represents the median rate of change between all mono-age populations. A consistently negative $\dot{\avg{r}}<0 \kpc \Gyr^{-1}$ is observed (dashed horizontal line), with the exception of a few transient positive values for the oldest population. This is indicative of continuous inward migration for all warp populations, regardless of their $\tform$.
}
\label{fig:r_settle}
\end{figure}

We, therefore, analyse how the radial positions of warp stars change with time. We use the spherical radius rather than the cylindrical one since the disc is warped. Fig.~\ref{fig:r_settle} presents the evolution of the population-averaged spherical radius, $\avg{r}$, for the mono-age warp populations. The top panel shows the evolution of $\avg{r}$, which clearly decreases at all times for all populations. This change implies that the decrease of the angular momentum of warp stars is accompanied by a net radial movement inwards and continues well after the population tilting has ended. The decrease of $\avg{r}$ is continuous for all populations which we confirm by plotting the rate of change for $\avg{r}$, $\dot{\avg{r}}$, (bottom panel) which is predominantly negative after $\dt=1 \Gyr$. The net inward movement of warp populations is a result of the fact that, by definition, they form only at large radii ($\geq 10\kpc$).

\begin{figure*}
\includegraphics[width=1\linewidth]{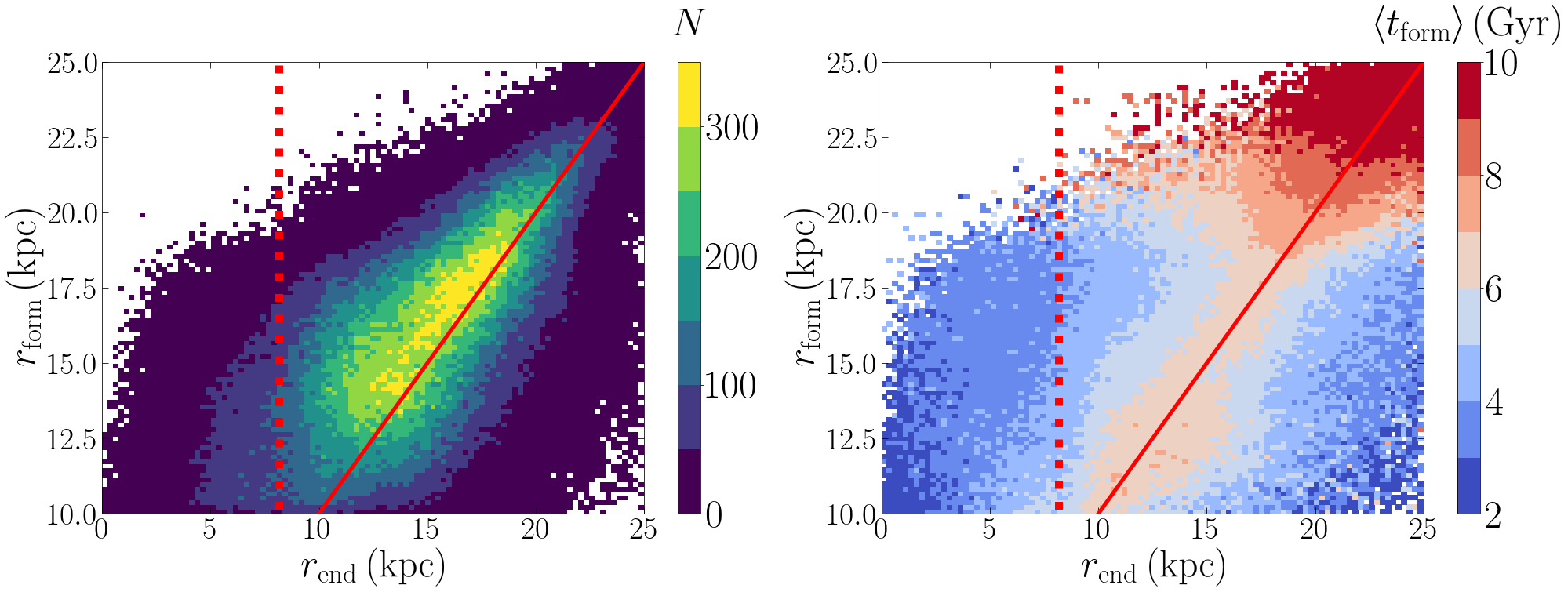}  
	\caption{Distribution of spherical formation radius, $\rform$, versus the spherical radius at the end of the simulation, $\re$, for warp stars coloured by the number (left) and by the mean time of formation, $\tform$ (right). The diagonal solid line indicates $\rform=\re$. Stars that are below the $\rform=\re$ line comprise $1/3$ of the total warp star sample. The vertical dotted line indicates the location of the Solar annulus. Warp stars born after $\tform \gtrsim 6\Gyr$ do not have enough time to migrate into the Solar annulus.
    }
\label{fig:r_rform}
\end{figure*}

In Fig.~\ref{fig:r_rform} we consider the relation between the formation radius, $\rform$, and the final radial position, $\re$, for all warp stars. The left panel shows that 66\% of warp stars move inwards. This movement inwards happens regardless of $\tform$ (right panel), with older populations experiencing the strongest inward movement (extending to $\rform-\re\sim15 \kpc$). This could indicate migration by spiral churning where the migration is characterised by a random walk \citep{SellwoodBinney}. A radial gradient of decreasing age is established in the inner disc, with warp stars at the smallest radius being the oldest ones. This is the mirror image of the usual outwardly increasing age gradient for stars formed within the main disc and migrating outwards \citep{Roskar+08a, beraldoesilva+20}. We note that the oldest populations also move outwards, which also hints at migration via spiral churning.

\begin{figure}
\includegraphics[width=1\linewidth]{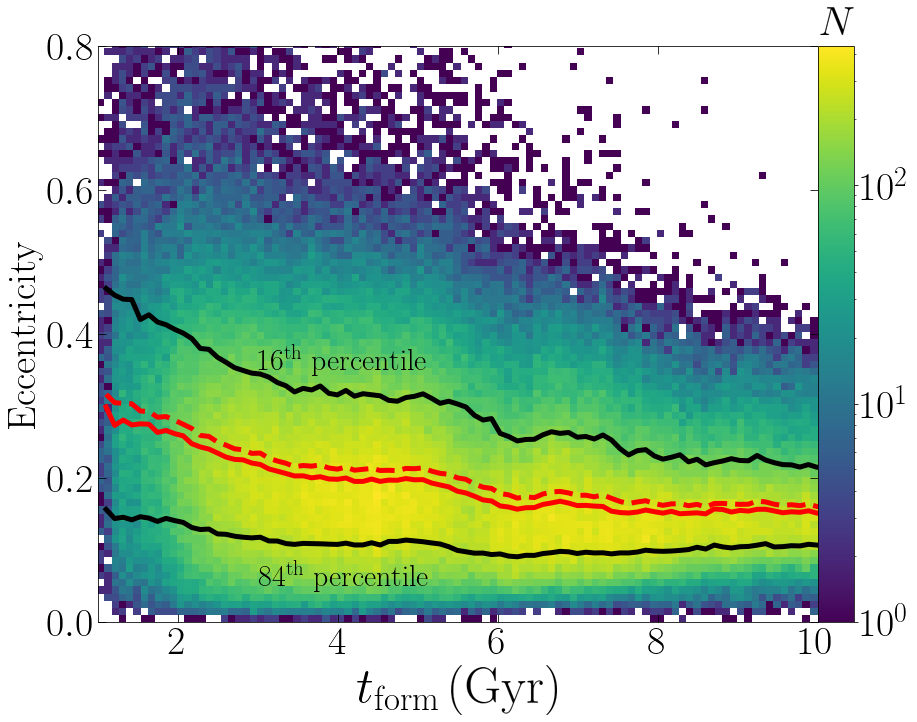}  
	\caption{Eccentricities of warp stars versus $\tform$. The lines indicate the median (red solid), mean (red dashed), and the $16^{\rm{th}}$ and $84^{\rm{th}}$ percentiles (black annotated) of the eccentricity for each $\tform$ bin. The younger warp stars having slightly more circular orbits. In general most orbits are fairly close to circular.
}
\label{fig:ecc}
\end{figure}

Finally, we explore the orbital parameters of warp stars. We integrate the orbits of settled warp stars in the interpolated potential derived using {\sc AGAMA} (see Section~\ref{subsection:preproc}). The initial conditions of the warp stars are set to their 6D coordinates at the end of the simulation at $12 \Gyr$. Because the youngest warp stars may not have had enough time to tilt into the disc, we limit our analysis to warp stars with $\tform\leq 10 \Gyr$.
After integrating each warp star for 10 orbital periods, we use the maximum and minimum cylindrical radii along the orbit to calculate the orbital eccentricities:
\begin{equation}
e = \frac{R_{max}-R_{min}}{R_{max}+R_{min}}
\end{equation}
Fig.~\ref{fig:ecc} presents the 2D histogram of orbital eccentricities plotted versus $\tform$. This distribution shows that a large fraction of warp stars have $0.1 \leq e \leq 0.4$. The lines indicate the median (red solid), mean (red dashed), and the $16^{\rm{th}}$ and $84^{\rm{th}}$ percentiles (black annotated) of the eccentricity in each $\tform$ bin; a weak decline of the mean eccentricity from $0.3$ for the oldest population to $0.2$ for the youngest is evident. These nearly circular orbits indicate that the radial migration is driven by spiral churning \citep{SellwoodBinney, Roskar+12} rather than by heating. The decreasing mean angular momentum amplitude is purely a result of the unbalanced distribution of formation radii of warp stars.


\section{Resulting disc structure}
\label{section:disc_structure}

The dynamical evolution of warp populations explored in Section~\ref{section:dyn_evo} showed that as soon as warp stars form they begin rapidly tilting and then phase-mixing in the galactic disc. These processes are accompanied by the slow but continuous inward (and outward) migration of the warp populations. We now explore the resulting disc structure of settled warp populations.

\begin{figure*}
\includegraphics[width=1\linewidth]{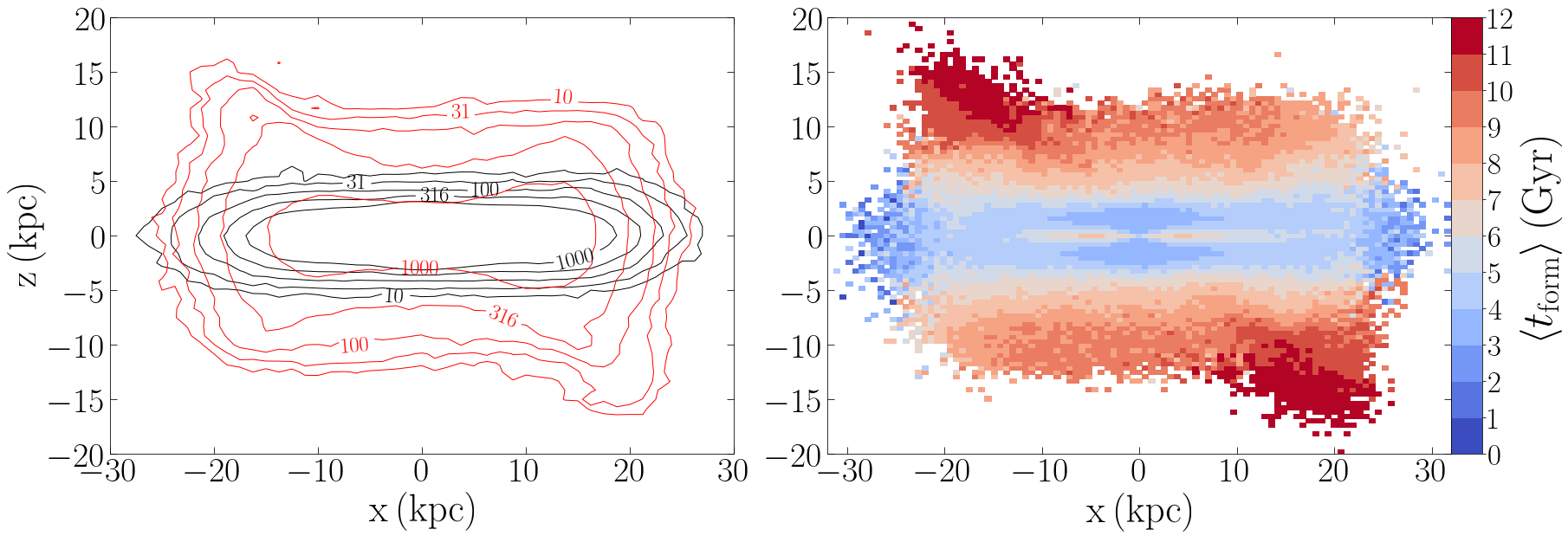}  
\caption{
Edge-on views of the simulation at $12 \Gyr$. Left column: number density contour plots of the warp (red contours) and main disc (black contours) populations. Warp stars occupy higher $\left| z\right|$ than stars formed in the disc and outnumber disc stars by a factor of 10 starting from at $|z| \sim 4 \kpc$. Right column: distribution of the mean formation time, $\left< \tform\right>$ for all stars formed throughout the simulation. There is a vertical gradient in $\left< \tform \right>$ and a young warp population that traces the gas warp starting from $|z|\geq 5 \kpc$. See Fig.~\ref{fig:edge_on_HT} in the Appendix for different trends in the supplementary simulation which implements more realistic star formation subgrid physics.
}
\label{fig:edgeon_warpndisk_age}
\end{figure*}	
	
Fig.~\ref{fig:edgeon_warpndisk_age} presents the edge-on distributions of warp and in-situ stars at $t=12\Gyr$. In the left panel, the contours show the number density distribution of warp (red) and in-situ (black) stars. Warp stars occupy the geometric thick disc with visible flaring at $|x|\geq15\kpc$ and a maximum vertical extent of $|z|\leq10 \kpc$. In the right column is the edge-on distribution of $\avgbin{\tform}$ for all the stars formed throughout the simulation. Starting from $|z|\geq 5\kpc$, newly formed warp stars can be observed tracing the gas warp, the major axis of which is along the $x$-axis (Section~\ref{section:Simulation}). A negative vertical gradient in the $\avgbin{\tform}$ distribution is visible. This gradient is due to the enhanced star formation in the warp, which we have tuned our simulation for by setting a low gas density threshold for star formation. To a lesser extent it also reflects the overall growth of the warp. While the negative age gradient in this simulation does not match the trends in the MW's outskirts \citep{Laporte+20, 2021MNRAS.502.5686I}, where older stars are observed at higher $|z|$, this incorrect vertical age gradient in the simulation only indicates that star formation is not efficient in the warp, rather than that the warp is not produced by gas inflows. Indeed in the supplemental simulation, which we present in the Appendix, the same initial conditions as our fiducial simulation result in a vertical age gradient with the opposite age trend when evolved with more realistic subgrid physics that form less stars in the warp. We remind the reader that our choice of subgrid physics for our fiducial simulation was motivated by the need to produce enough stars to be able to study their settling in detail. 

We now analyse in detail the resulting density distributions of stars formed in the warp. Aware that we are over-producing warp stars, our motivation here is not to predict in detail the density distribution of warp stars, but to demonstrate how they increasingly occupy a larger radial range while largely retaining their original vertical distribution. We start by selecting the warp stars currently located at $2 < R/\kpc < 25$ and $0 < |z|/\kpc < 15$. We define 30 broad mono-age populations (bin width ${\Delta \tf = 250\Myr}$) in the interval $2 < \tf/\Gyr < 9.5$, where the lower limit is chosen to avoid the stars formed in the early chaotic transient warp, and the upper limit chosen such that the youngest population considered has had enough time to settle. For each of these mono-age populations, we simultaneously fit the radial surface number density profile, $\Sigma(R)$, and the vertical number density profile, $\xi(z | R)$. 

The radial profile $\Sigma(R)$ is modelled as a skew-normal distribution \citep[][]{skewnorm}, which we found to be the best functional form after comparison with different models \citep[see e.g.][]{bovy+16, beraldoesilva+20}. In this model, the profile is given by
\begin{equation}
    \Sigma(R|\mu_R, h_R, \alpha) = \frac{1}{A} \phi(R|\mu_R, h_R)\Phi(\alpha R|\mu_R, h_R),
\end{equation}
where $\phi(x|\mu,h)$ is the normal (Gaussian) distribution with location $\mu$ and scale $h$, $\Phi(x)$ is its cumulative distribution function and $A$ is determined by the normalisation condition 
\begin{equation}
    \int_{\Rmin}^{\Rmax} \Sigma(R) 2\upi R\,\mathrm{d}R = 1.
\end{equation}
The parameter $\alpha$ controls the skewness and the Gaussian distribution is recovered for $\alpha=0$. Note that ${-\infty < \alpha < \infty}$, while the real skewness can be obtained from $\alpha$ and ranges from -1 to 1. Note also that $\mu_R$ and $h_R$ are close to, but not exactly, the position of the peak, $\Rpeak$, and the dispersion $\sigma_R$, respectively, which are also obtained by simple formulae from the parameters $\alpha$, $\mu_R$ and $h_R$ \citep[see][]{skewnorm}.

The vertical density profiles $\xi(z | R)$ are modelled with the so-called generalised normal distribution \citep[][]{Nadarajah_2005}:
\begin{equation}
\label{eq:xi}
    \xi(z | R, \mu_z, h_z, \beta) = \frac{1}{B} \exp\left[-\left|\frac{|z| - \mu_z}{h_z}\right|^\beta\right],
\end{equation}
where $\beta$ controls the kurtosis ($\beta=2$ for the Gaussian) and $B$ is obtained by imposing the condition
\begin{equation}
    \int_{\zmin}^{\zmax} \xi(z|R)\, \mathrm{d}z = 1.
\end{equation}
In the above expressions, all three parameters $\beta$, $\mu_z$ and $h_z$ depend on $R$ in a non-trivial way. After some experimentation we determined that each of these parameters needs to be modelled as a third-order polynomial in radius $R$. The position of the peak in $|z|$ is directly given by $\zpeak=\mu_z$.

Finally, the total number density profile is written as
\begin{equation}
    \nu(R, z|\theta) = \Sigma(R|\theta)\xi(z|R,\theta),
\end{equation}
where $\theta$ is the set of parameters. For each mono-age population, we first fit this model maximising the log-likelihood
\begin{equation}
    \ln \mathcal{L}(\theta) = \sum_i \ln\left[ \nu(R_i, z_i|\theta)\right]
\end{equation}
with a variant of Powell's method, which is a conjugate direction method \citep[][]{1964Powell, PresTeukVettFlan92}. Then, we use this first fitting result as input to MCMC-sample the posterior distribution function with the {\sc emcee} package \citep[][]{2013PASP..125..306F}, assuming flat priors for all parameters. Best fit parameters and uncertainties are estimated with the median and the 16 and 84 percentiles of the parameter samples, respectively.

\begin{figure}
\includegraphics[width=1\columnwidth]{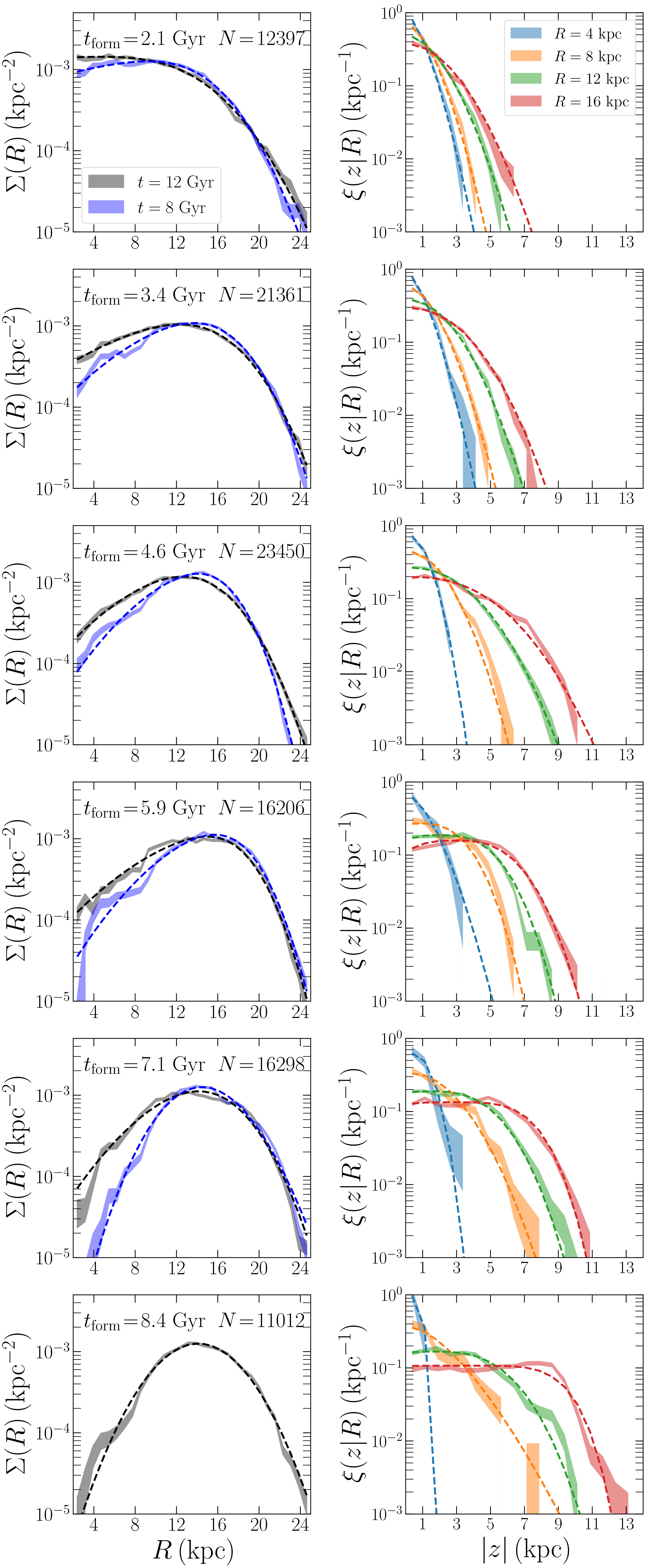}  

	\caption{Left: radial density profiles for different mono-age populations (rows), with formation times indicated. Black (blue) curves show the profiles at $t=12\Gyr$ ($t=8\Gyr$), which are well described by skew-normal distributions (dashed lines). Young populations (bottom) have approximately Gaussian profiles peaking at $\Rpeak\approx 15\kpc$. Due to inward migration, older populations (upper panels) are progressively negatively-skewed, peaking at smaller radii, and all populations evolve significantly over the last 4 Gyr (compare black and blue curves within each panel). Right: vertical profiles for different radii (colours) at $t=12\Gyr$. The profiles are well described by generalised normal distributions (dashed lines), and get more flattened and thicker for larger $R$ and larger $\tf$.}
\label{fig:Sigma_rho}
\end{figure}

For illustrative purposes, in Fig.~\ref{fig:Sigma_rho} we slice some of these populations into cylindrical shells, determining the surface number density profile $\Sigma(R)$ (left panels) and, for each shell, the vertical number density profiles $\xi(z | R)$ (right panels). In Fig.\ref{fig:Sigma_rho}, each row represents a different mono-age population, with the formation times indicated. The best fit models are represented by dashed lines.

\subsection{Radial density profiles}

The black shaded curves in the left column of Fig.~\ref{fig:Sigma_rho} show $\Sigma(R)$, with Poisson uncertainties, at the final snapshot (${t=12\Gyr}$). For young populations (bottom panels), $\Sigma(R)$ peaks at ${\Rpeak\approx 14\kpc}$ and is approximately symmetric around this peak. Older populations (top panels) get increasingly centrally concentrated, with $\Rpeak$ moving to lower values and $\Sigma(R)$ becoming increasingly skewed. In order to confirm that this is the consequence of a continuous evolution (as opposed to rather different initial conditions of different mono-age populations), we also show the profiles of the same populations, calculated at $t=8\Gyr$ (blue shaded curves). The apparent time evolution from the bottom to the top panels is confirmed within each panel, with each mono-age population (including the oldest one at the top) becoming more centrally concentrated over the $4\Gyr$ between $t=8\Gyr$ and $t=12\Gyr$, indicating the effect of continuous inward migration.

\begin{figure*}
\includegraphics[width=1\linewidth]{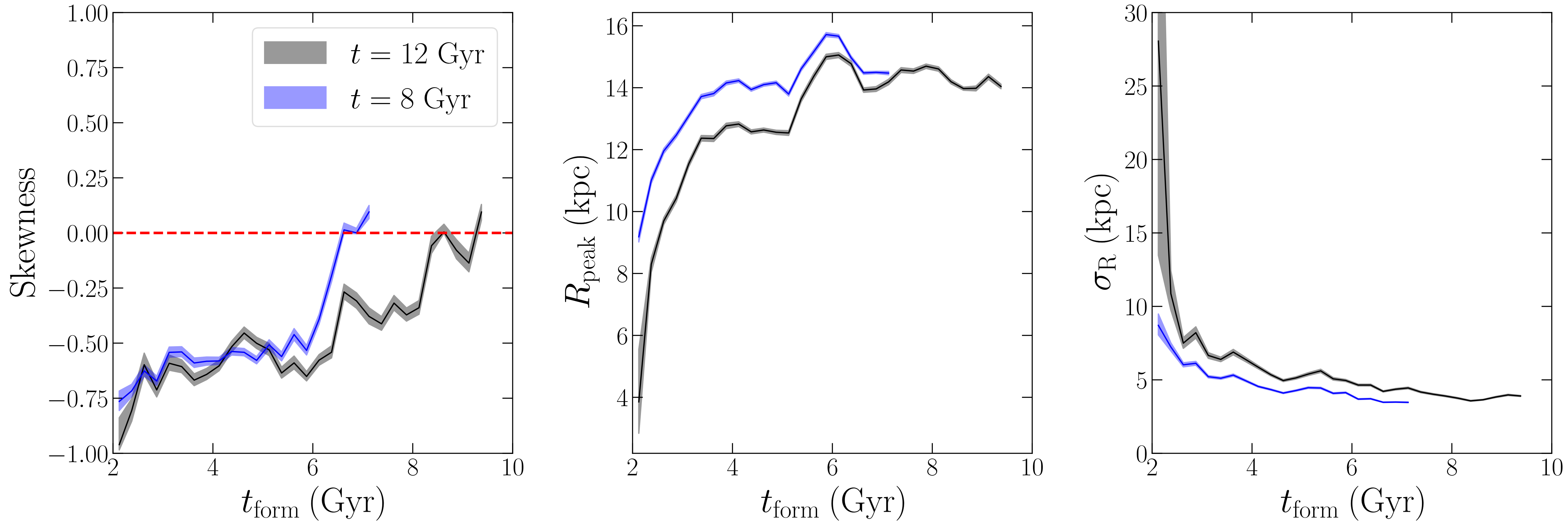}  

	\caption{Derived quantities from the best fit models of the radial density profiles for different mono-age populations, evaluated at $t=12\Gyr$ (black) and $t=8\Gyr$ (blue). The skewness (left panel) is approximately zero for young populations (large $\tf$) and gets progressively more negatively-skewed for older populations (small $\tf$). The central panel shows that the profiles peak at $\Rpeak\approx 14\Gyr$ for $\tf\gtrsim 5\kpc$, and drops quickly to $\Rpeak \approx 4\kpc$ at $\tf\approx 2 \Gyr$. The dispersion (right panel) is $\sigma_R\approx 5\kpc$ for $\tf\gtrsim 5\Gyr$ and increases rapidly for $\tf \lesssim 5\Gyr$.}
\label{fig:params_Sigma_skewnorm}
\end{figure*}

The best fit models (dashed lines) in the left column of Fig.~\ref{fig:Sigma_rho} show a good agreement with the empirical profiles. In Fig.~\ref{fig:params_Sigma_skewnorm}, we show the best fit parameters as a function of the formation time (black shaded curves, evaluated at $t=12\Gyr$). Instead of $\alpha$, $\mu_R$ and $h_R$, we show the derived quantities representing the skewness, the peak position, $\Rpeak$, and the dispersion of the radial coordinate, $\sigma_R$. The skewness (left panel) shows a clear trend, decreasing from $\approx 0$ for the youngest populations (large $\tf$) to $\sim -1$ for the oldest ones (small $\tf$), consistent with the strong change of slope of the inner part of $\Sigma(R)$ observed in Fig.~\ref{fig:Sigma_rho}. The position of the density peak is shown in the central panel. It  decreases mildly from $\Rpeak \approx 14\kpc$ for the youngest populations to $\Rpeak\approx 12\kpc$ for $\tf \approx 4\Gyr$, after which it strongly decreases (from right to left) to $\Rpeak \approx 4\kpc$ for the oldest populations. This strong decrease seems to be associated with the inner slope of $\Sigma(R)$ becoming close to zero for small $\tf$ (see Fig.~\ref{fig:Sigma_rho}), in which case a small change in this slope can imply large changes in the peak position. Finally, the radial dispersion (right-hand panel) is $\sigma_R\approx 5 \kpc$ for $\tf \gtrsim 5 \Gyr$, and increases rapidly for older populations, which is correlated with the behaviour of $\Rpeak$ just mentioned (a strictly horizontal inner $\Sigma(R)$ would imply an infinite dispersion).

As in Fig.~\ref{fig:Sigma_rho}, blue shaded curves represent the best fit models of the same mono-age populations, evaluated at $t=8\Gyr$. All parameters follow similar trends with \tf. Comparison of the black and blue curves shows that the profiles become more negatively skewed, more centrally concentrated and with larger dispersion over the last $4\Gyr$ of evolution. Interestingly, for both $t=8\Gyr$ and $t=12\Gyr$, the skewness is $\sim 0$ for those populations with $\tf \sim t$, \ie\ for the youngest populations at each snapshot. This suggests that all mono-age populations are formed with (or quickly evolve to) a Gaussian radial density profile, subsequently evolving towards negatively skewed distributions associated with inward migration. It is also interesting to note that, at $t=8\Gyr$, $\Rpeak$ and $\sigma_R$ do not show the strong gradients near $\tf \approx 2\Gyr$ observed at $t=12\Gyr$. The profiles become more negatively skewed, more centrally concentrated and with larger dispersion over the last 4 Gyr of evolution (compare black and blue curves).

\subsection{Vertical density profiles}
\label{section:vertical_profiles}

The right panels in Fig.~\ref{fig:Sigma_rho} show the vertical number density profiles $\xi(z|R)$, with Poisson uncertainties, within cylindrical shells of width $2\kpc$ at different radii (colours), for each mono-age population ($\tf$ indicated in the left panels), evaluated at $t=12\Gyr$. As a general trend, the vertical profiles get flatter and thicker, both as a function of $R$ (for a fixed $\tf$) and as a function of $\tf$ (for a fixed $R$). The dashed lines represent the best fit models and we observe a good agreement with the empirical profiles for all radii and formation times.

\begin{figure*}
\includegraphics[width=1\linewidth]{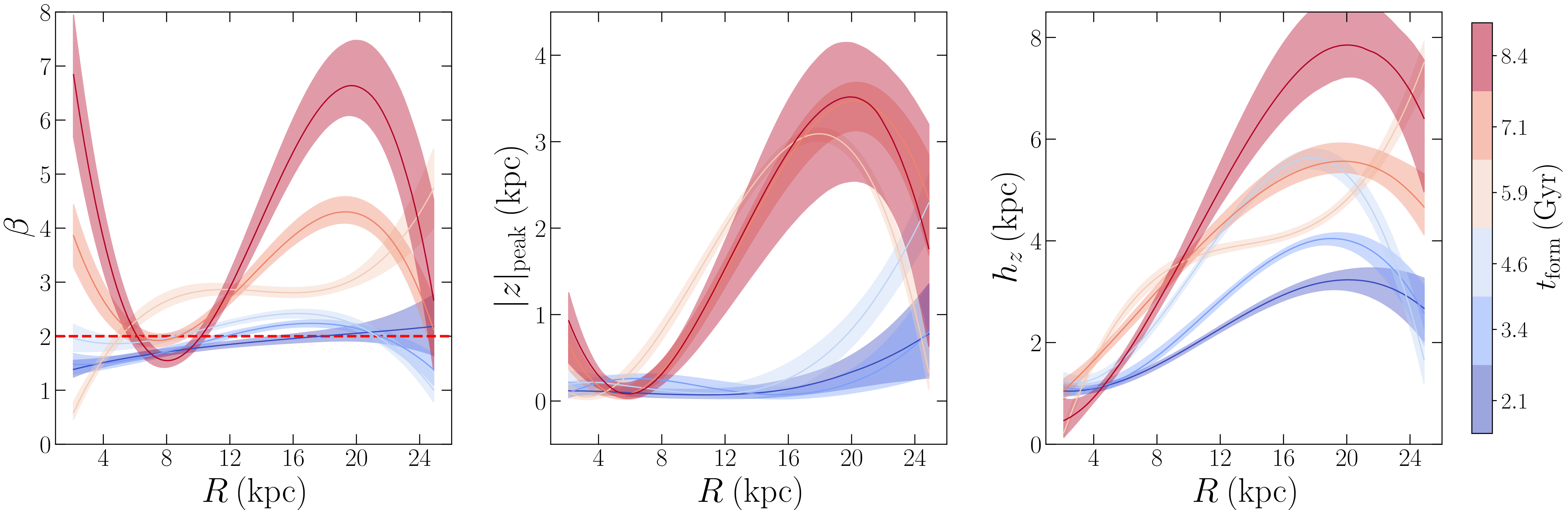} 
	\caption{Best fit parameters of the vertical density profiles, Eq.~\ref{eq:xi}, for different mono-age populations, colour-coded by the formation time. All quantities have complex radial dependencies, but are simpler in the restricted interval $8\lesssim R/\kpc \lesssim 20$, where, broadly speaking, all quantities increase monotonically with radius for most populations. The $\beta$ parameter (left panel) of no population is compatible with a Gaussian (horizontal dashed line) for all radii, while young populations (red) are highly non-Gaussian. The central panel shows the peak location, which seems to split into two groups: old populations ($\tf\lesssim 5\Gyr$) peak near the plane ($0\lesssim \zpeak\lesssim 1$), while for younger populations $\zpeak$ increases rapidly with radius, achieving $\zpeak\approx 4\kpc$. The right panel shows that, broadly speaking, the scale parameter $h_z$ increases with radius (flaring profiles), with younger populations flaring more than the older ones.}
	\label{fig:params_rho_gennorm}
\end{figure*}

Fig.~\ref{fig:params_rho_gennorm} shows the best fit values and uncertainties of parameters $\beta$, $\zpeak$ and $h_z$ (see Eq. \ref{eq:xi}) as functions of radius, for the same formation times shown in Fig.~\ref{fig:Sigma_rho}. As mentioned above, the radial variation of each of these parameters is modelled as a third-order polynomial, resulting in a total of 12 parameters.
The left panel shows the parameter $\beta$. The horizontal line at $\beta = 2$ represents a Gaussian distribution and we see that no mono-age population has $\xi(z|R)$ compatible with a Gaussian for all radii. The oldest population ($\tf\approx 2.1\Gyr$) is the only one for which $\beta$ increases monotonically as a function of radius, being compatible with a Gaussian for $R \gtrsim 16\kpc$, while $\xi(z|R)$ is more spiky than the Gaussian for $R\lesssim 16\kpc$. For all the other populations, $\beta$ depends non-trivially on radius. However, despite this complexity, broadly speaking in the range ${8\lesssim R/\kpc \lesssim 20}$, the parameter $\beta$ increases as a function of $R$ (for a fixed $\tf$) and as a function of $\tf$ (for a fixed $R$). The location of the peak, shown in the central panel, also has non-trivial dependencies on $R$, if we take into account all the radial interval. In the restricted interval ${8\lesssim R/\kpc \lesssim 20}$, we note two main groups, with the three oldest populations peaking at small $|z|$, with ${0\lesssim \zpeak/\kpc \lesssim 1}$, while for the three youngest populations $\zpeak$ increases fast with radius, achieving $\zpeak \approx 4\kpc$. This suggests some abrupt change in the final $\langle |z| \rangle$ for stars formed at $\tf \approx 5\Gyr$. Finally, the scale parameter $h_z$ is shown in the right panel. Once more, despite the complex radial dependence, if we restrict to the interval ${8\lesssim R/\kpc \lesssim 20}$, we note an approximately monotonic increase with $R$, \ie\ flaring vertical profiles, with younger populations flaring more than the old ones. 

In summary, stars formed in the warp make their way into the disc and settle into radial density profiles which can be described as skew-normal distributions \citep[][]{skewnorm}, with young stars approximately described by Gaussians peaking at a radius close to where the warp reaches its peak tilt, and which evolve to increasingly negatively skewed distributions and smaller $\Rpeak$ for older populations. The vertical density profiles are well described by the generalised normal distribution \citep[][]{Nadarajah_2005}, and become flatter and thicker, both as a function of $R$ and of $\tf$.


\section{Discussion}
\label{section:discussion}

We have studied the dynamical evolution of stars formed in the warp of an $N$-body$+$SPH simulation. We showed that the warp stars experience a rapid tilting, becoming more aligned with the disc (Figs.~\ref{fig:thetaf_v_thetaj} and \ref{fig:Mthetaj_evo}). The extent by which warp stars can tilt into the main plane is limited by the speed with which differential precession disrupts a coherent warped plane. Warp populations in the simulation tilt by $\lesssim 5\degrees$. Once they have tilted, warp stars continue to experience differential precession, which drives phase-mixing, a slower process that continues much longer. After settling, the warp stars populate the geometric thick disc (Fig.~\ref{fig:edgeon_warpndisk_age}), in good agreement with the results of \cite{roskar+10}, who showed that the warp stars in their cosmological simulation end up populating a geometric thick disc. 

In Figs.~\ref{fig:r_settle} and \ref{fig:r_rform} we found that the average radial positions of the warp stars are constantly decreasing even after settling. This decrease in the average radius is partly due to the fact that no warp stars, by our definition, are formed inside $10\kpc$, which means that there is a net inward migration of warp stars. This decrease by itself is not evidence of radial migration, since the stars might be reaching smaller radii via heating. Stars need to be on low-eccentricity orbits to migrate radially via spiral churning \citep{SellwoodBinney, Roskar+12}.
We found that warp stars have low eccentricities after they settle into the disc, making them susceptible to spiral churning which drives radial migration. Though warp stars populate the geometric thick disc (higher $|z|$), that does not exclude them from radial migration since thick disc stars can migrate \citep{solway+12, mikkola+19, beraldoesilva+21}.

\subsection{The role of subgrid physics}

The fiducial simulation presented in this work was run with a low gas density threshold to promote star formation in low-density regions, such as the warp. The resulting large number of warp stars ($\sim 6\times10^5$) allow us to examine the settling process in detail. At the same time, the total number of warp stars is still overall low ($\sim18\%$ of all stars) which means the low star formation threshold is not biasing the overall evolution of the galaxy, including the warp settling and migration processes themselves. We repeat part of the analysis of Section~\ref{section:dyn_evo} on the supplemental warped simulation described in Section~\ref{subsection:supp_sim}, which has different subgrid physics, including a higher density threshold for star formation. The result is presented in Appendix~\ref{sec:appendix}. The warp in the supplemental simulation wanes (Fig~\ref{fig:warp_briggs_HT}), with a $\sim 5\kpc$ decrease in its $|z_{\rm{max}}|$ over $6 \Gyr$. Throughout its $10\Gyr$ evolution, the simulation produces $2.5\times10^5$ warp stars (Fig~\ref{fig:HT_form}), which is $\sim10\%$ of all formed stars, while the fiducial simulation produces twice as many warp stars by $t=10\Gyr$.
We observe the same settling process in the warp populations of the supplemental simulation (Fig.~\ref{fig:HT_metheta}), however, the settling timescale is noticeably shorter, consistent with a smaller warp, resulting in warp stars forming closer to the galactic disc. Warp stars in the supplemental simulation also experience migration (Fig.~\ref{fig:HT_r_rform}), with $70\%$ of warp stars moving inwards. Thus the settling processes of warp stars are the same regardless of the subgrid physics.

What is strikingly different when the subgrid physics are changed is the vertical age profile. The negative vertical age gradient shown in Section~\ref{section:vertical_profiles} is not present in the supplemental simulation, but instead the distribution of \avgbin{\tform} appears to decrease with $|z|$ (Fig.~\ref{fig:edge_on_HT}, right panel). This is similar to the vertical age distribution observed in the MW \citep{Laporte+20}. Determining the cause of the opposing gradients is outside the scope of this work due to the parameter differences of the two simulations (see Section~\ref{section:Simulation}). However, we hypothesise that the lower gas density threshold (thus large number of warp stars) and growing warp play a role in forming a negative gradient as warp stars continue to form at increasingly higher $\thetaj$ and $|z|$ (Fig.~\ref{fig:z_settle}) and, subsequently, settle at higher $|z|$. Lastly, we find that the warp populations in the supplemental simulation also settle onto low-eccentricity orbits and are able to reach the Solar annulus (Fig.~\ref{fig:HT_r_rform}). As a result, the settling and inward migration of warp populations in our simulations appear to be generic for stars formed in the warp regardless of the subgrid physics employed.

\subsection{Consequences for the Milky Way}
\subsubsection{Stellar populations tracing the warp}
The study of stellar populations in NGC~4565 by \cite{Radburn+14} showed that the H{\sc i} warp is traced by young stars ($\rm{age}\leq600\Myr$), while older ($>1\Gyr$) stars are symmetrically distributed around the mid-plane. More recently, the \cite{chen} discovery of a warp signature in the MW's Cepheid distribution also reaffirmed that star formation can occur in the H{\sc i} warp. From the populations they considered, they found that only the youngest one, namely the Cepheids, closely traced the H{\sc i} warp. This suggests not only that the Cepheids formed in-situ, but also that the MW's warp cannot have a purely tidal origin, since a corresponding warp signature would have been observed in other stellar populations in a tidal scenario.

In Section~\ref{section:dyn_evo} we showed that, in our fiducial simulation, warp populations initially trace the gas warp (Fig.~\ref{fig:briggs_evo}) and present an $m=1$ distribution in angular momentum space (Figs.~\ref{fig:tau} and \ref{fig:thetajM_evo}). However, we also observed how the correspondence between the warp stars and the gas warp faded as these stars tilted and phase-mixed into the galactic disc. The decoupling happens rapidly during the first $0.3\Gyr$ of each warp population's evolution, indicating that only the youngest populations would trace the warp. This suggests that misaligned gas accretion, rather than tidal interaction, is the predominant cause of the Milky Way's warp. 

\subsubsection{Warp stars in the Solar annulus}
Another important question is whether warp populations can eventually reach the Solar Neighbourhood.
The warp stars in our simulation form at increasingly large radii, and in any case are all formed outside $R=10\kpc$, and therefore, in order to observe them in the Solar Neighbourhood, these stars must migrate radially. 
In Section \ref{subsection:migration} we have shown that warp stars migrate inwards to quite small radii via spiral churning. We conclude that stars forming in the Milky Way's warp can indeed be found in the Solar Neighbourhood, contaminating the thick disc. Note that our fiducial simulation is tuned to promote star formation in the warp, so we expect that warp stars in real galaxies, such as the MW, will form only a trace population in the geomtric thick disc. The age range of warp stars at the Solar annulus of our fiducial simulation shows that due to the diffusive nature of migration, the only young warp stars that can be observed are those born closest to the Solar annulus (see right panel of Fig.~\ref{fig:r_rform}). Young warp stars born further in the warp experience rapid tilting, but do not have enough time to migrate to the Solar Neighbourhood, unlike older populations which are observed throughout the disc.

Since the detailed history of the warp in our fiducial model is unlikely to match that of the Milky Way's warp, we refrain from more direct comparisons of our fiducial model to the Milky Way. Indeed we see that the negative age gradient in the outer disc shown in Fig.~\ref{fig:edgeon_warpndisk_age} is opposite to that observed in the MW \citep{Laporte+20, 2021MNRAS.502.5686I}. However, this age gradient is only a feature of this specific model, in which the subgrid physics have been set to promote copious star formation in the warp. Instead, the supplemental simulation in Appendix~\ref{sec:appendix}, evolved from the same initial conditions but with different subgrid physics which are less favourable to star formation in the low density gas of the warp, has a vertical age gradient of the same sign as the MW. Nonetheless, one observation that both models can match is the presence of younger ($\lesssim3\Gyr$) flaring populations  at higher $|z|$ on the outskirts of the MW \citep{Mackereth+17,Feuillet+19,Sharma+21}. Based on the flared warp populations in Section~\ref{section:disc_structure}, we speculate that if the MW warp formed via misaligned gas accretion, then warp stars that have tilted may explain, in part, the presence of young flaring populations on the outskirts of the MW.

\subsection{Summary}
\label{section:sum_con}

We have selected populations of stars formed in the warp
of an $N$-body+SPH simulation. We identified warp populations by first measuring the inclination of the angular momentum, $\thetaj$, and cylindrical radius, $R_{\rm{form}}$, of each star at formation. Then, we isolate a highly inclined stellar population that formed on the outskirts of the galactic disc, identifying these stars as a warp population. We proceeded to analyse the dynamical evolution of the warp stars and, in summary, have shown that:
    
\begin{itemize}
\item Most warp stars tilt to become more aligned with the galactic disc by $\sim 5\degrees$. Orbital tilting is evident from the mean tilt of the angular momenta, $\avg{\thetaj}$, and the mean absolute height above the mid-plane, $\avg{|z|}$, of mono-age warp populations, which experience rapid declines during the first $\sim 1 \Gyr$ before becoming roughly constant. Using the $\avg{\thetaj}$ and $\avg{|z|}$ rates of change, we found tilting times ranging from $0.25 \Gyr$ to $1.75\Gyr$. Warp tilting ends when differential precession of different radii disrupt the coherent plane of the warped population. Once tilting is over, the average height of a population remains approximately constant.
\item Mono-age warp populations phase-mix in angular momentum space via differential precession at different rates. Stars are completely homogeneous in the distribution of the angular momentum azimuthal angle $\phi_L$ after $\sim6\Gyr$.
\item The time derivative of the vertical angular momentum, $L_z$, along with that of the radial positions is negative after $1\Gyr$ for all warp populations and both decrease until the end of the simulation. This is suggestive of inward radial migration of warp populations.
\item We found that almost all warp stars that have settled are on close to circular orbits, with mean eccentricities ranging from $0.2$ to $0.3$ for all settled warp populations. These low eccentricities indicate that warp stars are able to migrate to the inner disc via spiral churning.
\item A detailed modelling of the density distribution of settled warp stars finds that their initial torus-shaped density distribution is slowly filled in as warp stars migrate to smaller radii.  Because the warp in the fiducial model grows with time, the warp populations settle to increasingly thicker tori/discs.
\item A settled mono-age warp population is radially flaring. In our model the flaring increases with the formation time of the population, an indication of the growing warp.
\item By means of a supplemental simulation which implements (more realistic) subgrid physics less favourable to star formation in the warp, we demonstrate that settling processes are unchanged. However the resulting reduced amount of star formation in the warp, and the smaller warp, result in a vertical age gradient which is opposite to that in the fiducial model. We conclude that age profiles cannot constrain the warp's formation mechanism but only its star formation history and evolution.

\item We find that, at formation, warp stars trace the gas warp but then quickly ($\sim 0.3\Gyr$) lose coherence as they settle into the disc. This result matches observations in the MW and NGC~4565 that only younger populations trace the H{\sc i} warp.
\end{itemize}

\section*{Acknowledgements}
We thank the anonymous referee for comments that improved the paper. V.P.D. and L.B.S. are supported by STFC Consolidated grant \#~ST/R000786/1. The simulation used in this paper was run at the DiRAC Shared Memory Processing system at the University of Cambridge, operated by the COSMOS Project at the Department of Applied Mathematics and Theoretical Physics on behalf of the STFC DiRAC HPC Facility (www.dirac.ac.uk). This equipment was funded by BIS National E-infrastructure capital grant ST/J005673/1, STFC capital grant ST/H008586/1 and STFC DiRAC Operations grant ST/K00333X/1. DiRAC is part of the National E-Infrastructure.

\section*{Data availability}

The simulation dataset can be shared on reasonable request.

\bibliographystyle{mnras}
\bibliography{references} 

\begin{thebibliography}{}
\makeatletter
\relax
\def\mn@urlcharsother{\let\do\@makeother \do\$\do\&\do\#\do\^\do\_\do\%\do\~}
\def\mn@doi{\begingroup\mn@urlcharsother \@ifnextchar [ {\mn@doi@}
  {\mn@doi@[]}}
\def\mn@doi@[#1]#2{\def\@tempa{#1}\ifx\@tempa\@empty \href
  {http://dx.doi.org/#2} {doi:#2}\else \href {http://dx.doi.org/#2} {#1}\fi
  \endgroup}
\def\mn@eprint#1#2{\mn@eprint@#1:#2::\@nil}
\def\mn@eprint@arXiv#1{\href {http://arxiv.org/abs/#1} {{\tt arXiv:#1}}}
\def\mn@eprint@dblp#1{\href {http://dblp.uni-trier.de/rec/bibtex/#1.xml}
  {dblp:#1}}
\def\mn@eprint@#1:#2:#3:#4\@nil{\def\@tempa {#1}\def\@tempb {#2}\def\@tempc
  {#3}\ifx \@tempc \@empty \let \@tempc \@tempb \let \@tempb \@tempa \fi \ifx
  \@tempb \@empty \def\@tempb {arXiv}\fi \@ifundefined
  {mn@eprint@\@tempb}{\@tempb:\@tempc}{\expandafter \expandafter \csname
  mn@eprint@\@tempb\endcsname \expandafter{\@tempc}}}

\bibitem[\protect\citeauthoryear{{Agertz}, {Teyssier}  \& {Moore}}{{Agertz}
  et~al.}{2009}]{Agertz+2009}
{Agertz} O.,  {Teyssier} R.,   {Moore} B.,  2009, \mn@doi [\mnras]
  {10.1111/j.1745-3933.2009.00685.x}, \href
  {https://ui.adsabs.harvard.edu/abs/2009MNRAS.397L..64A} {397, L64}

\bibitem[\protect\citeauthoryear{{Aumer} \& {White}}{{Aumer} \&
  {White}}{2013}]{AumerWhite2013}
{Aumer} M.,  {White} S. D.~M.,  2013, \mn@doi [\mnras] {10.1093/mnras/sts083},
  \href {https://ui.adsabs.harvard.edu/abs/2013MNRAS.428.1055A} {428, 1055}

\bibitem[\protect\citeauthoryear{Azzalini}{Azzalini}{1985}]{skewnorm}
Azzalini A.,  1985, Scandinavian Journal of Statistics, 12, 171

\bibitem[\protect\citeauthoryear{{Bailin}}{{Bailin}}{2003}]{bailin03}
{Bailin} J.,  2003, \mn@doi [\apjl] {10.1086/368160}, \href
  {https://ui.adsabs.harvard.edu/abs/2003ApJ...583L..79B} {583, L79}

\bibitem[\protect\citeauthoryear{{Bailin} et~al.,}{{Bailin}
  et~al.}{2005}]{Bailin+05}
{Bailin} J.,  et~al., 2005, \mn@doi [\apjl] {10.1086/432157}, \href
  {https://ui.adsabs.harvard.edu/abs/2005ApJ...627L..17B} {627, L17}

\bibitem[\protect\citeauthoryear{{Beraldo e Silva}, {de Siqueira Pedra},
  {Valluri}, {Sodr{\'e}}  \& {Bru}}{{Beraldo e Silva}
  et~al.}{2019a}]{beraldoesilva+19a}
{Beraldo e Silva} L.,  {de Siqueira Pedra} W.,  {Valluri} M.,  {Sodr{\'e}} L.,
   {Bru} J.-B.,  2019a, \mn@doi [\apj] {10.3847/1538-4357/aaf397}, \href
  {https://ui.adsabs.harvard.edu/abs/2019ApJ...870..128B} {870, 128}

\bibitem[\protect\citeauthoryear{{Beraldo e Silva}, {de Siqueira Pedra}  \&
  {Valluri}}{{Beraldo e Silva} et~al.}{2019b}]{beraldoesilva+19b}
{Beraldo e Silva} L.,  {de Siqueira Pedra} W.,   {Valluri} M.,  2019b, \mn@doi
  [\apj] {10.3847/1538-4357/aaf8a7}, \href
  {https://ui.adsabs.harvard.edu/abs/2019ApJ...872...20B} {872, 20}

\bibitem[\protect\citeauthoryear{{Beraldo e Silva}, {Debattista},
  {Khachaturyants}  \& {Nidever}}{{Beraldo e Silva}
  et~al.}{2020}]{beraldoesilva+20}
{Beraldo e Silva} L.,  {Debattista} V.~P.,  {Khachaturyants} T.,   {Nidever}
  D.,  2020, \mn@doi [\mnras] {10.1093/mnras/staa065}, \href
  {https://ui.adsabs.harvard.edu/abs/2020MNRAS.492.4716B} {492, 4716}

\bibitem[\protect\citeauthoryear{{Beraldo e Silva}, {Debattista}, {Nidever},
  {Amarante}  \& {Garver}}{{Beraldo e Silva} et~al.}{2021}]{beraldoesilva+21}
{Beraldo e Silva} L.,  {Debattista} V.~P.,  {Nidever} D.,  {Amarante} J. A.~S.,
    {Garver} B.,  2021, \mn@doi [MNRAS] {10.1093/mnras/staa3966}, \href
  {https://ui.adsabs.harvard.edu/abs/2021MNRAS.502..260B} {502, 260}

\bibitem[\protect\citeauthoryear{Biau \& Devroye}{Biau \&
  Devroye}{2015}]{10.5555/2875145}
Biau G.,  Devroye L.,  2015, Lectures on the Nearest Neighbor Method, 1st edn.
Springer Publishing Company, Incorporated

\bibitem[\protect\citeauthoryear{{Binney}}{{Binney}}{1992}]{binney92}
{Binney} J.,  1992, \mn@doi [\araa] {10.1146/annurev.aa.30.090192.000411},
  \href {https://ui.adsabs.harvard.edu/abs/1992ARA&A..30...51B} {30, 51}

\bibitem[\protect\citeauthoryear{{Binney}, {Dehnen}  \& {Bertelli}}{{Binney}
  et~al.}{2000}]{2000MNRAS.318..658B}
{Binney} J.,  {Dehnen} W.,   {Bertelli} G.,  2000, \mn@doi [\mnras]
  {10.1046/j.1365-8711.2000.03720.x}, \href
  {https://ui.adsabs.harvard.edu/abs/2000MNRAS.318..658B} {318, 658}

\bibitem[\protect\citeauthoryear{{Bosma}}{{Bosma}}{1991}]{bosma}
{Bosma} A.,  1991, in Warped Disks and Inclined Rings around Galaxies. p.~181

\bibitem[\protect\citeauthoryear{{Bovy}, {Rix}, {Schlafly}, {Nidever},
  {Holtzman}, {Shetrone}  \& {Beers}}{{Bovy} et~al.}{2016}]{bovy+16}
{Bovy} J.,  {Rix} H.-W.,  {Schlafly} E.~F.,  {Nidever} D.~L.,  {Holtzman}
  J.~A.,  {Shetrone} M.,   {Beers} T.~C.,  2016, \mn@doi [\apj]
  {10.3847/0004-637X/823/1/30}, \href
  {https://ui.adsabs.harvard.edu/abs/2016ApJ...823...30B} {823, 30}

\bibitem[\protect\citeauthoryear{{Briggs}}{{Briggs}}{1990}]{Briggs90}
{Briggs} F.~H.,  1990, \mn@doi [\apj] {10.1086/168512}, \href
  {https://ui.adsabs.harvard.edu/abs/1990ApJ...352...15B} {352, 15}

\bibitem[\protect\citeauthoryear{{Bullock}, {Dekel}, {Kolatt}, {Kravtsov},
  {Klypin}, {Porciani}  \& {Primack}}{{Bullock} et~al.}{2001}]{Bullock+2001}
{Bullock} J.~S.,  {Dekel} A.,  {Kolatt} T.~S.,  {Kravtsov} A.~V.,  {Klypin}
  A.~A.,  {Porciani} C.,   {Primack} J.~R.,  2001, \mn@doi [\apj]
  {10.1086/321477}, \href
  {https://ui.adsabs.harvard.edu/abs/2001ApJ...555..240B} {555, 240}

\bibitem[\protect\citeauthoryear{{Chen}, {Jing}  \& {Yoshikaw}}{{Chen}
  et~al.}{2003}]{Chen+03}
{Chen} D.~N.,  {Jing} Y.~P.,   {Yoshikaw} K.,  2003, \mn@doi [\apj]
  {10.1086/378379}, \href
  {https://ui.adsabs.harvard.edu/abs/2003ApJ...597...35C} {597, 35}

\bibitem[\protect\citeauthoryear{{Chen}, {Wang}, {Deng}, {de Grijs}  \&
  {Yang}}{{Chen} et~al.}{2018}]{2018ApJS..237...28C}
{Chen} X.,  {Wang} S.,  {Deng} L.,  {de Grijs} R.,   {Yang} M.,  2018, \mn@doi
  [\apjs] {10.3847/1538-4365/aad32b}, \href
  {https://ui.adsabs.harvard.edu/abs/2018ApJS..237...28C} {237, 28}

\bibitem[\protect\citeauthoryear{{Chen}, {Wang}, {Deng}, {de Grijs}, {Liu}  \&
  {Tian}}{{Chen} et~al.}{2019}]{chen}
{Chen} X.,  {Wang} S.,  {Deng} L.,  {de Grijs} R.,  {Liu} C.,   {Tian} H.,
  2019, \mn@doi [Nature Astronomy] {10.1038/s41550-018-0686-7}, \href
  {https://ui.adsabs.harvard.edu/abs/2019NatAs...3..320C} {3, 320}

\bibitem[\protect\citeauthoryear{{Debattista}, {Ro{\v{s}}kar}, {Valluri},
  {Quinn}, {Moore}  \& {Wadsley}}{{Debattista} et~al.}{2013}]{Debattista+13}
{Debattista} V.~P.,  {Ro{\v{s}}kar} R.,  {Valluri} M.,  {Quinn} T.,  {Moore}
  B.,   {Wadsley} J.,  2013, \mn@doi [\mnras] {10.1093/mnras/stt1217}, \href
  {https://ui.adsabs.harvard.edu/abs/2013MNRAS.434.2971D} {434, 2971}

\bibitem[\protect\citeauthoryear{{Debattista}, {van den Bosch}, {Ro{\v{s}}kar},
  {Quinn}, {Moore}  \& {Cole}}{{Debattista} et~al.}{2015a}]{Debattista+15}
{Debattista} V.~P.,  {van den Bosch} F.~C.,  {Ro{\v{s}}kar} R.,  {Quinn} T.,
  {Moore} B.,   {Cole} D.~R.,  2015a, \mn@doi [\mnras] {10.1093/mnras/stv1563},
  \href {https://ui.adsabs.harvard.edu/abs/2015MNRAS.452.4094D} {452, 4094}

\bibitem[\protect\citeauthoryear{{Debattista}, {van den Bosch}, {Ro{\v{s}}kar},
  {Quinn}, {Moore}  \& {Cole}}{{Debattista} et~al.}{2015b}]{vpd2015}
{Debattista} V.~P.,  {van den Bosch} F.~C.,  {Ro{\v{s}}kar} R.,  {Quinn} T.,
  {Moore} B.,   {Cole} D.~R.,  2015b, \mn@doi [\mnras] {10.1093/mnras/stv1563},
  \href {https://ui.adsabs.harvard.edu/abs/2015MNRAS.452.4094D} {452, 4094}

\bibitem[\protect\citeauthoryear{{Djorgovski} \& {Sosin}}{{Djorgovski} \&
  {Sosin}}{1989}]{Djorgovski+89}
{Djorgovski} S.,  {Sosin} C.,  1989, \mn@doi [\apjl] {10.1086/185446}, \href
  {https://ui.adsabs.harvard.edu/abs/1989ApJ...341L..13D} {341, L13}

\bibitem[\protect\citeauthoryear{{Drimmel} \& {Spergel}}{{Drimmel} \&
  {Spergel}}{2001}]{Drimmel+01}
{Drimmel} R.,  {Spergel} D.~N.,  2001, \mn@doi [\apj] {10.1086/321556}, \href
  {https://ui.adsabs.harvard.edu/abs/2001ApJ...556..181D} {556, 181}

\bibitem[\protect\citeauthoryear{{Duckworth}, {Tojeiro}  \&
  {Kraljic}}{{Duckworth} et~al.}{2020}]{Duckworth+20}
{Duckworth} C.,  {Tojeiro} R.,   {Kraljic} K.,  2020, \mn@doi [\mnras]
  {10.1093/mnras/stz3575}, \href
  {https://ui.adsabs.harvard.edu/abs/2020MNRAS.492.1869D} {492, 1869}

\bibitem[\protect\citeauthoryear{{Earp}, {Debattista}, {Macci{\`o}}, {Wang},
  {Buck}  \& {Khachaturyants}}{{Earp} et~al.}{2019}]{earp+19}
{Earp} S. W.~F.,  {Debattista} V.~P.,  {Macci{\`o}} A.~V.,  {Wang} L.,  {Buck}
  T.,   {Khachaturyants} T.,  2019, \mn@doi [\mnras] {10.1093/mnras/stz2109},
  \href {https://ui.adsabs.harvard.edu/abs/2019MNRAS.488.5728E} {488, 5728}

\bibitem[\protect\citeauthoryear{{Feuillet}, {Frankel}, {Lind}, {Frinchaboy},
  {Garc{\'\i}a-Hern{\'a}ndez}, {Lane}, {Nitschelm}  \&
  {Roman-Lopes}}{{Feuillet} et~al.}{2019}]{Feuillet+19}
{Feuillet} D.~K.,  {Frankel} N.,  {Lind} K.,  {Frinchaboy} P.~M.,
  {Garc{\'\i}a-Hern{\'a}ndez} D.~A.,  {Lane} R.~R.,  {Nitschelm} C.,
  {Roman-Lopes} A.,  2019, \mn@doi [\mnras] {10.1093/mnras/stz2221}, \href
  {https://ui.adsabs.harvard.edu/abs/2019MNRAS.489.1742F} {489, 1742}

\bibitem[\protect\citeauthoryear{{Foreman-Mackey}, {Hogg}, {Lang}  \&
  {Goodman}}{{Foreman-Mackey} et~al.}{2013}]{2013PASP..125..306F}
{Foreman-Mackey} D.,  {Hogg} D.~W.,  {Lang} D.,   {Goodman} J.,  2013, \mn@doi
  [PASP] {10.1086/670067}, \href
  {http://adsabs.harvard.edu/abs/2013PASP..125..306F} {125, 306}

\bibitem[\protect\citeauthoryear{{Fraternali} \& {Binney}}{{Fraternali} \&
  {Binney}}{2008}]{Fraternali}
{Fraternali} F.,  {Binney} J.~J.,  2008, \mn@doi [\mnras]
  {10.1111/j.1365-2966.2008.13071.x}, \href
  {https://ui.adsabs.harvard.edu/abs/2008MNRAS.386..935F} {386, 935}

\bibitem[\protect\citeauthoryear{{Freudenreich}}{{Freudenreich}}{1998}]{Freudenreich+98}
{Freudenreich} H.~T.,  1998, \mn@doi [\apj] {10.1086/305065}, \href
  {https://ui.adsabs.harvard.edu/abs/1998ApJ...492..495F} {492, 495}

\bibitem[\protect\citeauthoryear{{Gaia Collaboration} et~al.,}{{Gaia
  Collaboration} et~al.}{2018}]{2018A&A...616A...1G}
{Gaia Collaboration} et~al., 2018, \mn@doi [\aap]
  {10.1051/0004-6361/201833051}, \href
  {https://ui.adsabs.harvard.edu/abs/2018A&A...616A...1G} {616, A1}

\bibitem[\protect\citeauthoryear{{Garc{\'\i}a-Ruiz}, {Sancisi}  \&
  {Kuijken}}{{Garc{\'\i}a-Ruiz} et~al.}{2002}]{garciaruiz02}
{Garc{\'\i}a-Ruiz} I.,  {Sancisi} R.,   {Kuijken} K.,  2002, \mn@doi [\aap]
  {10.1051/0004-6361:20020976}, \href
  {https://ui.adsabs.harvard.edu/abs/2002A&A...394..769G} {394, 769}

\bibitem[\protect\citeauthoryear{{G{\'o}mez}, {Minchev}, {O'Shea}, {Beers},
  {Bullock}  \& {Purcell}}{{G{\'o}mez} et~al.}{2013}]{2013MNRAS.429..159G}
{G{\'o}mez} F.~A.,  {Minchev} I.,  {O'Shea} B.~W.,  {Beers} T.~C.,  {Bullock}
  J.~S.,   {Purcell} C.~W.,  2013, \mn@doi [\mnras] {10.1093/mnras/sts327},
  \href {https://ui.adsabs.harvard.edu/abs/2013MNRAS.429..159G} {429, 159}

\bibitem[\protect\citeauthoryear{{G{\'o}mez}, {White}, {Grand }, {Marinacci},
  {Springel}  \& {Pakmor}}{{G{\'o}mez} et~al.}{2017}]{Gomez+17}
{G{\'o}mez} F.~A.,  {White} S. D.~M.,  {Grand } R. J.~J.,  {Marinacci} F.,
  {Springel} V.,   {Pakmor} R.,  2017, \mn@doi [\mnras]
  {10.1093/mnras/stw2957}, \href
  {https://ui.adsabs.harvard.edu/abs/2017MNRAS.465.3446G} {465, 3446}

\bibitem[\protect\citeauthoryear{{Herbert-Fort}, {Zaritsky}, {Christlein}  \&
  {Kannappan}}{{Herbert-Fort} et~al.}{2010}]{zaritsky}
{Herbert-Fort} S.,  {Zaritsky} D.,  {Christlein} D.,   {Kannappan} S.~J.,
  2010, \mn@doi [\apj] {10.1088/0004-637X/715/2/902}, \href
  {https://ui.adsabs.harvard.edu/abs/2010ApJ...715..902H} {715, 902}

\bibitem[\protect\citeauthoryear{{Iorio} \& {Belokurov}}{{Iorio} \&
  {Belokurov}}{2021}]{2021MNRAS.502.5686I}
{Iorio} G.,  {Belokurov} V.,  2021, \mn@doi [\mnras] {10.1093/mnras/stab005},
  \href {https://ui.adsabs.harvard.edu/abs/2021MNRAS.502.5686I} {502, 5686}

\bibitem[\protect\citeauthoryear{{Jiang} \& {Binney}}{{Jiang} \&
  {Binney}}{1999}]{1999MNRAS.303L...7J}
{Jiang} I.-G.,  {Binney} J.,  1999, \mn@doi [\mnras]
  {10.1046/j.1365-8711.1999.02333.x}, \href
  {https://ui.adsabs.harvard.edu/abs/1999MNRAS.303L...7J} {303, L7}

\bibitem[\protect\citeauthoryear{{Kalberla}, {Dedes}, {Kerp}  \&
  {Haud}}{{Kalberla} et~al.}{2007}]{Kalberla}
{Kalberla} P.~M.~W.,  {Dedes} L.,  {Kerp} J.,   {Haud} U.,  2007, \mn@doi
  [\aap] {10.1051/0004-6361:20066362}, \href
  {https://ui.adsabs.harvard.edu/abs/2007A&A...469..511K} {469, 511}

\bibitem[\protect\citeauthoryear{{Kerr}}{{Kerr}}{1957}]{Kerr}
{Kerr} F.~J.,  1957, \mn@doi [\aj] {10.1086/107466}, \href
  {https://ui.adsabs.harvard.edu/abs/1957AJ.....62...93K} {62, 93}

\bibitem[\protect\citeauthoryear{{Kuijken} \& {Garcia-Ruiz}}{{Kuijken} \&
  {Garcia-Ruiz}}{2001}]{kuijken+01}
{Kuijken} K.,  {Garcia-Ruiz} I.,  2001, in {Funes} J.~G.,  {Corsini} E.~M.,
  eds,  Astronomical Society of the Pacific Conference Series Vol. 230, Galaxy
  Disks and Disk Galaxies. pp 401--408 (\mn@eprint {arXiv} {astro-ph/0011345})

\bibitem[\protect\citeauthoryear{{Laporte}, {G{\'o}mez}, {Besla}, {Johnston}
  \& {Garavito-Camargo}}{{Laporte} et~al.}{2018}]{2018MNRAS.473.1218L}
{Laporte} C. F.~P.,  {G{\'o}mez} F.~A.,  {Besla} G.,  {Johnston} K.~V.,
  {Garavito-Camargo} N.,  2018, \mn@doi [\mnras] {10.1093/mnras/stx2146}, \href
  {https://ui.adsabs.harvard.edu/abs/2018MNRAS.473.1218L} {473, 1218}

\bibitem[\protect\citeauthoryear{{Laporte}, {Belokurov}, {Koposov}, {Smith}  \&
  {Hill}}{{Laporte} et~al.}{2020}]{Laporte+20}
{Laporte} C. F.~P.,  {Belokurov} V.,  {Koposov} S.~E.,  {Smith} M.~C.,   {Hill}
  V.,  2020, \mn@doi [\mnras] {10.1093/mnrasl/slz167}, \href
  {https://ui.adsabs.harvard.edu/abs/2020MNRAS.492L..61L} {492, L61}

\bibitem[\protect\citeauthoryear{{Levine}, {Blitz}  \& {Heiles}}{{Levine}
  et~al.}{2006}]{levine06}
{Levine} E.~S.,  {Blitz} L.,   {Heiles} C.,  2006, \mn@doi [\apj]
  {10.1086/503091}, \href
  {https://ui.adsabs.harvard.edu/\#abs/2006ApJ...643..881L} {643, 881}

\bibitem[\protect\citeauthoryear{{Li}, {Wang}, {Yang}, {Chen}, {Xie}  \&
  {Wang}}{{Li} et~al.}{2013}]{Li+13}
{Li} Z.,  {Wang} Y.,  {Yang} X.,  {Chen} X.,  {Xie} L.,   {Wang} X.,  2013,
  \mn@doi [\apj] {10.1088/0004-637X/768/1/20}, \href
  {https://ui.adsabs.harvard.edu/abs/2013ApJ...768...20L} {768, 20}

\bibitem[\protect\citeauthoryear{{L{\'o}pez-Corredoira}, {Cabrera-Lavers},
  {Garz{\'o}n}  \& {Hammersley}}{{L{\'o}pez-Corredoira} et~al.}{2002}]{rcwarp}
{L{\'o}pez-Corredoira} M.,  {Cabrera-Lavers} A.,  {Garz{\'o}n} F.,
  {Hammersley} P.~L.,  2002, \mn@doi [\aap] {10.1051/0004-6361:20021175}, \href
  {https://ui.adsabs.harvard.edu/abs/2002A&A...394..883L} {394, 883}

\bibitem[\protect\citeauthoryear{{Macci{\`o}}, {Moore}  \&
  {Stadel}}{{Macci{\`o}} et~al.}{2006}]{2006ApJ...636L..25M}
{Macci{\`o}} A.~V.,  {Moore} B.,   {Stadel} J.,  2006, \mn@doi [\apjl]
  {10.1086/499778}, \href
  {https://ui.adsabs.harvard.edu/abs/2006ApJ...636L..25M} {636, L25}

\bibitem[\protect\citeauthoryear{{Mackereth} et~al.,}{{Mackereth}
  et~al.}{2017}]{Mackereth+17}
{Mackereth} J.~T.,  et~al., 2017, \mn@doi [\mnras] {10.1093/mnras/stx1774},
  \href {https://ui.adsabs.harvard.edu/abs/2017MNRAS.471.3057M} {471, 3057}

\bibitem[\protect\citeauthoryear{{Mikkola}, {McMillan}  \& {Hobbs}}{{Mikkola}
  et~al.}{2020}]{mikkola+19}
{Mikkola} D.,  {McMillan} P.~J.,   {Hobbs} D.,  2020, \mn@doi [\mnras]
  {10.1093/mnras/staa1223}, \href
  {https://ui.adsabs.harvard.edu/abs/2020MNRAS.495.3295M} {495, 3295}

\bibitem[\protect\citeauthoryear{{Miller} \& {Scalo}}{{Miller} \&
  {Scalo}}{1979}]{MillerScalo1979}
{Miller} G.~E.,  {Scalo} J.~M.,  1979, \mn@doi [\apjs] {10.1086/190629}, \href
  {https://ui.adsabs.harvard.edu/abs/1979ApJS...41..513M} {41, 513}

\bibitem[\protect\citeauthoryear{{Mondal}, {Subramaniam}  \& {George}}{{Mondal}
  et~al.}{2019}]{mondal}
{Mondal} C.,  {Subramaniam} A.,   {George} K.,  2019, \mn@doi [Journal of
  Astrophysics and Astronomy] {10.1007/s12036-019-9603-4}, \href
  {https://ui.adsabs.harvard.edu/abs/2019JApA...40...35M} {40, 35}

\bibitem[\protect\citeauthoryear{Nadarajah}{Nadarajah}{2005}]{Nadarajah_2005}
Nadarajah S.,  2005, \mn@doi [Journal of Applied Statistics]
  {10.1080/02664760500079464}, 32, 685

\bibitem[\protect\citeauthoryear{{Navarro}, {Frenk}  \& {White}}{{Navarro}
  et~al.}{1996}]{NFW1996}
{Navarro} J.~F.,  {Frenk} C.~S.,   {White} S. D.~M.,  1996, \mn@doi [\apj]
  {10.1086/177173}, \href
  {https://ui.adsabs.harvard.edu/abs/1996ApJ...462..563N} {462, 563}

\bibitem[\protect\citeauthoryear{{Nierenberg}, {Auger}, {Treu}, {Marshall}  \&
  {Fassnacht}}{{Nierenberg} et~al.}{2011}]{Nierenberg+11}
{Nierenberg} A.~M.,  {Auger} M.~W.,  {Treu} T.,  {Marshall} P.~J.,
  {Fassnacht} C.~D.,  2011, \mn@doi [\apj] {10.1088/0004-637X/731/1/44}, \href
  {https://ui.adsabs.harvard.edu/abs/2011ApJ...731...44N} {731, 44}

\bibitem[\protect\citeauthoryear{{Ostriker} \& {Binney}}{{Ostriker} \&
  {Binney}}{1989}]{1989MNRAS.237..785O}
{Ostriker} E.~C.,  {Binney} J.~J.,  1989, \mn@doi [\mnras]
  {10.1093/mnras/237.3.785}, \href
  {https://ui.adsabs.harvard.edu/abs/1989MNRAS.237..785O} {237, 785}

\bibitem[\protect\citeauthoryear{{Porcel} \& {Battaner}}{{Porcel} \&
  {Battaner}}{1995}]{Porcel+95}
{Porcel} C.,  {Battaner} E.,  1995, \mn@doi [\mnras]
  {10.1093/mnras/274.4.1153}, \href
  {https://ui.adsabs.harvard.edu/abs/1995MNRAS.274.1153P} {274, 1153}

\bibitem[\protect\citeauthoryear{Powell}{Powell}{1964}]{1964Powell}
Powell M. J.~D.,  1964, \mn@doi [The Computer Journal]
  {10.1093/comjnl/7.2.155}, 7, 155

\bibitem[\protect\citeauthoryear{Press, Teukolsky, Vetterling  \&
  Flannery}{Press et~al.}{1992}]{PresTeukVettFlan92}
Press W.~H.,  Teukolsky S.~A.,  Vetterling W.~T.,   Flannery B.~P.,  1992,
  Numerical Recipes in C, second edn.
Cambridge University Press, Cambridge, USA

\bibitem[\protect\citeauthoryear{{Purcell}, {Bullock}, {Tollerud}, {Rocha}  \&
  {Chakrabarti}}{{Purcell} et~al.}{2011}]{2011Natur.477..301P}
{Purcell} C.~W.,  {Bullock} J.~S.,  {Tollerud} E.~J.,  {Rocha} M.,
  {Chakrabarti} S.,  2011, \mn@doi [\nat] {10.1038/nature10417}, \href
  {https://ui.adsabs.harvard.edu/abs/2011Natur.477..301P} {477, 301}

\bibitem[\protect\citeauthoryear{{Radburn-Smith} et~al.,}{{Radburn-Smith}
  et~al.}{2014}]{Radburn+14}
{Radburn-Smith} D.~J.,  et~al., 2014, \mn@doi [\apj]
  {10.1088/0004-637X/780/1/105}, \href
  {https://ui.adsabs.harvard.edu/abs/2014ApJ...780..105R} {780, 105}

\bibitem[\protect\citeauthoryear{{Reshetnikov}, {Battaner}, {Combes}  \&
  {Jim{\'e}nez-Vicente}}{{Reshetnikov} et~al.}{2002}]{reshetnikov}
{Reshetnikov} V.,  {Battaner} E.,  {Combes} F.,   {Jim{\'e}nez-Vicente} J.,
  2002, \mn@doi [\aap] {10.1051/0004-6361:20011672}, \href
  {https://ui.adsabs.harvard.edu/abs/2002A&A...382..513R} {382, 513}

\bibitem[\protect\citeauthoryear{{Ro{\v{s}}kar}, {Debattista}, {Quinn},
  {Stinson}  \& {Wadsley}}{{Ro{\v{s}}kar} et~al.}{2008}]{Roskar+08a}
{Ro{\v{s}}kar} R.,  {Debattista} V.~P.,  {Quinn} T.~R.,  {Stinson} G.~S.,
  {Wadsley} J.,  2008, \mn@doi [\apjl] {10.1086/592231}, \href
  {https://ui.adsabs.harvard.edu/abs/2008ApJ...684L..79R} {684, L79}

\bibitem[\protect\citeauthoryear{{Ro{\v{s}}kar}, {Debattista}, {Brooks},
  {Quinn}, {Brook}, {Governato}, {Dalcanton}  \& {Wadsley}}{{Ro{\v{s}}kar}
  et~al.}{2010}]{roskar+10}
{Ro{\v{s}}kar} R.,  {Debattista} V.~P.,  {Brooks} A.~M.,  {Quinn} T.~R.,
  {Brook} C.~B.,  {Governato} F.,  {Dalcanton} J.~J.,   {Wadsley} J.,  2010,
  \mn@doi [\mnras] {10.1111/j.1365-2966.2010.17178.x}, \href
  {https://ui.adsabs.harvard.edu/abs/2010MNRAS.408..783R} {408, 783}

\bibitem[\protect\citeauthoryear{{Ro{\v{s}}kar}, {Debattista}, {Quinn}  \&
  {Wadsley}}{{Ro{\v{s}}kar} et~al.}{2012}]{Roskar+12}
{Ro{\v{s}}kar} R.,  {Debattista} V.~P.,  {Quinn} T.~R.,   {Wadsley} J.,  2012,
  \mn@doi [\mnras] {10.1111/j.1365-2966.2012.21860.x}, \href
  {https://ui.adsabs.harvard.edu/abs/2012MNRAS.426.2089R} {426, 2089}

\bibitem[\protect\citeauthoryear{{Sancisi}}{{Sancisi}}{1976}]{sancisi}
{Sancisi} R.,  1976, \aap, \href
  {https://ui.adsabs.harvard.edu/abs/1976A&A....53..159S} {53, 159}

\bibitem[\protect\citeauthoryear{{Sancisi}, {Fraternali}, {Oosterloo}  \& {van
  der Hulst}}{{Sancisi} et~al.}{2008}]{2008A&ARv..15..189S}
{Sancisi} R.,  {Fraternali} F.,  {Oosterloo} T.,   {van der Hulst} T.,  2008,
  \mn@doi [\aapr] {10.1007/s00159-008-0010-0}, \href
  {https://ui.adsabs.harvard.edu/abs/2008A&ARv..15..189S} {15, 189}

\bibitem[\protect\citeauthoryear{Sellwood}{Sellwood}{2013}]{Sellwood+13}
Sellwood J.~A.,  2013, Dynamics of Disks and Warps.
Springer Netherlands, Dordrecht, pp 923--983,
  \mn@doi{10.1007/978-94-007-5612-0_18}, \url
  {https://doi.org/10.1007/978-94-007-5612-0_18}

\bibitem[\protect\citeauthoryear{{Sellwood} \& {Binney}}{{Sellwood} \&
  {Binney}}{2002}]{SellwoodBinney}
{Sellwood} J.~A.,  {Binney} J.~J.,  2002, \mn@doi [\mnras]
  {10.1046/j.1365-8711.2002.05806.x}, \href
  {https://ui.adsabs.harvard.edu/abs/2002MNRAS.336..785S} {336, 785}

\bibitem[\protect\citeauthoryear{{Semczuk}, {{\L}okas}, {D'Onghia},
  {Athanassoula}, {Debattista}  \& {Hernquist}}{{Semczuk}
  et~al.}{2020}]{Semczuk+20}
{Semczuk} M.,  {{\L}okas} E.~L.,  {D'Onghia} E.,  {Athanassoula} E.,
  {Debattista} V.~P.,   {Hernquist} L.,  2020, \mn@doi [\mnras]
  {10.1093/mnras/staa2609}, \href
  {https://ui.adsabs.harvard.edu/abs/2020MNRAS.498.3535S} {498, 3535}

\bibitem[\protect\citeauthoryear{{Sharma} \& {Steinmetz}}{{Sharma} \&
  {Steinmetz}}{2005}]{Sharma+05}
{Sharma} S.,  {Steinmetz} M.,  2005, \mn@doi [\apj] {10.1086/430660}, \href
  {https://ui.adsabs.harvard.edu/abs/2005ApJ...628...21S} {628, 21}

\bibitem[\protect\citeauthoryear{{Sharma} et~al.,}{{Sharma}
  et~al.}{2021}]{Sharma+21}
{Sharma} S.,  et~al., 2021, \mn@doi [\mnras] {10.1093/mnras/stab1086}, \href
  {https://ui.adsabs.harvard.edu/abs/2021MNRAS.506.1761S} {506, 1761}

\bibitem[\protect\citeauthoryear{{Shen}, {Wadsley}  \& {Stinson}}{{Shen}
  et~al.}{2010}]{Shen+2010}
{Shen} S.,  {Wadsley} J.,   {Stinson} G.,  2010, \mn@doi [\mnras]
  {10.1111/j.1365-2966.2010.17047.x}, \href
  {https://ui.adsabs.harvard.edu/abs/2010MNRAS.407.1581S} {407, 1581}

\bibitem[\protect\citeauthoryear{{Solway}, {Sellwood}  \&
  {Sch{\"o}nrich}}{{Solway} et~al.}{2012}]{solway+12}
{Solway} M.,  {Sellwood} J.~A.,   {Sch{\"o}nrich} R.,  2012, \mn@doi [\mnras]
  {10.1111/j.1365-2966.2012.20712.x}, \href
  {https://ui.adsabs.harvard.edu/abs/2012MNRAS.422.1363S} {422, 1363}

\bibitem[\protect\citeauthoryear{{Spavone}, {Iodice}, {Arnaboldi}, {Gerhard},
  {Saglia}  \& {Longo}}{{Spavone} et~al.}{2010}]{Spavone+10}
{Spavone} M.,  {Iodice} E.,  {Arnaboldi} M.,  {Gerhard} O.,  {Saglia} R.,
  {Longo} G.,  2010, \mn@doi [\apj] {10.1088/0004-637X/714/2/1081}, \href
  {https://ui.adsabs.harvard.edu/abs/2010ApJ...714.1081S} {714, 1081}

\bibitem[\protect\citeauthoryear{{Starkenburg}, {Sales}, {Genel},
  {Manzano-King}, {Canalizo}  \& {Hernquist}}{{Starkenburg}
  et~al.}{2019}]{Starkenburg+19}
{Starkenburg} T.~K.,  {Sales} L.~V.,  {Genel} S.,  {Manzano-King} C.,
  {Canalizo} G.,   {Hernquist} L.,  2019, \mn@doi [\apj]
  {10.3847/1538-4357/ab2128}, \href
  {https://ui.adsabs.harvard.edu/abs/2019ApJ...878..143S} {878, 143}

\bibitem[\protect\citeauthoryear{{Stewart}, {Kaufmann}, {Bullock}, {Barton},
  {Maller}, {Diemand}  \& {Wadsley}}{{Stewart} et~al.}{2011}]{Stewart+11}
{Stewart} K.~R.,  {Kaufmann} T.,  {Bullock} J.~S.,  {Barton} E.~J.,  {Maller}
  A.~H.,  {Diemand} J.,   {Wadsley} J.,  2011, \mn@doi [\apj]
  {10.1088/0004-637X/738/1/39}, \href
  {https://ui.adsabs.harvard.edu/abs/2011ApJ...738...39S} {738, 39}

\bibitem[\protect\citeauthoryear{{Stinson}, {Seth}, {Katz}, {Wadsley},
  {Governato}  \& {Quinn}}{{Stinson} et~al.}{2006}]{Stinson+2006}
{Stinson} G.,  {Seth} A.,  {Katz} N.,  {Wadsley} J.,  {Governato} F.,   {Quinn}
  T.,  2006, \mn@doi [\mnras] {10.1111/j.1365-2966.2006.11097.x}, \href
  {https://ui.adsabs.harvard.edu/abs/2006MNRAS.373.1074S} {373, 1074}

\bibitem[\protect\citeauthoryear{{Thilker} et~al.,}{{Thilker}
  et~al.}{2005}]{2005ApJ...619L..79T}
{Thilker} D.~A.,  et~al., 2005, \mn@doi [\apjl] {10.1086/425251}, \href
  {https://ui.adsabs.harvard.edu/abs/2005ApJ...619L..79T} {619, L79}

\bibitem[\protect\citeauthoryear{{Twarog}}{{Twarog}}{1980}]{1980ApJ...242..242T}
{Twarog} B.~A.,  1980, \mn@doi [\apj] {10.1086/158460}, \href
  {https://ui.adsabs.harvard.edu/abs/1980ApJ...242..242T} {242, 242}

\bibitem[\protect\citeauthoryear{{Vasiliev}}{{Vasiliev}}{2019}]{agama}
{Vasiliev} E.,  2019, \mn@doi [\mnras] {10.1093/mnras/sty2672}, \href
  {https://ui.adsabs.harvard.edu/abs/2019MNRAS.482.1525V} {482, 1525}

\bibitem[\protect\citeauthoryear{{Wadsley}, {Stadel}  \& {Quinn}}{{Wadsley}
  et~al.}{2004}]{Wadsley+2004}
{Wadsley} J.~W.,  {Stadel} J.,   {Quinn} T.,  2004, \mn@doi [\na]
  {10.1016/j.newast.2003.08.004}, \href
  {https://ui.adsabs.harvard.edu/abs/2004NewA....9..137W} {9, 137}

\bibitem[\protect\citeauthoryear{{Wang}, {Yang}, {Mo}, {Li}, {van den Bosch},
  {Fan}  \& {Chen}}{{Wang} et~al.}{2008}]{Wang+08}
{Wang} Y.,  {Yang} X.,  {Mo} H.~J.,  {Li} C.,  {van den Bosch} F.~C.,  {Fan}
  Z.,   {Chen} X.,  2008, \mn@doi [\mnras] {10.1111/j.1365-2966.2008.12927.x},
  \href {https://ui.adsabs.harvard.edu/abs/2008MNRAS.385.1511W} {385, 1511}

\bibitem[\protect\citeauthoryear{{Wang}, {Park}, {Hwang}  \& {Chen}}{{Wang}
  et~al.}{2010}]{Wang+10}
{Wang} Y.,  {Park} C.,  {Hwang} H.~S.,   {Chen} X.,  2010, \mn@doi [\apj]
  {10.1088/0004-637X/718/2/762}, \href
  {https://ui.adsabs.harvard.edu/abs/2010ApJ...718..762W} {718, 762}

\bibitem[\protect\citeauthoryear{{Weaver} \& {Williams}}{{Weaver} \&
  {Williams}}{1974}]{weaver_williams}
{Weaver} H.,  {Williams} D.~R.~W.,  1974, \aaps, \href
  {https://ui.adsabs.harvard.edu/abs/1974A&AS...17..251W} {17, 251}

\bibitem[\protect\citeauthoryear{{Weinberg}}{{Weinberg}}{1998}]{Weinberg98}
{Weinberg} M.~D.,  1998, \mn@doi [\mnras] {10.1046/j.1365-8711.1998.01456.x},
  \href {https://ui.adsabs.harvard.edu/abs/1998MNRAS.297..101W} {297, 101}

\bibitem[\protect\citeauthoryear{{Westmeier}, {Braun}  \&
  {Koribalski}}{{Westmeier} et~al.}{2011}]{westmeier}
{Westmeier} T.,  {Braun} R.,   {Koribalski} B.~S.,  2011, \mn@doi [\mnras]
  {10.1111/j.1365-2966.2010.17596.x}, \href
  {https://ui.adsabs.harvard.edu/abs/2011MNRAS.410.2217W} {410, 2217}

\bibitem[\protect\citeauthoryear{{Zaritsky} \& {Christlein}}{{Zaritsky} \&
  {Christlein}}{2007}]{zaritsky+07}
{Zaritsky} D.,  {Christlein} D.,  2007, \mn@doi [\aj] {10.1086/518238}, \href
  {https://ui.adsabs.harvard.edu/abs/2007AJ....134..135Z} {134, 135}

\bibitem[\protect\citeauthoryear{{Zschaechner}, {Rand}  \&
  {Walterbos}}{{Zschaechner} et~al.}{2015}]{Zschaechner}
{Zschaechner} L.~K.,  {Rand} R.~J.,   {Walterbos} R.,  2015, \mn@doi [\apj]
  {10.1088/0004-637X/799/1/61}, \href
  {https://ui.adsabs.harvard.edu/abs/2015ApJ...799...61Z} {799, 61}

\bibitem[\protect\citeauthoryear{{van de Voort}, {Davis}, {Kere{\v{s}}},
  {Quataert}, {Faucher-Gigu{\`e}re}  \& {Hopkins}}{{van de Voort}
  et~al.}{2015}]{vandeVoort+15}
{van de Voort} F.,  {Davis} T.~A.,  {Kere{\v{s}}} D.,  {Quataert} E.,
  {Faucher-Gigu{\`e}re} C.-A.,   {Hopkins} P.~F.,  2015, \mn@doi [\mnras]
  {10.1093/mnras/stv1217}, \href
  {https://ui.adsabs.harvard.edu/abs/2015MNRAS.451.3269V} {451, 3269}

\bibitem[\protect\citeauthoryear{{van den Bosch}, {Abel}, {Croft}, {Hernquist}
  \& {White}}{{van den Bosch} et~al.}{2002}]{bosch+02}
{van den Bosch} F.~C.,  {Abel} T.,  {Croft} R. A.~C.,  {Hernquist} L.,
  {White} S. D.~M.,  2002, \mn@doi [\apj] {10.1086/341619}, \href
  {https://ui.adsabs.harvard.edu/abs/2002ApJ...576...21V} {576, 21}

\makeatother
\end{thebibliography}
	
	
\appendix
\section{Supplemental simulation}
\label{sec:appendix}
The fiducial simulation has a low star formation threshold to promote the formation of stars in the warp. To reaffirm that stars can still form in the warp, settle into the disc, and reach the Solar Neighbourhood regardless of the star formation threshold, we repeat the analysis performed in Section~\ref{section:dyn_evo} on a supplemental simulation. This simulation has the same initial conditions as the fiducial simulation, however, it uses different subgrid physics (absence of metal-line cooling), has a higher density threshold (by two orders of magnitude), and the stellar feedback has more energy coupling to the gas ($4\times10^{50} \rm{erg}$).

\subsection{Warp evolution}
\begin{figure}
\includegraphics[width=1\columnwidth]{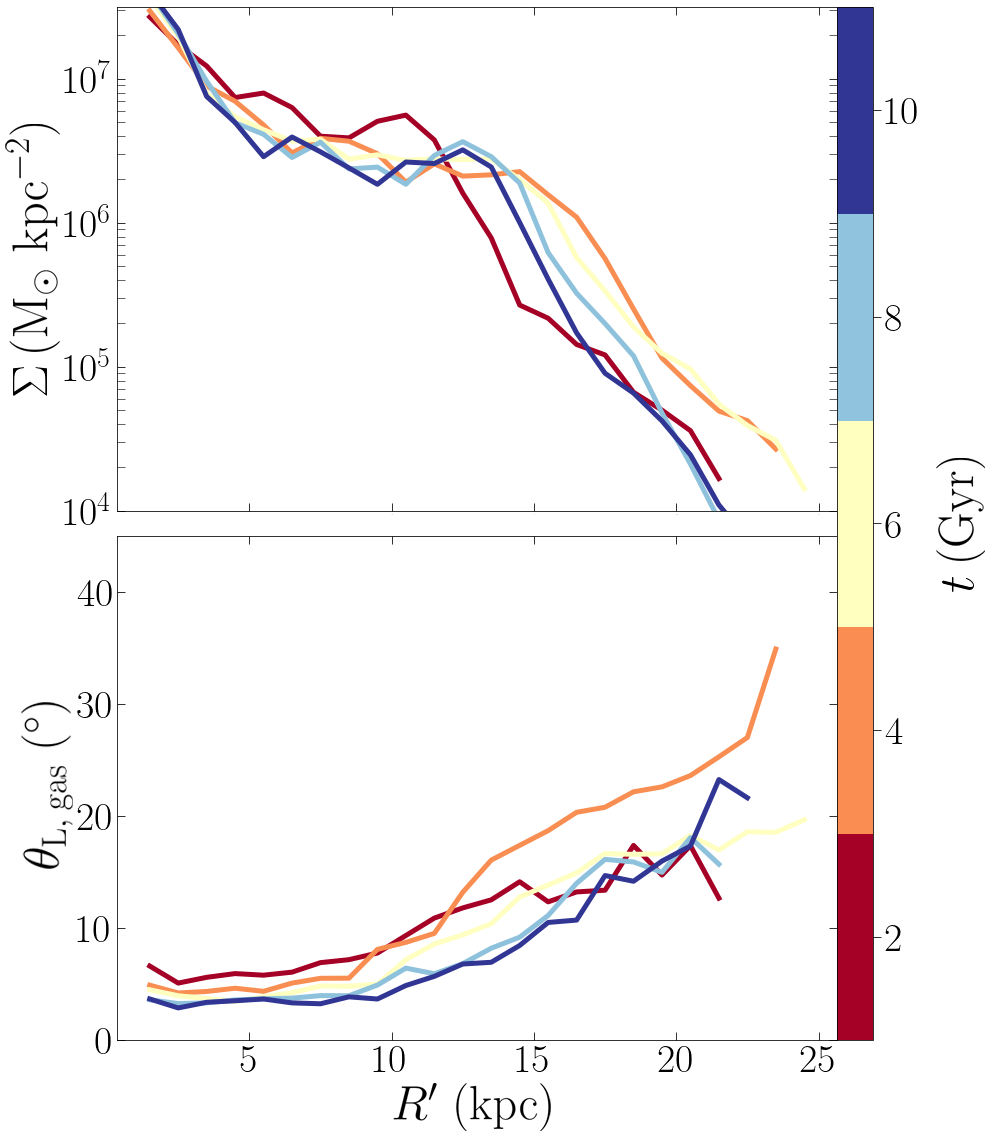}  
\caption{Profiles of the surface density, $\Sigma$, (top) and $\thetajg$ (bottom) in the cold gas of the supplemental simulation at different times (colour), where $R^{\prime}$ is the cylindrical radius in the cold gas plane at each annulus. 
}
\label{fig:warp_evo_HT}
\end{figure}
In Fig.~\ref{fig:warp_evo_HT} we show the profiles of the surface density, $\Sigma$, (top) and of $\thetajg$ (bottom) for the cold gas disc in the supplemental simulation at five different times (colour), where $R^{\prime}$ is defined as the cylindrical radius in the cold gas plane at each annulus. Over the model's evolution, the inclination of the cold gas warp decreases by a factor $\sim 2$ at $R=15\kpc$.
The warp appears to also decrease in mass and size as the $\Sigma$ profile decreases beyond $15 \kpc$ and drops from $R^{\prime} \sim 25\kpc$ to $R^{\prime} \sim 20\kpc$ by $t=10\Gyr$.
\begin{figure*}
\includegraphics[width=1\linewidth]{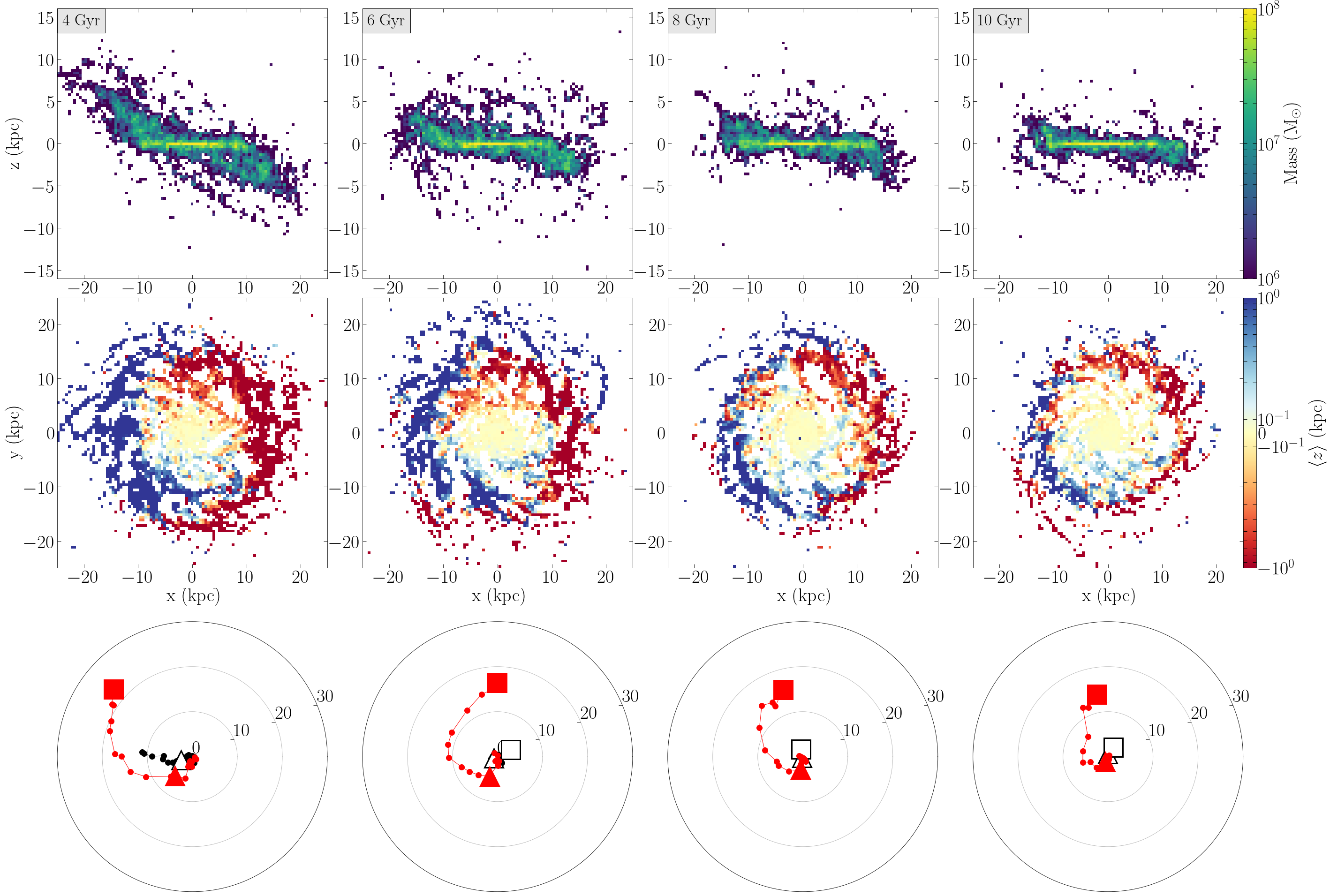}  
\caption{The structure of the gas warp at different times (upper left corner) in the supplemental simulation (see Section~\ref{section:Simulation}). Top row: The edge-on column density distribution of cold gas (T$\leq 50,000$ K) in the simulation. Middle row: the face-on mean height, $\left<z\right>$, distribution of cold gas (T$\leq 50,000$ K) in the simulation. Bottom row: The Briggs figures for the cold gas (red) and stellar (black) discs. There are two distinct markers that show values at $R=10 \kpc$ (triangle marker) and at $R=20 \kpc$ (square marker).
}
\label{fig:warp_briggs_HT}
\end{figure*}
The extent of the gas warp in the supplemental simulation is shown in the top rows Fig.~\ref{fig:warp_briggs_HT} where we present the edge-on column density of cold gas at four different times. In the span of $6 \Gyr$ the gas warp decreases in both radial and vertical extent, becoming less inclined relative to the disc. In the bottom row of Figs.~\ref{fig:warp_briggs_HT} we present Briggs figures for the stellar (black) and cold gas (red) discs, where the triangle (square) marker represents $R=10 \kpc$ ($R=20 \kpc$). The $\thetaj$ and $\phi_L$ angles are calculated for the mean angular momentum vector in each bin of a cylindrical grid with $0 \leq R / \kpc \leq 20 \kpc$ and $\Delta R=1 \kpc$. The cold gas warp experiences a $\sim6\degrees$ decrease in inclination over the $6 \Gyr$, while the stellar warp decreases in extent and then flattens similar to the stellar disc in the fiducial simulation.

\begin{figure}
\includegraphics[width=1\columnwidth]{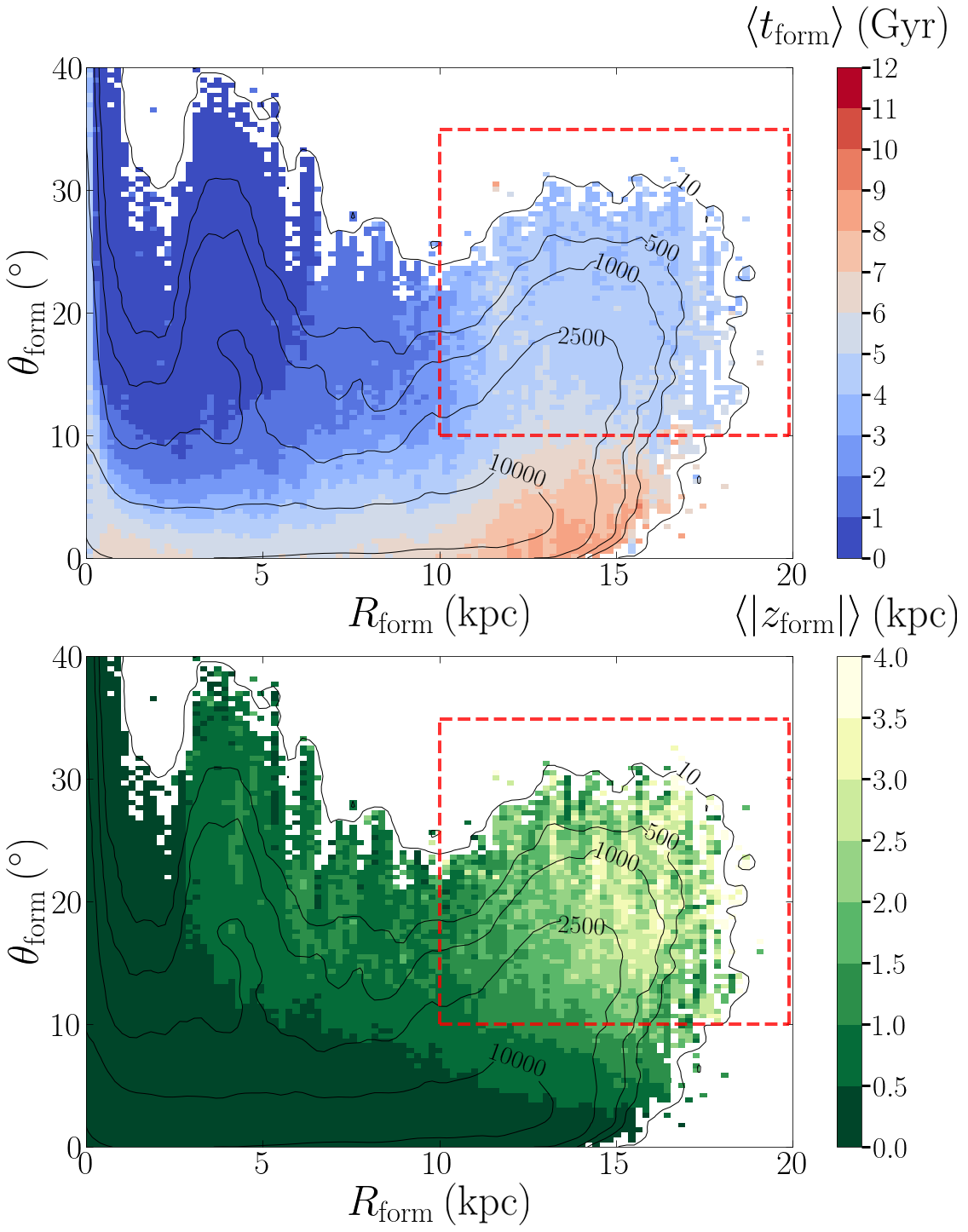}  
\caption{The distribution of stars in the formation space of the supplemental simulation, coloured by the mean time of formation (top) and by the mean absolute formation height, $\avgbin{|\zf|}$, (bottom). Bins that contain less than 10 stellar particles are not shown. The black lines show the number counts in the formation space for both panels. We define stars formed in the warp as those with $\thetaf\geq 10\degrees$ and $\rxyf \geq 10 \kpc$, the "tail-like" region outlined by the red square. A population of stars that was formed in an early, transient warp at low radii ($\rxyf \leq 5 \kpc$) and high inclinations relative to the disc ($\thetaf\geq 10\degrees$) is not included in our warp star population.
}
\label{fig:HT_form}
\end{figure}

Similar to the fiducial simulation, we record the phase-space coordinates and time at formation, $\tform$, for every star in the supplemental simulation to compute $\rxyf$ and $\thetaf$. We use the same warp star definition as in Section~\ref{subsection:warp_stars} to define the primary warp population in the formation space. Fig.~\ref{fig:HT_form} shows the distribution of $\avgbin{\tform}$ (top) and $\avgbin{|\zf|}$ (bottom) in the formation space.  We observe a familiar "tail-like" region at $\rxyf > 10\kpc$ (outlined by a red square). However the $|\zf|$ of the enclosed population increases with decreasing \avgbin{\tform}. This population forms throughout the model's evolution starting from $2 \Gyr$ and lasting till the end of the simulation, at $10 \Gyr$, however, the star formation rate in the warp greatly decreases after $t=6\Gyr$. This population is highly inclined ($\thetaf>10\degrees$) and is formed on the outskirts of the disc; thus we define the primary warp population in the supplemental simulation with the same conditions as in the fiducial simulation: $\rxyf\geq10 \kpc$ and $\thetaf\geq 10 \degrees$. We observe the in-situ main disc population ($\rxyf \leq 10 \kpc$ and $\thetaf \leq 10 \degrees$), and a similar old warp population in the "hump-like" region ($2.5 \leq \rxyf / \kpc \leq 7$ and $\thetaf \geq 15 \degrees$). This early warp population derives from a short-lived warp epoch when the model is still settling, and we do not include it in our analysis of the warp as we did in the fiducial simulation. We do not observe any changes in the following results when neglecting this population.

\subsection{Warp star settling and distribution}
\begin{figure}
\includegraphics[width=1\linewidth]{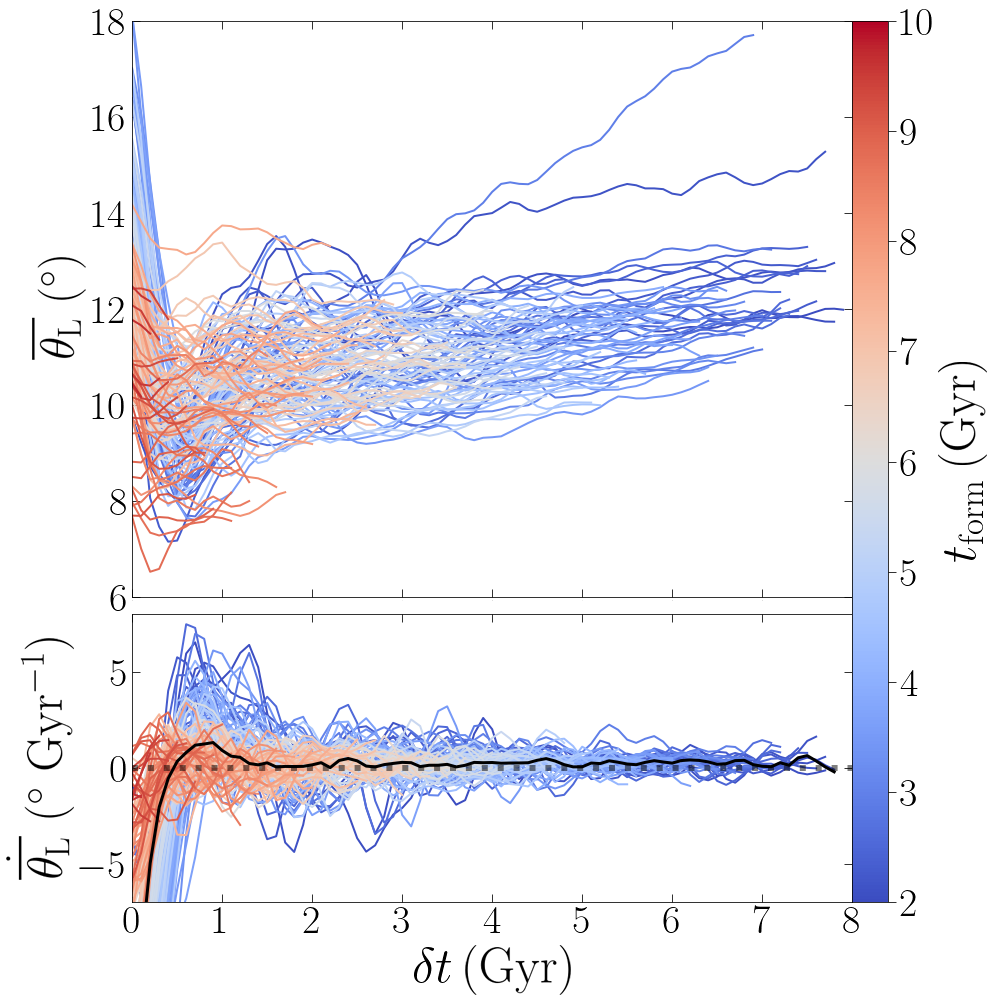}
\caption{Top: evolution of $\avg{\thetaj}$ for all mono-age warp populations formed in the supplemental simulation before $\tform\leq 8 \Gyr$, where $\dt$ is the time since a population's formation. Each curve is coloured by $\tform$. A 1D Gaussian filter with a mask size of $w=0.5\Gyr$ and standard deviation of $\sigma=0.1\Gyr$ is applied to the evolution at each $\dt$. Bottom: evolution of the rate of change of $\avg{\thetaj}$, $\Dthetaj$, for the same mono-age warp populations. The rates of change are calculated from the smoothed evolution curves. The solid black line represents the median rate of change between all mono-age populations which has a tilting time of $\ttilt\sim0.5\Gyr$. The dotted horizontal line indicates $\Dthetaj=0\degrees\Gyr^{-1}$.
}
\label{fig:HT_metheta}
\end{figure}
The top panel of Fig.~\ref{fig:HT_metheta} presents the $\avg{\thetaj}$ (Eq.~\ref{eq:theta_avg_L}) evolution for all mono-age warp populations in the supplemental simulation that formed during $2\leq \tform / \Gyr\leq 8$, in bins of $\Delta \tform=50 \Myr$. 
All warp populations experience a rapid drop in $\avg{\thetaj}$ by $\dt \sim 0.7 \Gyr$, followed by a smaller and gentler rise. The bottom panel shows the rate of change of $\avg{\thetaj}$, $\Dthetaj$, for the same populations. The horizontal dotted line represents $\Dthetaj=0\degrees \Gyr^{-1}$. We observe that $\Dthetaj$ starts out negative for all populations and quickly plateaus at a nearly constant value of $\Dthetaj\sim0.5\deg \Gyr^{-1}$. The tilting of the warp stars in the supplemental simulation appear to be very similar to the ones in the fiducial simulation. However, the settling timescale is noticeably faster as the median rate of change reaches $0\degrees\Gyr^{-1}$ by $\delta t=0.5\Gyr$ and a positive gradient can be observed with younger stars forming less inclined to the disc.
\begin{figure*}
\includegraphics[width=1\linewidth]{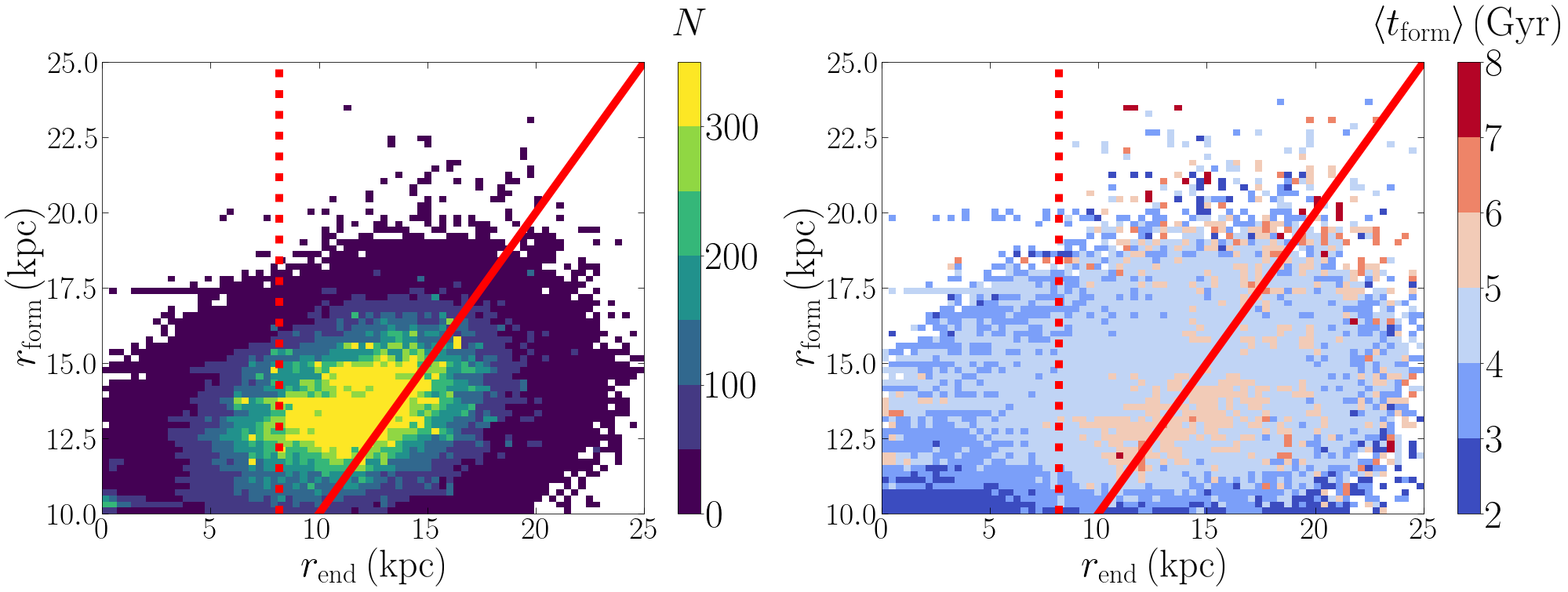}  
\caption{Distribution of spherical formation radius, $\rform$, versus the spherical radius at the end of the supplemental simulation, $\re$, for warp stars coloured by the number (left) and by the mean time of formation, $\tform$ (right). The diagonal solid line indicates $\rform=\re$. Stars that are below the $\rform=\re$ line comprise $30\%$ of the total warp star sample. The vertical dotted line indicates the location of the Solar annulus.}
\label{fig:HT_r_rform}
\end{figure*}
In Fig~\ref{fig:HT_r_rform} we look at the relation between the formation radius, $\rform$, and the final radius, $\re$, for all warp stars, to confirm that inward migration still takes place in the supplemental simulation. The left panel shows that 70\% of warp stars move inwards, which is similar to the fiducial simulation. The \avgbin{\tform} distribution in the right panel shows that stars older than $\avgbin{\tform}\leq6\Gyr$ are able to reach the Solar annulus.
\begin{figure*}
\includegraphics[width=1\linewidth]{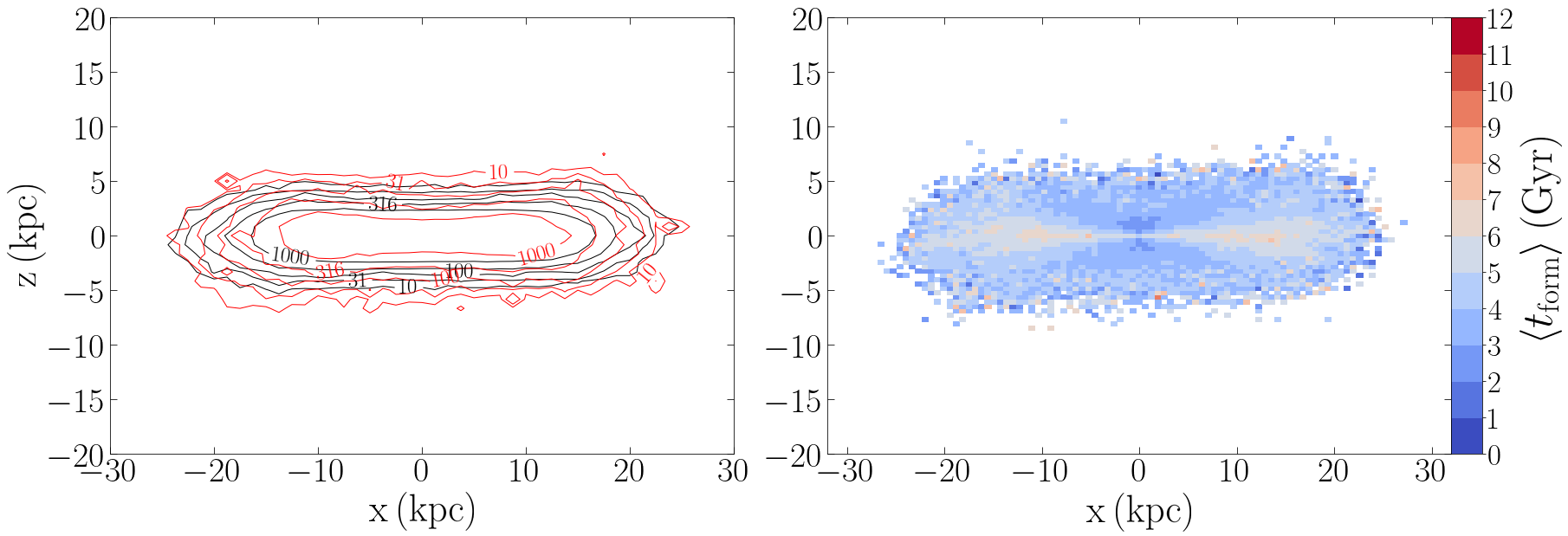}
\caption{Edge-on views of the supplemental simulation with a high star formation threshold (see Section~\ref{section:Simulation}) at $10 \Gyr$. Left column: number density contour plots of the warp (red contours) and main disc (black contours) populations. Right column: distribution of the mean formation time, $\left< \tform\right>$ for all stars formed throughout the simulation. We observe a positive vertical gradient in $\left< \tform \right>$.
}
\label{fig:edge_on_HT}
\end{figure*}

Fig.~\ref{fig:edge_on_HT} presents the edge-on distributions of warp and in-situ stars at $t=10\Gyr$ in the supplemental simulation. In the left panel, the contours show the number density distribution of warp (red) and in-situ (black) stars. Similar to the fiducial simulation, the warp stars in the supplemental simulation occupy the geometric thick disc, however, we observe an inversion of the vertical age profile as the $\avgbin{\tform}$ appears to decrease with $|z|$; this profile is contrary to the one in the fiducial simulation.

We demonstrated that warp stars form, settle, and migrate inwards in simulations regardless of their stellar feedback, gas density threshold, warp morphology, and gas cooling physics. We conclude that the negative age gradient observed in the fiducial simulation occurs due to the relatively high star formation rate in the warp, as well as the growing warp.

\bsp	

\end{document}